\documentclass[]{aastex701}

\newcommand{\ilam}{erg cm$^{-2}$ s$^{-1}$ sr$^{-1}$ \AA$^{-1}$}
\begin{document}

\title{The Atmospheric Response to Large Electron Beam Fluxes in Solar Flares III: Comprehensive  Modeling of the Brightest Observed Near-Ultraviolet Continuum Source in an X9 Solar Flare}

\author[0000-0001-7458-1176]{Adam F. Kowalski}
\affiliation{National Solar Observatory, University of Colorado Boulder, 3665 Discovery Drive, Boulder, CO 80303, USA}
\affiliation{Department of Astrophysical and Planetary Sciences, University of Colorado, Boulder, 2000 Colorado Ave, CO 80305, USA}
\affiliation{Laboratory for Atmospheric and Space Physics, University of Colorado Boulder, 3665 Discovery Drive, Boulder, CO 80303, USA.}
\email{adam.f.kowalski@colorado.edu}

%% Use the \collaboration command to identify collaborations. This command
%% takes an optional argument that is either a number or the word "all"
%% which tells the compiler how many of the authors above the command to
%% show. For example "\collaboration[all]{(DELVE Collaboration)}" wil include
%% all the authors above this command.
%%
%% Mark off the abstract in the ``abstract'' environment. 
\begin{abstract}
I report on the high resolution spectra of the remarkable X9 solar flare of 2024 Oct 03 (SOL2024-10-03T12:08) and evaluate the extent to which nonthermal electron beams that generate dense chromospheric condensations  can power very bright kernels in solar flares. 
 1D Radiative-hydrodynamic models predict extreme H$\alpha$ near-wing broadening, bright continuum intensities, and  a rapid Fe II red wing asymmetry evolution at the brightest NUV continuum source in the flare.   Detailed comparisons to the spectral observations reveal that the H$\alpha$ line is too broad, the Fe II red wing is too bright, and the NUV continuum decays too slowly in a fiducial high-flux beam model.  However, chromospheric condensations with maximum electron densities of $n_e \approx 5 \times 10^{14}$ cm$^{-3}$ and optical depths $\tau \approx 1$ in the near wing of H$\alpha$ are consistent with the observed intensity of a broad spectrum in the Southern ribbon.    Model similarities demonstrate that Fe \textsc{i} emission lines and the FUV continuum intensity can form at chromospheric heights during flares, but I find that the ratios of the NUV to FUV continuum intensities are generally too large in the models. This suggests that radiative-hydrodynamic models of chromospheric condensations cool through $T \approx 30,000$ K too rapidly.  The larger than expected FUV continuum intensities are not nearly bright enough to explain recent stellar megaflare spectra from the Hubble Space Telescope.
\end{abstract}

% Atmospheres with densities as high as n_e 5e14 are consistent with the broadening.

%% Keywords should appear after the \end{abstract} command. 
%% The AAS Journals now uses Unified Astronomy Thesaurus (UAT) concepts:
%% https://astrothesaurus.org
%% You will be asked to selected these concepts during the submission process
%% but this old "keyword" functionality is maintained in case authors want
%% to include these concepts in their preprints.
%%
%% You can use the \uat command to link your UAT concepts back its source.
\keywords{\uat{Solar flares}{1496} --- \uat{Solar flare spectra}{1982} --- \uat{Solar ultraviolet emission}{1533} --- \uat{Hydrodynamical simulations}{767} ---  \uat{Solar physics}{1476}}

%% From the front matter, we move on to the body of the paper.
%% Sections are demarcated by \section and \subsection, respectively.
%% Observe the use of the LaTeX \label
%% command after the \subsection to give a symbolic KEY to the
%% subsection for cross-referencing in a \ref command.
%% You can use LaTeX's \ref and \label commands to keep track of
%% cross-references to sections, equations, tables, and figures.
%% That way, if you change the order of any elements, LaTeX will
%% automatically renumber them.

\section{Introduction} \label{sec:intro}

Solar flares are explosive brightenings that occur in sunspot groups with compact, mixed magnetic polarity distributions \citep{Toriumi2019}.   Non-potential magnetic energy above interacting sunspots is stored at lower coronal heights; when released catastrophically into the atmosphere, flare radiation occurs across the electromagnetic spectrum.  Most flares begin with an impulsive phase of nonthermal particle excitation of the relatively dense chromosphere. The resulting collection of individual chromospheric brightenings are organized into two or more ``ribbons'' that straddle a magnetic polarity inversion  \citep[e.g.,][]{Kosovichev2001, Qiu2010,Fletcher2011,Qiu2017, Kazachenko2022}.   ``Kernels'' with  enhanced brightness are often embedded within these ribbons \citep[e.g.,][]{Neidig1994,Asai2002, Krucker2011,KCF15, Namekata2022,Pietrow2024, Faber2025}.

  Kernels are especially prominent in  UV and optical continuum radiation \citep[e.g.,][]{Neidig1993, Neidig1994, Liu2007, Radz2007, Coyner2009, Maurya2009, KCF15}, and they exhibit small projected areas at the Sun \citep[$\approx 10^{16}$ cm$^2$ or smaller, e.g.,][]{Jess2008}.  They are sources of ``white-light'' continuum radiation and thus constitute an important contribution to the bolometric energy of solar flares \citep{Woods2004, Fletcher2007, Kretzschmar2011}.  Within the NUV wavelength range, they may account for a significant fraction ($\gtrsim 1/3$) of the spectral luminosity in solar and stellar flares \citep{Kowalski2017Broadening, Kowalski2022Frontiers, Kowalski2025}.   Slit spectra of the cores of kernels are unfortunately rare due to their spatial and temporal intermittency.  Furthermore, solar instrumentation does not yet fully resolve their fundamental scales  \citep{Dennis2009, Hudson2016}.  

It is far from understood how energy transport from coronal magnetic field reconnection region high above the brightest kernels can deliver the requisite power \citep{Fletcher2007, Krucker2011, Russell2024}.  Strong particle beams are a sensible explanation for kernel heating given that there are common similarities between the observed locations of nonthermal hard X-ray/gamma-ray radiation and the most prominent sources of optical and UV radiation \citep{Fletcher2007, Kleint2016, Battaglia2025}.  Powerful beams of electrons are thought to also produce bright and broad redshifted components as red-wing asymmetries, which appear in many chromospheric spectral lines  \citep{Ichimoto1984, Wuelser1989, Canfield1990,Wuelser1994, Graham2015, Graham2020, Namekata2022,Butler2024}.  High-velocity redshifts\footnote{I distinguish between red-wing asymmetries and red-core asymmetries, which can result from upflows \citep{Kuridze2015}.} likely originate in chromospheric condensations \citep{Livshits1981, Fisher1985V, Fisher1985VI, Fisher1985VII, Kowalski2017Mar29, Kowalski2022}, whereby nonthermal particle energy deposition generates a field-aligned thermal blast wave and causes hot chromospheric material to compress and cool as it propagates at $\approx20-100$ km s$^{-1}$ into the lower atmosphere.  Models without any nonthermal particle beams can also produce chromospheric condensations through thermal conductive transport \citep{Fisher1989,Kowalski2017Mar29,Ashfield2021}, which is initiated through a source of energy deposition that is presumably related to  magnetic reconnection processes \citep{Longcope2011, Haggerty2015}.  However, faint NUV and even fainter optical emergent continuum intensities are produced without  nonthermal particle heating at deeper chromospheric depths within and below the condensation \citep{Kowalski2017Mar29}.  Additionally, \citet{Canfield1984} argue that the Stark broadening of hydrogen lines is more sensitive to electron beam heating than thermal conduction.  Thus, it is critical that models of flare kernel heating are comprehensively tested against observations of both continuum intensities and  emission line shapes.  

Significant progress has been made in forward modeling optical and UV spectra of solar flares with electron beam heating, but several glaring discrepancies remain.  \citet{Kowalski2017Mar29} (hereafter Paper I) models several bright NUV continuum sources in the X1 solar flare of 2014-Mar-29 with the RADYN code.  This paper explores how  well high-flux density ($5\times 10^{11}$ erg cm$^{-2}$ s$^{-1}$; hereafter, 5F11) electron beam heating explains NUV flare spectra from the Interface Region Imaging Spectrograph \citep[IRIS;][]{DePontieu2014}.  In the models,  there are two Fe II emission line components: one at the rest wavelength and one redshifted satellite component. The redshifted component originates from a chromospheric condensation, and the rest-wavelength component originates from the stationary chromospheric flare layers just below the condensation.  They suggest that the prominence of the redshifts could be related to the relative optical depths within the chromospheric condensation:  an Fe II line with a larger optical depth displays a more prominent redshifted component.  A temporal average matching the exposure time of $\sim 8$~s is adequate to reproduce many of the details of the line profile and the flare NUV continuum intensity.  \citet{Graham2020} model another X1 flare using IRIS data with much higher temporal cadence.  At higher time-resolution, the model redshifted components of the Fe II lines are found to evolve too quickly (by factors of $2-3$).  The redshifted line component never attains a brighter peak than the component around rest wavelength, whereas the models predict a much brighter redshifted component.  Additionally, it is widely known that the widths of the Fe II and Mg II lines are poorly accounted for by self-consistent RHD models \citep{Kleint2017, Zhu2019}.  Either supersonic nonthermal broadening (``microturbulence'') parameters  \citep{Kowalski2017Mar29, SainzDalda23}, macroscale velocity gradients \citep{Kleint2017, SainzDalda23} that are not consistent with mass, momentum, and energy flow, or large pressure broadening enhancements \citep{Zhu2019} must be introduced without physical explanation.   

\citet{Kowalski2022} (hereafter Paper II) extends several RADYN model predictions to the hydrogen Balmer series in the hope that future spectral observations farther into the wings of these lines would clarify where the problems lie in the models.  Are 1D chromospheric condensations too dense, or are the stationary flare layers just below not heated enough?  The pressure broadening theory of the hydrogen Balmer series wing broadening is now rather well understood in the conditions of flare atmospheres \citep{Tremblay2009, Kowalski2017Broadening}.  Paper II shows how dense chromospheric condensations in the 1D RADYN models produce very optically thick and highly broadened Balmer H$\gamma$ and H$\alpha$ line profiles.   Detailed comparisons of models and observations of the brightest kernels in solar flares would constrain whether 1D models of chromospheric condensations heated by high fluxes of nonthermal electrons are too dense \citep{Graham2020}. However, all current Fabry-Perot H$\alpha$ imaging-spectroscopic data \citep[e.g.,][]{Kuridze2015,Rubio2016} are obtained with relatively low temporal resolution and very narrow wavelength coverage around the rest wavelength, and the higher order Balmer lines are rarely observed anymore \citep{Neidig1983, Acampa1982, Donati1985, KCF15, Ondrej1}.

This is the third paper in the series, which culminates in comparisons of models from Paper II to a remarkable data set of a large X-class flare during Solar Max 25.  This flare is unique in that both H$\alpha$ spectra in the wings and IRIS spectra of Fe II are available for several very bright kernels.  The paper is organized as follows.  In Section \ref{sec:data}, I describe the data and calibration.  In Section \ref{sec:rhd}, I summarize the radiative-hydrodynamic flare models that are used from Papers I and II.  Section \ref{sec:analysis} compares the models to the  light curves, chromospheric line profiles, and the NUV to FUV continuum ratios. Section \ref{sec:discussion} discusses the context of the results, and  Section \ref{sec:conclusions} summarizes and concludes.

\section{Data} \label{sec:data}

  We present the NUV spectral and imaging data from IRIS (Section \ref{sec:iris}) and the optical spectra from the Chinese H$\alpha$ Solar Explorer \citep[CHASE;][]{CHASEOverview1} (Section \ref{sec:chase}) during the X9 solar flare on 2024-Oct-03 (Section \ref{sec:overview}).

\subsection{IRIS Data} \label{sec:iris}

IRIS observes the Sun at high spatial, temporal, and spectral resolution in the NUV and FUV.  I retrieved the Level 2 calibrated data of the X9 flare from the Lockheed Martin Solar Astrophysics Laboratory archive\footnote{\url{https://iris.lmsal.com/search/}}. The spectral data\footnote{ The calibration of the IRIS data is the same as described in Namekata et al.\ 2025, in press.} are part of a high-cadence flare program (OBSID 4204700237) in sit-and-stare mode with a 0.8~s cadence and 0.3~s exposure times. Slit-jaw images at 1330 \AA\ (SJI 1330) have an 8~s cadence.  The spatial and spectral directions are binned by 2 pixels, thus giving a linear spatial dispersion of 0.\arcsec334 pix$^{-1}$ and a linear spectral dispersion of 0.02596 \AA\ pix$^{-1}$ (FUV) and 0.0509 \AA\ pix$^{-1}$ (NUV).  Spectral coverage spans 4 wavelength windows:  $\lambda = 1332.7 - 1336.8$ \AA\ (centered on C II 1336), 1400.3 -- 1406.3 \AA\ (centered on Si IV $\lambda$1403), 2793.6 -- 2800.0 \AA\ (centered on Mg II k $\lambda$2796), and 2813.3 -- 2816.3 \AA\ (centered on Fe II $\lambda$2814.5).  Level 3 data are produced using the standard SolarSoft IDL routines.  Inspection of the fiducial marks suggests, if any, a sub-pixel relative spatial offset between the FUV and NUV channels; thus I do not shift the spectra during generation of the Level 3 files. I check the accuracy of the wavelength calibration using the Fe I line with  $\lambda_0=2815.846$ \AA, which is prominently in absorption in the quiet Sun.  Intensity calibration of the spectra and slit-jaw images converts from corrected DN s$^{-1}$ to erg cm$^{-2}$ s$^{-1}$ sr$^{-1}$ \AA$^{-1}$; the procedure follows previous works \citep{Kowalski2019IRIS} and IRIS Technical Note 24 (Tian et al. 2014) and accounts for the binning factors and the time-dependent sensitivity with \texttt{iris\_get\_response.pro} (v005). Uncertainties are calculated with standard error propagation using the gain and read noise values reported in ITN 24 \citep[see also][]{Wulser2018}. 

According to ITN 22 and Section 1.3 of ITN 32, the header keywords for XCEN and YCEN may be inaccurate to about $\pm 5$\arcsec.  Thus, I adjust the IRIS spatial coordinates to those in 1600\AA\ images from the Atmospheric Imaging Assembly \citep[AIA;][]{Lemen2012} on the Solar Dynamics Observatory \citep[SDO;][]{Pesnell2012}. The level 2 SDO/AIA contextual data from the IRIS webpage are used to redefine the reference coordinates in the IRIS data.  Using a shift of $-3.\arcsec5$ in IRIS $x$ and a $-3.\arcsec0$ shift in IRIS $y$, I find that the flare ribbon features in SJI 1330 at UTC 12:13:49.5 align well with similar features in SDO/AIA 1600 at UTC 12:13:50.1, which is just before the SDO data saturate during the flare.

\subsection{CHASE Data} \label{sec:chase}
The recently deployed H$\alpha$ Imaging Spectrograph (HIS) on the CHASE satellite \citep{CHASEOverview1, CHASEOverview2, CHASEOverview3} makes rapid scans of the entire solar disk at $\lambda = 6559.43-6565.11$ \AA\ (around H$\alpha$ $\lambda 6562.82$) and at $\lambda = 6567.53-6569.71$ \AA\ (around Fe I $\lambda 6569.22$).
The Raster Scanning Mode (RSM) Level 1 \texttt{.fits} files of the X9 flare are downloaded from the CHASE public archive\footnote{\url{https://ssdc.nju.edu.cn/home}, \url{https://ssdc.nju.edu.cn/NdchaseSatellite}}.  The data were obtained in 2-pixel binning mode, resulting in a linear spectral dispersion of $0.04849$ \AA\ pix$^{-1}$ and a FWHM resolution of $0.072$ \AA.  The X9 flare is present in several full-disk scans, but the main impulsive phase observations occur during a scan from UTC 12:15:58 to 12:16:44.  Each scan with 2313 steps takes 46~s starting at the solar North pole \citep{CHASEOverview1}, with a $\sim 70$~s cadence.  The linear spatial dispersion is $1.\arcsec04$ per binned pixel in solar $x$ and $y$.
The Level 1 files are spectral-curvature corrected, dark subtracted, and flat-field corrected through the pipeline, but I take additional steps that are necessary for intensity calibration.

\begin{figure}
	\begin{center}
		\includegraphics[scale=0.15]{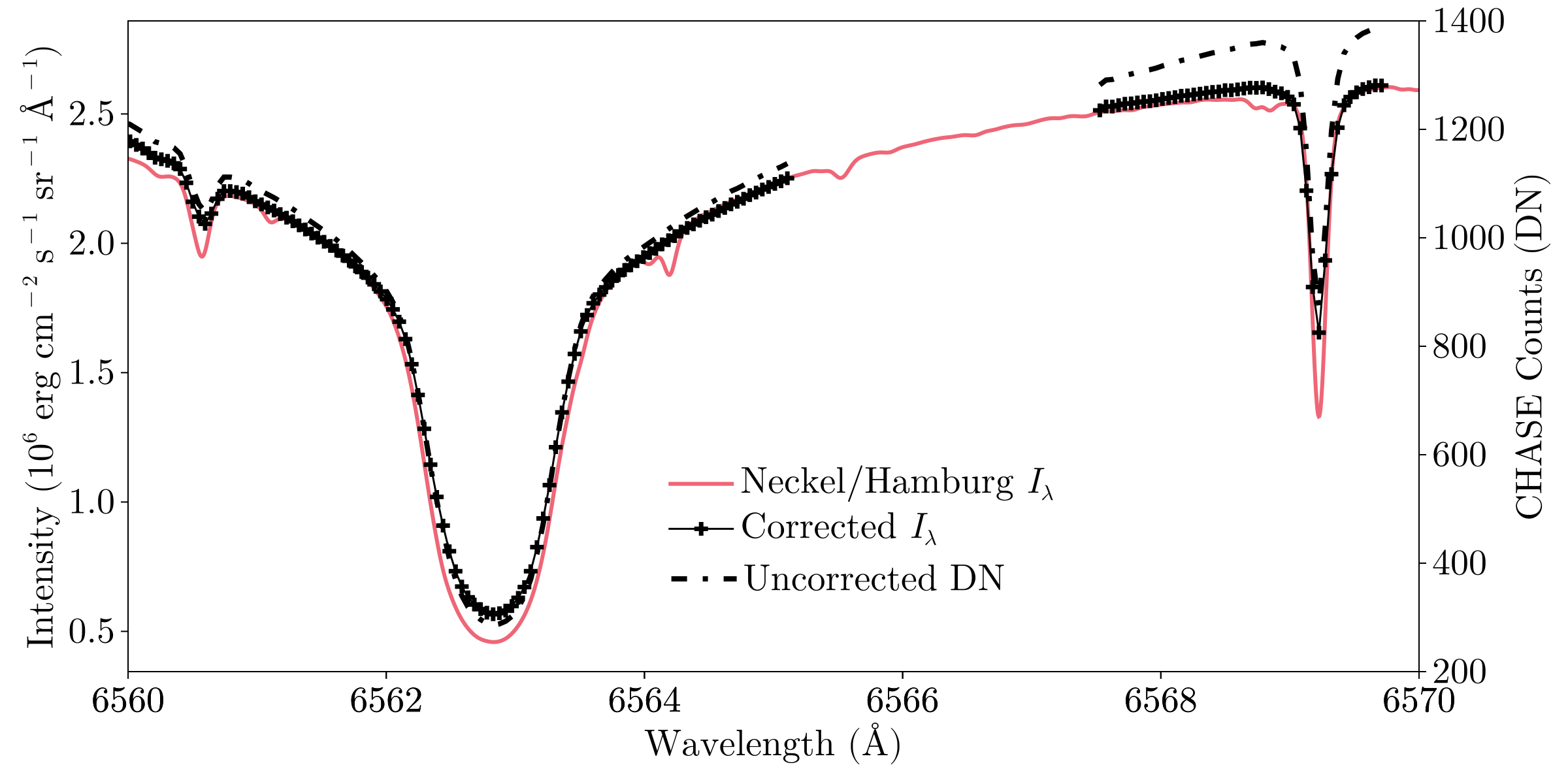}
		\caption{Calibration of the CHASE H$\alpha$ and Fe I spectra to specific intensity.  The corrections reproduce the detailed shape of the H$\alpha$ wings from the Neckel-Hamburg disk-center intensity atlas.   } \label{fig:chasecal}
	\end{center}
\end{figure}

\cite{CHASECal} describe a correction for sensitivity gradients over both the H$\alpha$ and Fe I spectral windows.  However, I use data after August 2022, when a rather significant optics correction was made\footnote{To my knowledge, this has not been formally documented in a peer-reviewed paper, but the changes are mentioned here: \url{https://ssdc.nju.edu.cn/web-service/software/Demo/CHASE_Lev1_Data_Processing_Demonstration_IDL.pdf} .}.  I examine the sensitivity correction by comparing median disk-center spectra at the time of the flare to the Neckel-Hamburg disk-center intensity spectrum  \citep{Neckel1999} (Figure \ref{fig:chasecal}).   A gradient is only present in the Fe I spectrum, whereas H$\alpha$ is satisfactorily corrected with a constant factor.  Thus I apply the following conversions from CHASE counts (DN) to spectral intensity. In the H$\alpha$ window, I divide by $0.000505$; in the Fe I window, I divide the counts with a wavelength dependent correction of $0.00051 + (\lambda - \lambda[1]) \times 0.00001$, where $\lambda[1]$ is the wavelength of the first pixel. This gradient correction has little effect on the comparisons to models that follow.

\subsection{Overview of the X9 Flare} \label{sec:overview}
The 2024-10-03 flare (NOAA identifier SOL2024-10-03T12:08) is one of the most luminous soft X-ray solar flares within the past two decades.  It peaked in the Extreme Ultraviolet and X-ray Sensor (EUXS) of GOES-16 on 2024-Oct-03 at UTC 12:18:52.  The background-subtracted peak irradiance corresponds to a GOES ($\lambda = 1-8$ \AA) class X9.0 flare.  The flare occurred in NOAA active region (AR) 13842 at a solar $\mu$ value of 0.92 (solar $x \approx 75\arcsec$, solar $y \approx -375\arcsec$).  \citet{Ding25} describe the trigger and detailed magnetic topology of non-conjugate colliding polarities within this active region.

\begin{figure}
			\gridline{\fig{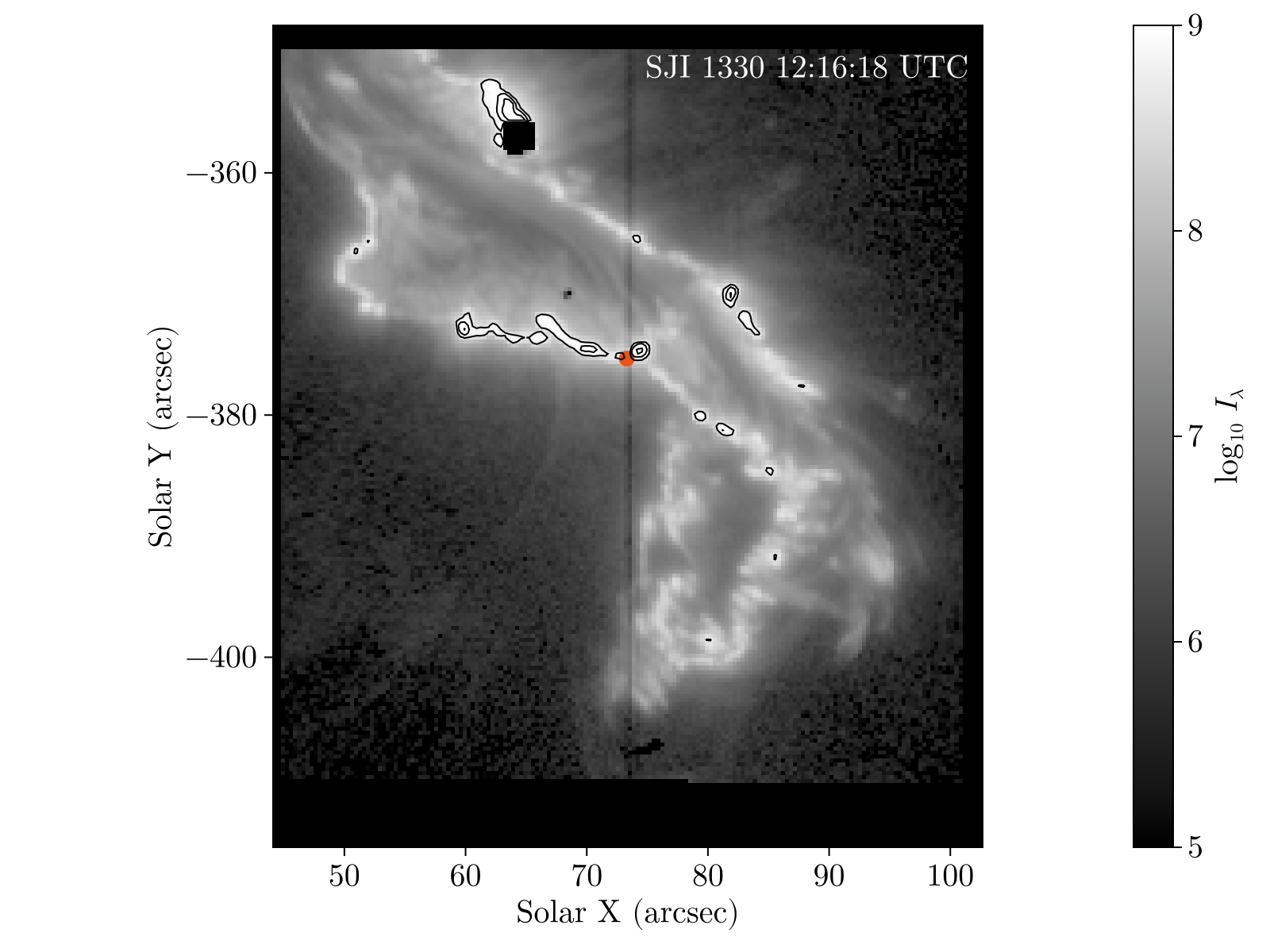}{0.6\textwidth}{(a)}}
	\gridline{\fig{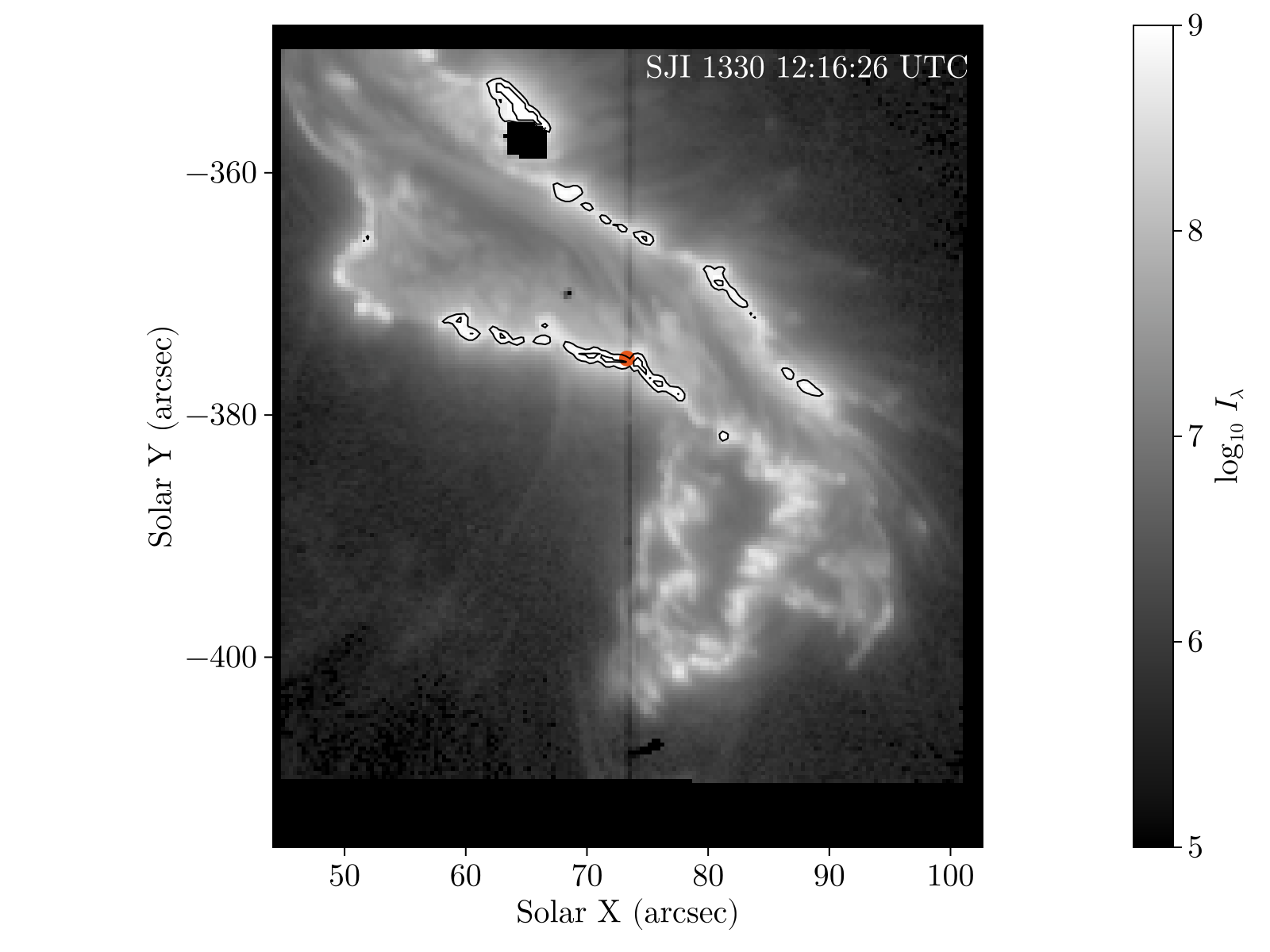}{0.6\textwidth}{(b)}}
	\caption{Contextual images of the flare ribbons  in IRIS SJI 1330.  The contours of $5\times 10^8$, $1.25 \times 10^9$, and $2.5 \times 10^9$ \ilam\ emphasize the locations of the kernels against the background intensity scaling, which is logarithmic (base-10). These images span the nearest times to the source of the brightest peak continuum intensity in the NUV spectra, indicated by a red circle.  The dark features are dust specks that are incorrectly removed during flat fielding (IRIS Technical Note 16).   A movie is available online.  The movie shows the bright Southern ribbon moving downward as it crosses the slit at the location of the red dot.  The same contours as in this figure are included in the movie of the flare.  The time range for the movie is 12:07:55 to  12:28:55.    \label{fig:context}}
\end{figure}

Figure \ref{fig:context}(a-b) contextualizes the brightest sources in IRIS FUV (SJI 1330) within the flare ribbons that cross the spectrograph slit.  At UTC 12:16:18, a bright circular source is located just to the upper right of the red dot, which indicates the slit position of the brightest NUV continuum source in the IRIS spectra (Section \ref{sec:brightest}).  By 8~s later, several bright sources cross the slit as the ribbons spread apart from a compact configuration along the polarity inversion line.

The two images in Figure \ref{fig:context}(a)-(b) span the times within the impulsive phase of the X9 flare:  The $E=50-84$ keV hard X-ray quicklook light curves from the Spectrometer and Telescope for Imaging X-rays \citep[STIX;][]{STIX1, STIX2} aboard Solar Orbiter \citep{Solo} exhibit several major bursts, and one burst peaks at UTC 12:16:24  \citep[accounting for the light travel time of 350~s; see also][]{Li2025}.  The $E=80-300$ keV hard X-ray images from the Hard X-ray Imager \citep{Su2019} on ASO-S \citep{Gan2019} in \citet{Li2025} indicate that one of the sources is located in the Southern ribbon close to the location of the bright NUV source in IRIS (Section \ref{sec:brightest}). Hard X-ray emission is understood to be nonthermal bremsstrahlung radiation from power-law electrons interacting with the chromospheric plasma \citep[e.g.,][]{Holman2011}.  Thus, it is reasonable to expect that the atmospheric heating is driven by strong electron beams at the times of the IRIS slit crossings.

\section{RHD Models and Calculated Quantities} \label{sec:rhd}

I use the RADYN code \citep{Carlsson1992B,Carlsson1995,Carlsson1997,Allred2015,Carlsson2023} solar flare models from Paper II to compare to the observations (Section \ref{sec:analysis}).  RADYN is a one-dimensional radiative-hydrodynamic code that has been widely used to model the impulsive phase atmospheric response to Coulomb heating from power-law electrons in solar and stellar flares \citep[e.g.,][]{Allred2005, Allred2006}.   Paper II presents several flare heating models with power-law indices and low-energy cutoffs that have been constrained by HXR observations of events around Solar Max 24.  To briefly summarize, Paper II discusses a model from Paper I with a constant injection of energy flux density of $5\times10^{11}$ erg cm$^{-2}$ s$^{-1}$ (5F11) for a duration of 15~s (hereafter, \texttt{c15s-5F11-25-4.2}), a model from \citet{Graham2020} with a constant injection of energy flux density of $10^{11}$ erg cm$^{-2}$ s$^{-1}$ (F11) for a duration of 20~s, a F11 model from \citet{Kuridze2015}, and a 5F11 model from \citep{Zhu2019} driven by a variable energy injection with a FWHM duration of 20~s (hereafter, \texttt{m20s-5F11-25-4}). In this model, the energy injection ramps up to a maximum flux density of $5\times10^{11}$ erg cm$^{-2}$ s$^{-1}$ at $t=9$~s, and the injection ramps down exponentially until the end of the simulation.  The energy injection follows the prescription in \citet{Aschwanden2004} \citep[see also][]{KAC24}.  These simulations are recalculated in Paper II using the newest hydrogen pressure broadening line profile functions \citep{Tremblay2009, Kowalski2017Broadening}.  

I focus the analysis on several calculated quantities from the \texttt{m20s-5F11-25-4} and the \texttt{c15s-5F11-25-4.2} models, which are driven by electron beam flux densities that are among the largest  in RHD models of solar flares.  These models also predict extreme Balmer line broadening  that I seek to test against the observations. The detailed continuum intensity predictions in the NUV at $\lambda=2830$ \AA\ (hereafter, C2830) are compared to the brightest NUV sources in the data (Sections \ref{sec:brightest}-\ref{sec:deconv}) in Section \ref{sec:lcanalysis}.  Spectra of Fe I and Fe II in the NUV are calculated from the models following Paper I and \citet{Kerr2024A}, and they are compared to the IRIS spectra in Section \ref{sec:feanalysis}.  The detailed spectra of H$\alpha$ are compared to the CHASE data of the brightest sources in the Southern ribbon in Section \ref{sec:haanalysis}.  The ratios of the model NUV continuum (C2830) intensity to the model FUV continuum intensity around $\lambda = 1407$ \AA\ (hereafter, C1407) are discussed in Section \ref{sec:ratioanalysis}.  All model intensity spectra are calculated along an emerging ray at a viewing angle $\theta$ that corresponds to  $\mu =0.95$ where $\mu = \cos \theta$.

\section{Analysis} \label{sec:analysis}

\subsection{The Brightest Flaring NUV Continuum Source} \label{sec:brightest}
\begin{figure}
	\begin{center}
\includegraphics[width=0.5\textwidth]{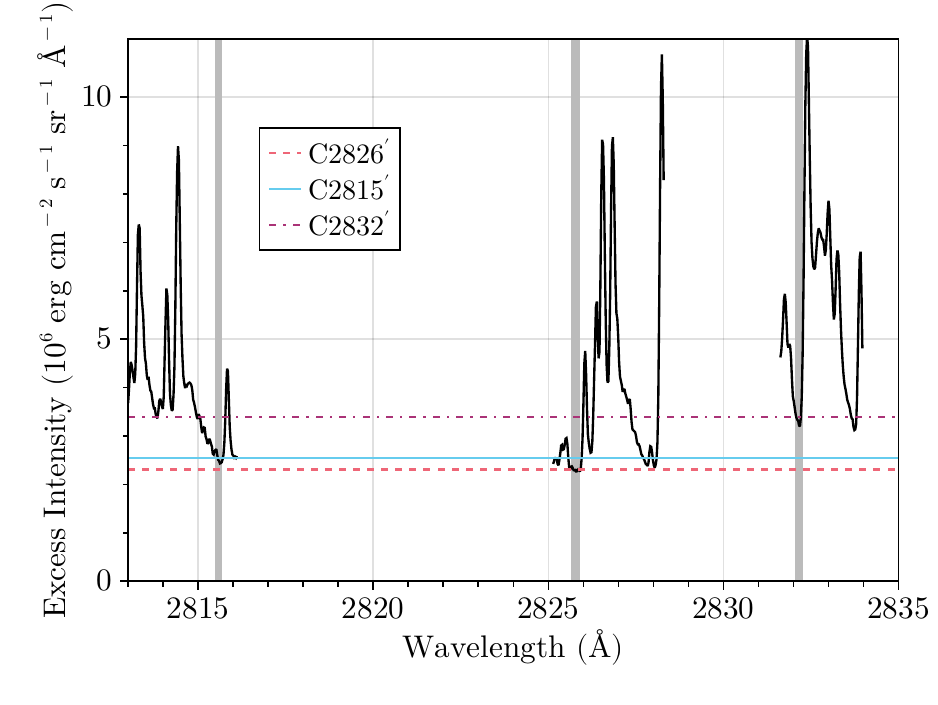}
	\caption{ Flare-only spectrum from the 2014-Mar-29 bright footpoint \#1 (``BFP1''; Paper I).  Horizontal lines show intensity levels within the C2815, C2826, and C2832 wavelength windows, which are indicated by grey shaded areas.  In this paper, we adopt C2815$^{\prime}$ as a proxy for bona-fide NUV flare continuum intensity.  \label{fig:mar29cont}}
	\end{center}
\end{figure}

The high-cadence IRIS data of the X9 flare have very limited spectral coverage (Section \ref{sec:iris}) with which to identify bona-fide continuum regions that are free of minor line emission.  Previously, the narrow wavelength regions near $\lambda \approx 2826$ \AA\  \citep[C2826;][]{Kleint2016,Kowalski2017Mar29, Kowalski2019IRIS} or $\lambda \approx 2832$ \AA\ \citep[C2832;][]{GarciaRivas2024}  have been used in other flares.  
I find that the spectral region from $\lambda = 2815.48 - 2815.68$ \AA\ (hereafter, C2815) is close to a measure of bona-fide flare continuum radiation.  I confirm that the excess (denoted with a prime-symbol) intensity given by C2815$^{\prime}$ is bracketed by C2826$^\prime$ and C2832$^\prime$  in other flares with wider spectral coverage.  In Figure \ref{fig:mar29cont}, I show the three wavelength regions for one of the two brightest footpoint sources (Paper I) in IRIS/NUV spectra of the 2014-Mar-29 X1 flare.
The wavelength window of C2815 has relatively minor contributions from extremely redshifted emission from Fe II $\lambda 2815.445$ (see Section \ref{sec:feanalysis}) and blueshifted emission from Fe I $\lambda 2815.846$ \AA\  that cause the C2815 window to be systematically 5-10\% brighter than the C2826 window.  The  nearby lines in the 2014-Mar-29 flare cause the C2832 continuum estimate from \citet{GarciaRivas2024} to be systematically 50\% larger than the C2826 window. The C2815 wavelength window is the  farthest from Mg II in the NUV in the 2024-Oct-03 dataset; at most, the average continuum intensity over-estimates the bona-fide NUV continuum radiation by 10\%. 

Thus, I search all IRIS NUV spectra on 2024-Oct-03 spanning the flare times for average C2815 intensities that far exceed ($I_{\lambda \sim 2815 \AA} \gg 10^6$ erg cm$^{-2}$ s$^{-1}$ sr$^{-1}$ \AA$^{-1}$) the nearby granulation intensity.
I find several notable brightenings as both flare ribbons cross the IRIS slit, but the Southern ribbon produces the brightest C2815 intensity at 12:16:21.3 UTC at the location of (solar $x$, solar $y$)$=(73\arcsec.300, -375\arcsec.337)$\footnote{This y-pixel corresponds to $y=122$, counting from $y=1$ northward along the slit.}. 
Hereafter, I label this brightening  as ``BFP1'' (bright footpoint 1), following  the \citet{Kowalski2017Mar29} nomenclature for the bright flaring footpoints in the 2014-Mar-29 flare.  The red dot in Figure \ref{fig:context} indicates the location of BFP1, which nearly coincides with a brighter-than-average circular kernel in the SJI 1330 intensity contours present at 12:16:18 - 12:16:34 UTC.

Of the solar flares of Cycle 24, the excess intensity values given by C2826$^{\prime}$ typically occur in the range of $0.1-1.5 \times 10^{6}$ erg cm$^{-2}$ s$^{-1}$ sr$^{-1}$ \AA$^{-1}$ \citep{Butler2022PhD}; many more flares probably produce values of $< 0.1\times 10^{6}$ erg cm$^{-2}$ s$^{-1}$ sr$^{-1}$ \AA$^{-1}$ that are typically within the variations due to granulation.  Two very bright NUV continuum sources occur in the 2014-Mar-29 flare with intensity values of $2-2.5 \times 10^6$ \ilam\ (Paper I).
 Table \ref{table:brightkernels} compares  the excess intensities at BFP1 in the 2024-Oct-03 flare to these two continuum sources (labeled BFP1 and BFP2 in Paper I) in the 2014-Mar-29 flare.  The lower limit on the excess NUV intensity of BFP1 in the 2024-Oct-03 flare is a factor of 1.3 brighter than BFP1 in the 2014-Mar-29 flare.  However, the spatial profile of the continuum at BFP1 in the 2024-Oct-03 data is unresolved at an onboard pixel binning of $0.33$\arcsec. Thus, I deconvolve the intensity to find a reasonable upper limit (6th row of Table \ref{table:brightkernels}) on the excess for this remarkably bright footpoint.

\begin{figure}
	\begin{center}
\includegraphics[width=0.5\textwidth]{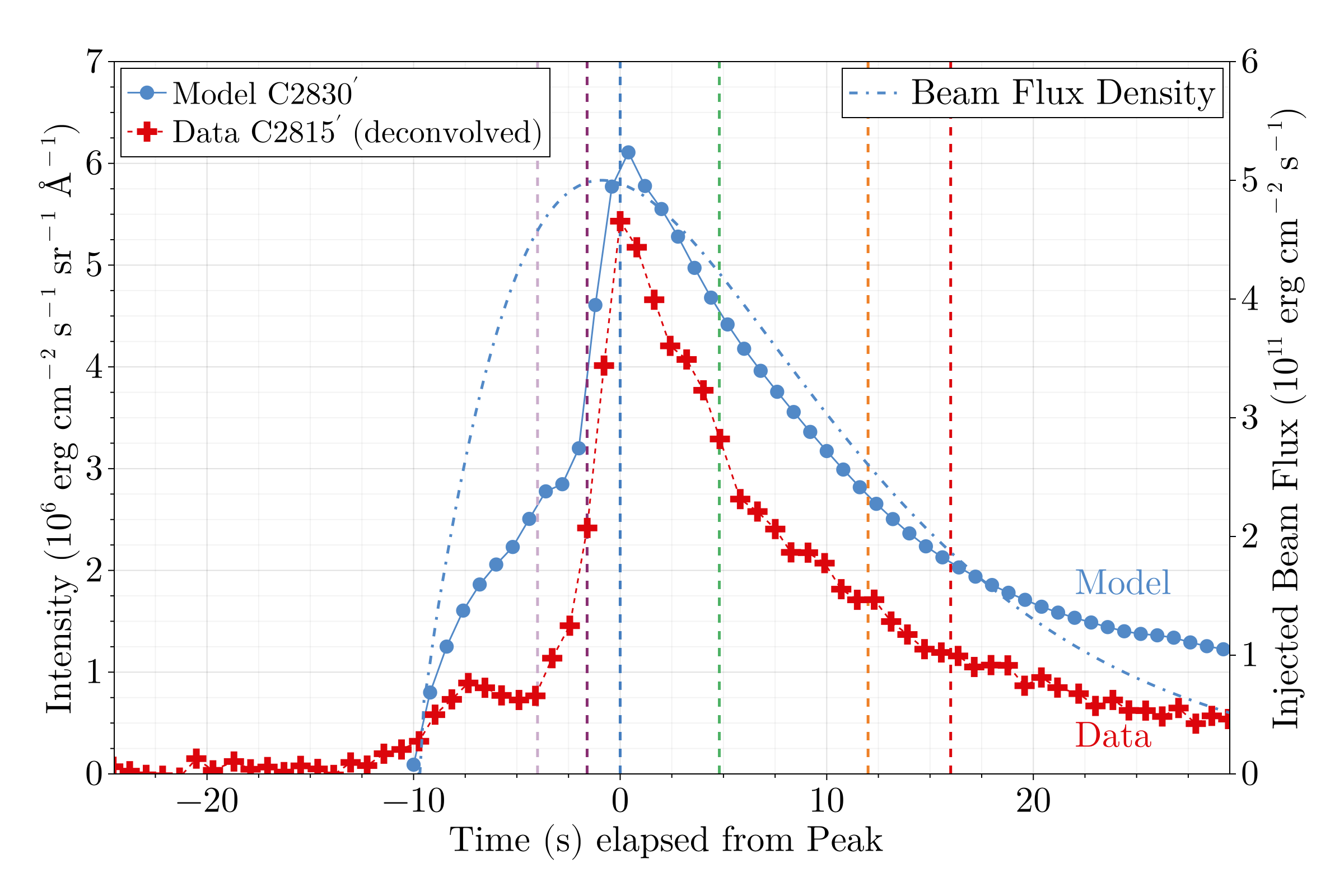}
	\caption{Deconvolved C2815$^{\prime}$ continuum intensity at the bright source (BFP1) indicated by the red circle in Figure \ref{fig:context}.  The injected flux density of an electron beam in model \texttt{m20s-5F11-25-4} is shown on the right axis compared to the calculated C2830$^{\prime}$ light curve.  The peak intensities and the general temporal morphologies are similar.  Vertical dashed lines indicate times that are color-coded to the six spectra in Figure \ref{fig:FeIISpec}(a). The model light curve has been offset by 10~s to align the peak NUV continuum intensities. \label{fig:lcs}}
	\end{center}
\end{figure}

\subsubsection{Intensity Deconvolution} \label{sec:deconv}
Robust deconvolution requires accurate knowledge of an instrument's point spread function (PSF), which \citet{Courrier2018} derive for the IRIS mission. Deconvolution leverages the convolution theorem to infer the original intensity ($I_{\rm{true}}(x,y)$) from the 2D PSF kernel ($K(x,y)$) and the observed image intensity ($I_{\rm{obs}}(x,y)$):

\begin{equation}	
	I_{\rm{true}}(x,y) = \mathcal{FT}^{-1} \{ \frac{\mathcal{FT}\{ I_{\rm{obs}}(x,y)\}}{\mathcal{FT}\{K(x,y)\}} \}
\end{equation}

\noindent where $\mathcal{FT}\{\}$ is the Fourier Transform operation.  I use the \texttt{iris\_sg\_deconvolve.pro} routine \citep{Courrier2018}, which employs iterations of the Richardson-Lucy algorithm for deconvolution.  However, the deconvolution routine is only described for single pixel binning.  Thus, I take extra steps to ensure a robust result with 2-pixel binning.  First, the spectral data are rebinned (using \texttt{congrid.pro}) in the spatial direction to twice as many pixels.  This produces a hypothetical unbinned image with every pair of pixels having the same intensity.  This image is an approximation to the intensity that would be observed at the original binning of $0.\arcsec167$ pix$^{-1}$.  The deconvolution routine is performed with 30 iterations.  Then, the spectra are binned down to the original number of pixels to give the final deconvolved intensity of $\approx 6\times10^{6}$  \ilam\ for BFP1 at 2-pixel binning. The deconvolved 2014-Mar-29 spectral intensities at C2815 (Table \ref{table:brightkernels}) are significantly less bright. Table \ref{table:brightkernels} lists the original and deconvolved C2826$^{\prime}$ intensities in the 2014-Mar-29 flare; these are 5-10\% fainter than the C2815$^\prime$ intensities  (Section \ref{sec:brightest}).

A \emph{gedanken} experiment ensures this procedure gives an accurate estimate of the original true intensity.  First, I assume that the true flare continuum source is spread over 5 pixels with $0.\arcsec 167$ pix$^{-1}$.  The five ``true'' intensity values in the model are assumed to be $0.8\times10^{6}$, $4\times10^6$, $8\times10^6$,$4\times10^6$, and $0.8\times10^{6}$  \ilam.  I convolve the model with the IRIS PSF and bin to two pixels. This gives the ``observed'' intensity. I then perform the same operations (rebinning, deconvolving, and binning) as on the actual data  and arrive at a sufficiently accurate retrieval of the intensity:  for a maximum ``observed'' (non-deconvolved) 2-pixel binned intensity of $3 \times 10^6$ \ilam, I obtain a deconvolved 2-pixel binned intensity of $5.8 \times 10^6$ \ilam, which is close to the maximum of the ``true'' binned-by-2 intensity  ($6 \times  10^6$ \ilam)\footnote{Note, this \emph{gedanken} experiment implies that the true single pixel intensity could be larger than $6\times10^6$ \ilam.} in the \emph{gedanken} experiment.

\begin{deluxetable}{lllllllll}[ht!]
	\tablecaption{Summary of Bright Footpoints in the 2014-Mar-29 and 2024-Oct-03 Flares} \label{table:brightkernels}
	\tablewidth{700pt}
	\tabletypesize{\scriptsize}
	\tablehead{
		\colhead{Flare} & \colhead{Label} &   \colhead{Preflare C2815} & \colhead{Deconvolved Preflare C2815} & \colhead{Total C2815} & \colhead{Deconvolved Total C2815} & \colhead{C2815 Excess (range)} & \colhead{C2826 Excess (range)} & \colhead{\arcsec/pix}} 
	\startdata
	2014-Mar-29 & BFP1 &  0.4 & 0.4 & 3.0 & 3.7   & $2.6 - 3.3$ & $2.3 - 3.0$ & 0.167 \\
	2014-Mar-29 & BFP2 &  0.7 & 0.8 & 3.0 & 4.1   & $2.3 - 3.3$ & $2.2 - 3.2$ & 0.167 \\
	2024-Oct-03 & BFP1  & 0.4 & 0.5 & 3.9 & 6.0   & $3.5 - 5.5$ & \nodata &  0.33  \\
	\enddata
	\tablecomments{Intensities are given in units of $10^6$ \ilam.  The C2815 Excess (range) indicates the range of values given by Total C2815 minus  Preflare C2815 to Deconvolved Total C2815 minus Deconvolved Preflare C2815. The C2826 Excess (range) reports a similar range of excess intensities in the C2826 window, which is available only for the 2014-Mar-29 flare data.}
\end{deluxetable}

\subsection{Light Curve Analysis} \label{sec:lcanalysis}

In Figure \ref{fig:lcs},  the deconvolved light curve of C2815$^{\prime}$ is compared to the \texttt{m20s-5F11-25-4} model values of C2830$^{\prime}$.  Although separated in wavelength by 15 \AA, the expected difference in excess continuum intensities at C2815 and C2830 is less than 2\% in the models, while the differences could be as much as 5-10\% in the observational measures (Section \ref{sec:brightest}).  The peak brightness and the timescales, such as the FWHM of the light curve and the decay time, are reasonably well represented by the model's evolution.   The peak intensities are in agreement to within 10-20\%, which is satisfactory given that no fine-tuning of the electron-beam flux densities in the models have been performed.  The model light curve consists of a two-stage rise phase:  first, the emergent intensity attains a plateau before the chromospheric condensation has cooled to the temperatures that emit copious continuum radiation in the NUV.  This plateau is due to beam heating of the stationary flare layers below the condensation, which is too hot to produce relatively large amounts of NUV continuum radiation at early times.  The data show a fainter and flatter plateau than the model. After several seconds, the beam-generated condensation is dense and cool enough to emit more prominently at these wavelengths (see Papers I and II for extensive discussions).  The decay of the light curve is largely determined by the choice of the beam flux injection function, which I adopt from \citet{Aschwanden2004}.  The decay phase in the observation consists of a fast decay that transitions into a slower decay one spectrum after the green vertical line.  This feature is not reproduced in the model light curve.   Thus, there is less agreement than at peak in that the model outshines the observations by a factor of 1.6-1.8 at the times indicated by the vertical orange and red lines. The step-function energy injection time-profile in the \texttt{c15s-5F11-25-4.2} (not shown, see Paper I) produces a roughly similar rise phase, albeit on shorter timescales. However, the continuum intensity drops much more rapidly when the beam energy injection is removed at $t \ge 15$~s in the model.

\subsection{Fe Emission Line Analysis}  \label{sec:feanalysis}
The Fe II emission lines in the NUV are useful diagnostics of chromospheric condensation evolution \citep{Graham2020,Kerr2024A, Butler2024}.  They are formed at cool temperatures that overlap with the formation heights and temperatures of the NUV hydrogen Balmer continuum in solar flares \citep[Paper I,][]{Kowalski2019IRIS}.  Moreover, the optical depth is low enough in several of the lines such that they constrain the velocity gradients from within the condensation into the layers below that are (mostly) at rest (Paper I).  Following previous works, I analyze the Fe II $\lambda_0= 2814.445$ \AA\ line, which is intentionally included in the small readout window of the IRIS high cadence modes.  Other prominent lines in this range are an Fe I line with rest wavelength $\lambda_0= 2814.115$ \AA\ and Fe II with $\lambda_0 = 2813.322$ \AA\ \citep{Kowalski2019IRIS}.  I also identify a new Fe I line with $\lambda_0=2815.846$ \AA\ in emission that is not in the NUV flare atlas of \citet{Kowalski2019IRIS}.

\begin{figure}
	\gridline{\fig{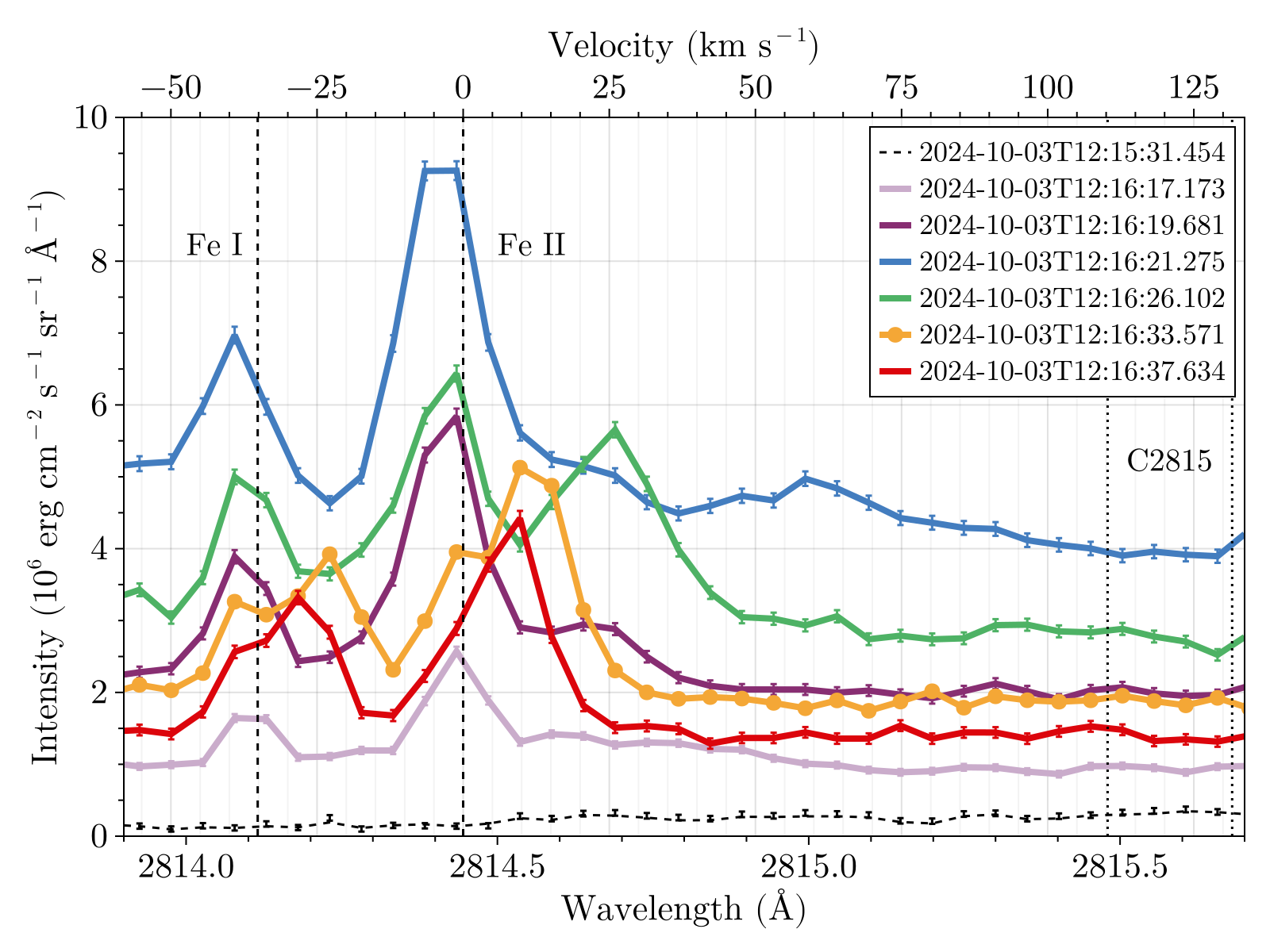}{0.45\textwidth}{(a)} \fig{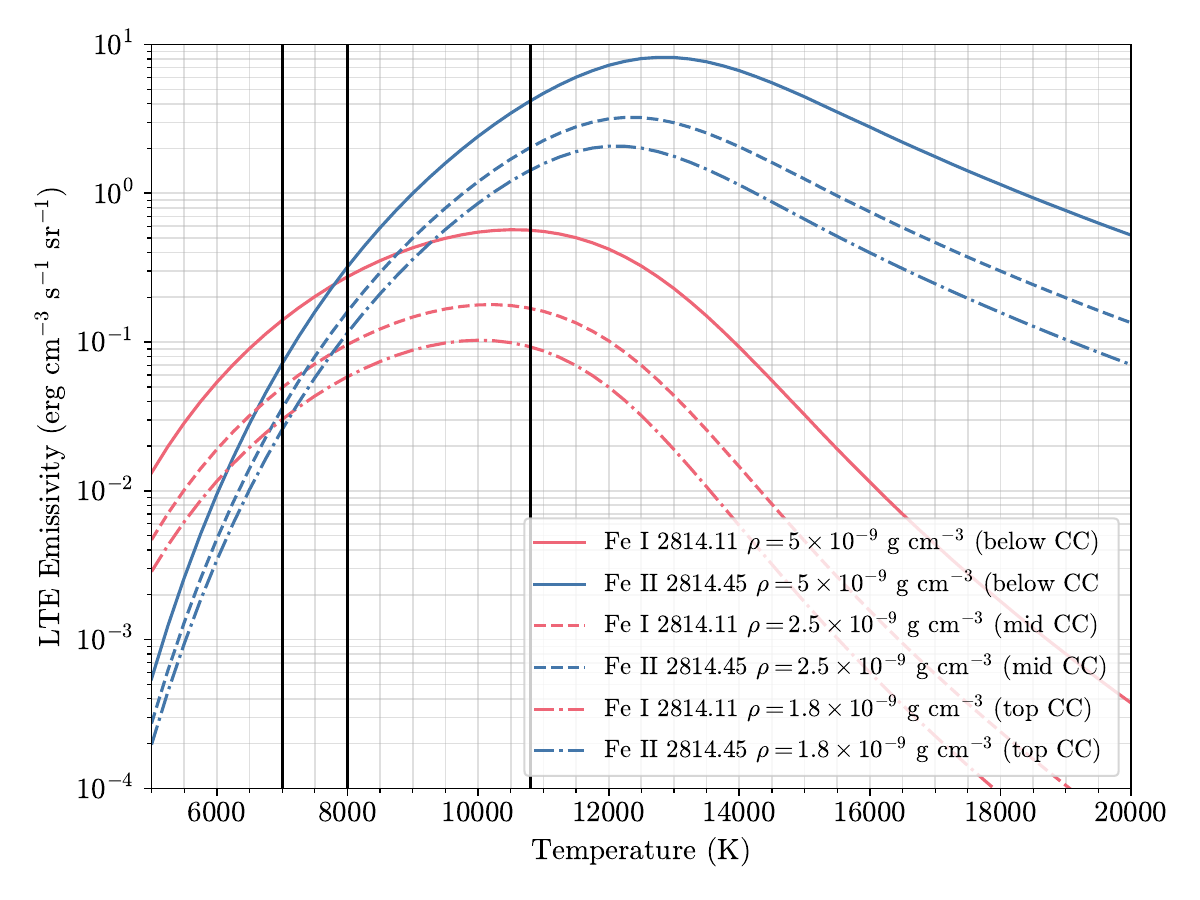}{0.45\textwidth}{(c)}}
	\gridline{\fig{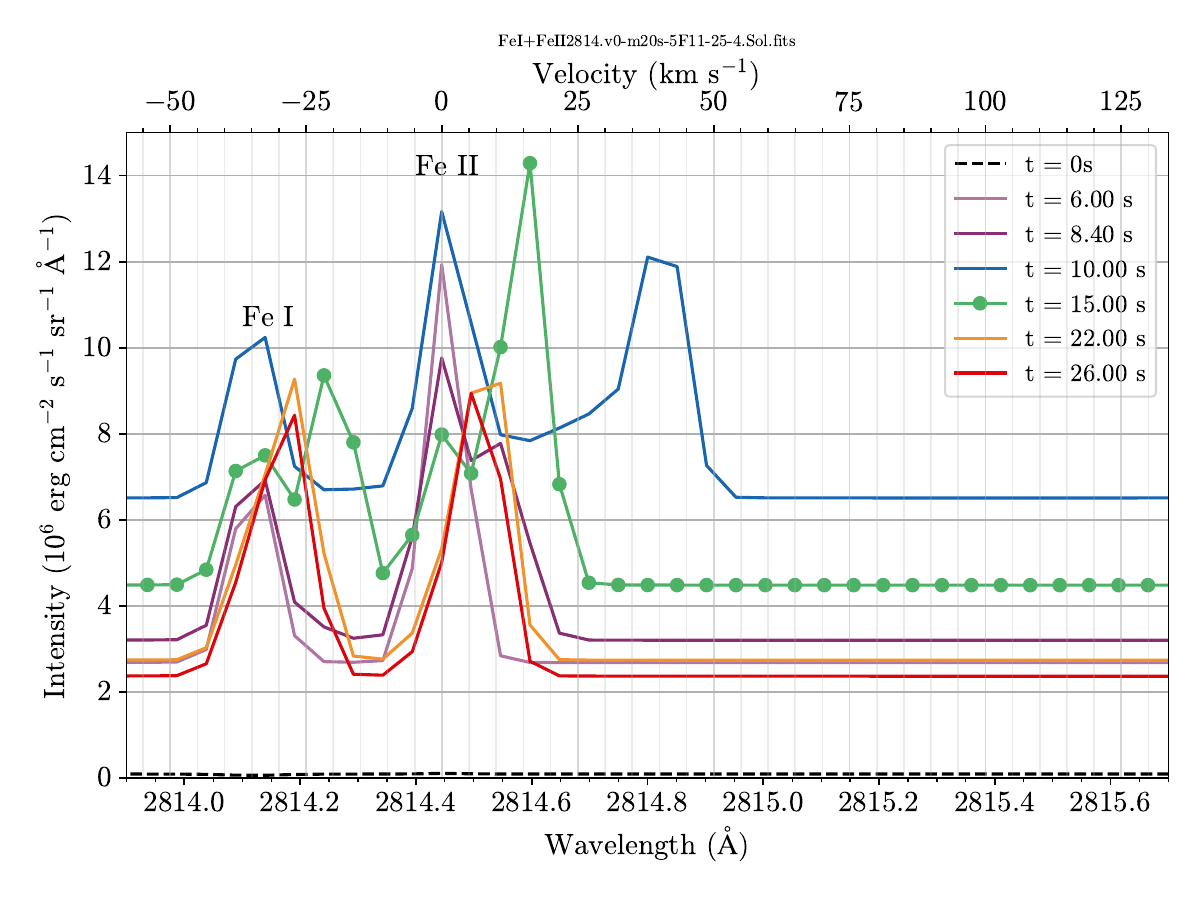}{0.45\textwidth}{(b)}
		\fig{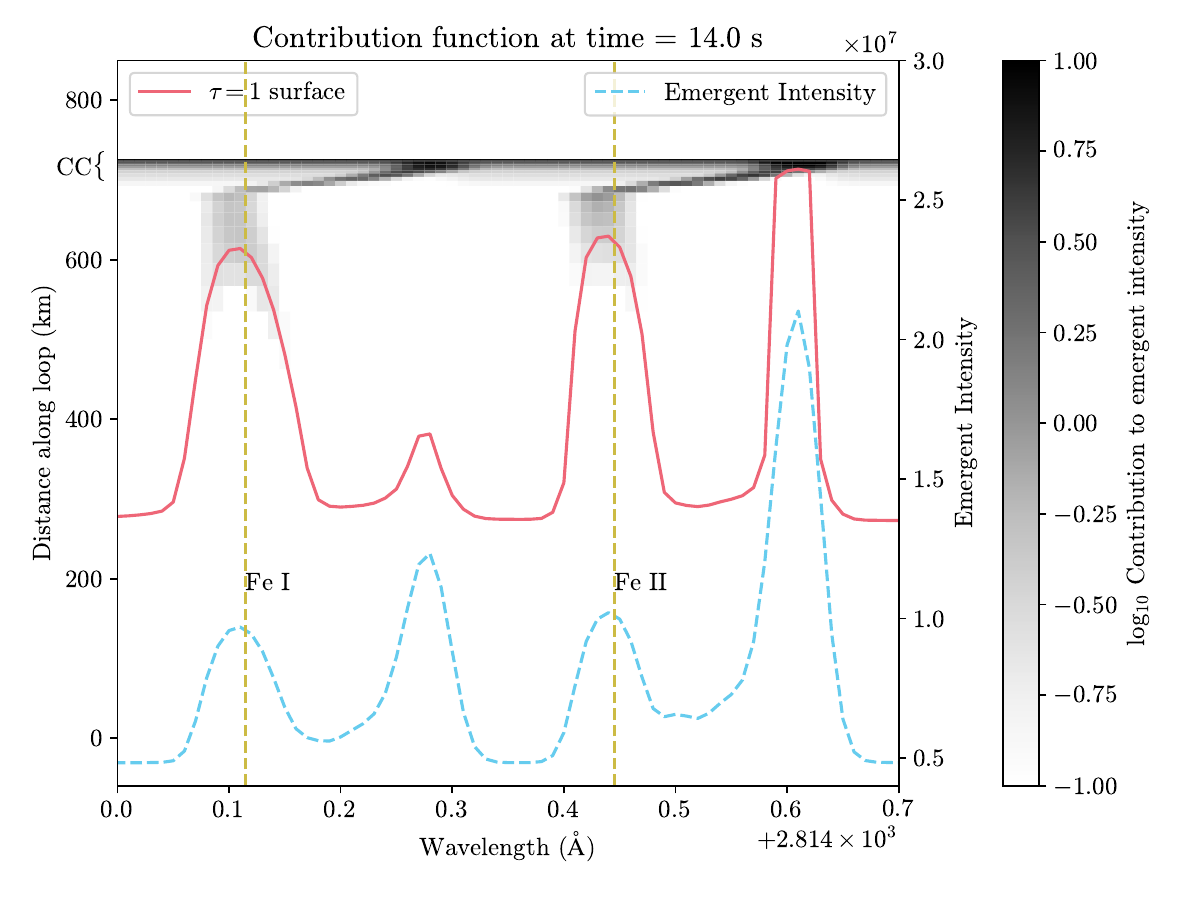}{0.45\textwidth}{(d)}}
	\gridline{\fig{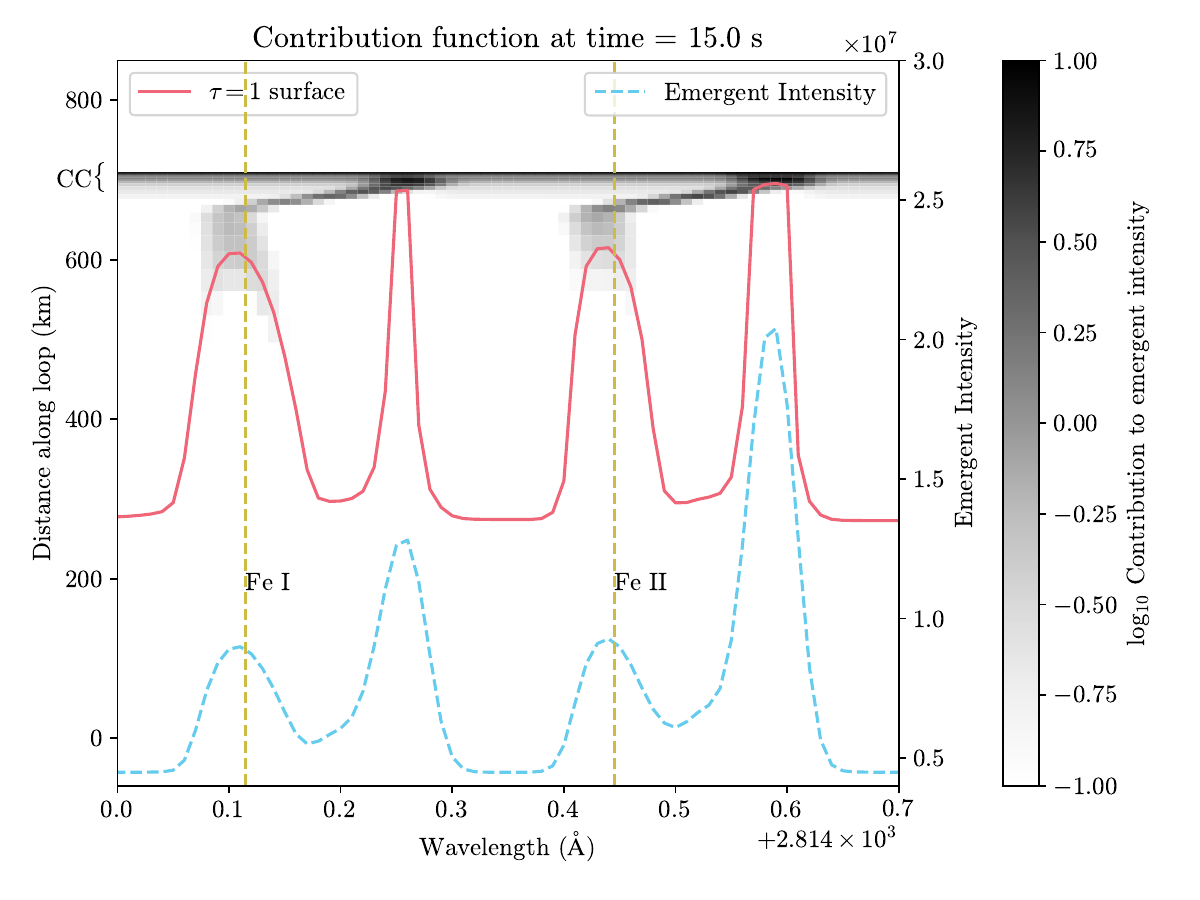}{0.45\textwidth}{(e)}}
	\caption{(a) Spectra of Fe I and Fe II (with rest wavelengths indicated by vertical dashed lines) color-coded to the times indicated in Figure \ref{fig:lcs}.  The vertical dotted lines indicate the wavelengths used in the C2815 continuum intensity calculation. (b) Spectra calculated from the \texttt{m20s-5F11-25-4} model evolution at similar temporal spacings as in panel (a).  (c) LTE emissivity calculations of the Fe I and Fe II lines in panel (a). Vertical lines correspond to the temperatures at representative locations within the model flare atmosphere.  Fe II greatly outshines Fe I at $T > 10^4$ K, but they have comparable emissivities at lower temperatures. (d) Contribution function to the emergent intensity at $t=14.0$~s in the model evolution in panel (b).  At this time, the Fe II $\tau(\lambda)=1$ surface is located in the chromospheric condensation, while the Fe I is just starting to build up optical depth at redshifted wavelengths. There is no contribution from deeper layers near the photosphere.  Vertical dashed lines indicate the rest wavelengths of Fe I $\lambda 2814.115$ and Fe II $\lambda 2814.445$. The approximate height range of the chromospheric condensation is annotated as ``CC'' on the left axis. (e) The contribution function to the emergent intensity at $t=15.0$~s in the model evolution of panel (b).  \label{fig:FeIISpec}}
\end{figure}

The observed temporal sequence around Fe II $\lambda 2814.445$ is shown in Figure \ref{fig:FeIISpec}(a).
 The spectra are color-coded to the times indicated by the vertical dashed lines in Figure \ref{fig:lcs}.  The temporal sequence of the two line components is generally similar to those in  the X1.6 2014-Sep-10 flare \citep{Graham2020}, but there are several notable differences.  First, the evolution in the 2014-Oct-03 spectra is much quicker; most evolution occurs over about 20~s.  This contrasts to the evolution over $\approx 45$~s in the 2014-Sep flare.  Second, the NUV continuum excess intensity is a factor of $3-6$ brighter (Section \ref{sec:deconv}).  Third, the redshifted components from both the Fe I and Fe II lines become brighter than the intensity around the rest wavelengths, whereas the red shifted component never becomes significantly brighter  in the 2014-Sep flare.  In the 2014-Mar-29 Fe II $\lambda 2814.445$ spectra, the redshifted components are well below the intensity at the rest wavelength. I also note what appears to be a minor blueshift of the intensity peak around the rest wavelength.  A minor blueshift appears in the 2014-Mar-29 flare as well; blueshifts have been discussed in \citet{Libbrecht, Kerr2024RibbonsII} in other flares and models.

 Snapshots of the atmospheric variables from each RADYN simulation are used to calculate Fe II $\lambda 2814.445$ and Fe I $\lambda 2814.115$ in LTE, following Paper I, \citet{Graham2020}, and \citet{Kerr2024A}.
 The calculation is performed with LTE excitation and ionization population densities, but the electron densities are obtained from the RADYN simulation at the respective time.  A nonthermal broadening (``microturbulence'') parameter is not included in these calculations, following \citet{Graham2020}.  The emergent intensity is calculated using a Feautrier solver \citep[][see also \citealt{Mihalas1985}]{Feautrier1964}.  The Feautrier solver is translated into Python from the MULTI \citep{MULTI1} and RADYN Fortran source codes.  The line profiles are convolved by the IRIS spectral resolution and binned to two pixels.  An error in the statistical weight of the ground state Fe III ionization stage has been corrected in \citet{Kowalski2022Frontiers} and \citet{Kerr2024A}.  In Paper I, the statistical weight of the ground state of Fe III was set to 7, which was obtained from a model iron atom file in the standard RH code distribution.  The total statistical weight of the ground state of Fe III is 30 in the National Institute of Standards and Technology (NIST).  Instead of using the Saha-Boltzmann distribution, which depends on this value, I write the Saha and Boltzmann equations separately.  The Saha equation is

\begin{equation}
	\frac{n_{FeII}}{n_{FeI}} = \frac{2 U_{II}(T)}{n_e U_I(T)} \big( \frac{2 \pi m_e k_B T}{h^2} \big)^{3/2} e^{-\frac{\chi_I h c}{k_B T}}
\end{equation}  

\noindent where $U_I(T)$ and $U_{II}(T)$ are the partition functions \citep{Halenka1984} for Fe I and Fe II, respectively, and $\chi_I$ is the ionization energy (in units of cm$^{-1}$) of the ground state of Fe I.  Atomic data are obtained through NIST and references therein \citep{Nave1994, Nave}.  The same equation for $\frac{n_{FeIII}}{n_{FeII}}$ is combined with the total Fe abundance to solve for $n_{FeI}$, $n_{FeII}$, and $n_{FeIII}$.  The upper-level  population densities for a Fe II transition are given by the Boltzmann excitation formula:

\begin{equation}
n_{u} = n_{FeII} \frac{g_u}{U_{II}(T)} e^{-\frac{E_u h c}{k_B T}}  
\end{equation}

\noindent where $n_u$ is the upper level population density, $E_u$ is the excitation energy (in units of cm$^{-1}$) of the upper level relative to the ground state, and $g_u$ is the statistical weight of the upper level.  The general formula for the bound-bound opacity is \citep{Mihalas1978}

\begin{equation}
	\chi(\nu) = \frac{\pi e^2}{m_e c} f_{osc} \phi_{\nu}(\nu) (n_l - n_u \frac{g_l}{g_u})
\end{equation}

\noindent where $\phi_{\nu}(\nu)$ is the normalized line profile (Voigt) function that includes thermal, van der Waals, and Stark broadening. The NLTE continuum opacity from RADYN and the LTE Mg II $h$ and $k$ damping wing opacities are added to the bound-bound opacities from Fe I and Fe II.  The line emissivity is obtained from $\chi(\nu) B_{\nu}(\nu)$, whereas the NLTE emissivity \citep{Mihalas1978} is used for the continuum (since population densities from detailed elements are known from RADYN).  The ratio of total emissivity to total opacity gives the source function, which, along with the optical depth, $\tau$, and opacity, are fed into the Feautrier solution to the emergent intensity.

The calculations from the \texttt{m20s-5F11-25-4} model are shown in Figure \ref{fig:FeIISpec}(b) using snapshots spaced at the same temporal cadence as in the observations (Figure \ref{fig:FeIISpec}(a)).   The model Fe II spectral evolution is similar to the observations, but the peak intensities of the lines are about $1.4-1.5$x too bright.  This discrepancy can be attributed to the spatial resolution (Section \ref{sec:deconv}), as I choose to not deconvolve the observations in Figure \ref{fig:FeIISpec}(a).  A qualitative similarity between the models and the observations is emphasized in the orange spectrum $\approx 12$~s after peak in panel (a) and the green spectrum $\approx 5$~s after peak in panel (b) when the respective redshifted components become brighter than the components around the rest wavelength.  
The peak of the red-wing asymmetry becomes progressively less redshifted over time. The two components of the line merge into an apparently single, redshifted line within $\approx 20$~s.  The models produce similar late phase spectra at $t=22-26$~s.

 There are several important differences between the model and the observations. The ratio of the peak of the continuum-subtracted red-wing component to the peak intensity near the rest wavelength is about 1.3 in the observation (orange spectrum), and it is nearly a factor of 2.7 in the model (green spectrum).   Additionally, a minor blueshift in the component around the rest wavelength does not occur in the models, and the model spectrum around the rest wavelength (light purple) is too bright initially.  On the red side, the red-wing asymmetry is much broader in the data.  Specifically, in the dark blue spectrum, the redshifted intensity appears to extend out to $80-100$ km s$^{-1}$ (see below), whereas in the model there is no such high velocity, red-wing tail. In the observations (dark blue spectrum), the NUV continuum intensity peaks before the red-wing asymmetry becomes as bright as the rest-wavelength component, whereas in the model dark blue spectrum the red-shifted component and the rest-wavelength component are comparable at this  time, which corresponds to the peak NUV continuum intensity.  The two line components attain comparable peak intensities in the observations only after the peak NUV continuum intensity.  Although both model and data then show a red-wing asymmetry that outshines the component around the rest wavelength, this occurs about 7~s too soon in the model.  In summary, the red-wing asymmetry becomes too bright too quickly after the peak NUV continuum intensity.

\begin{figure}
\fig{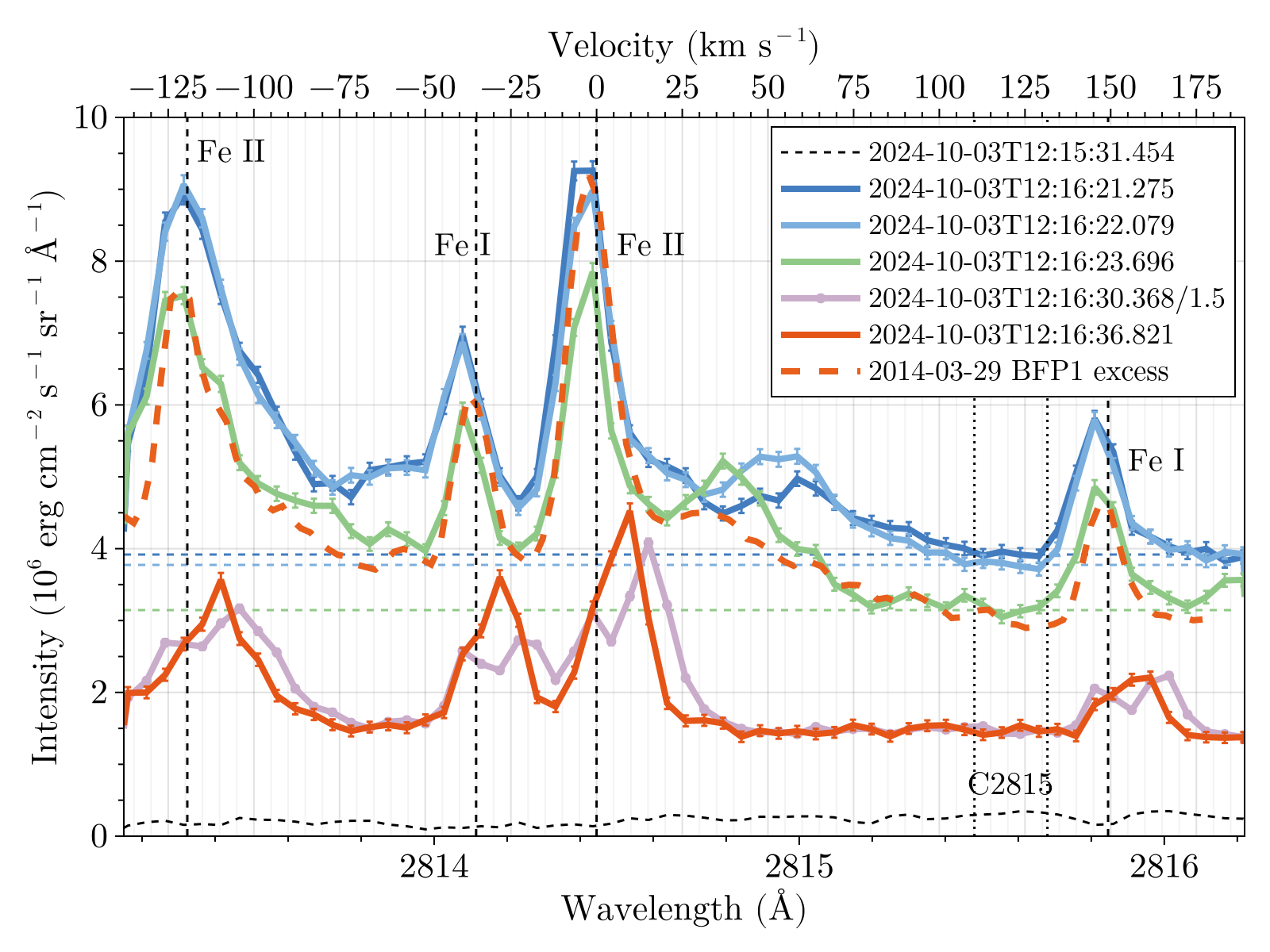}{0.5\textwidth}{}
\caption{ The blue and green spectra are three nearly sequential spectra around the times of peak NUV continuum intensity in the 2024-Oct-03 flare.  The green spectrum most closely resembles the brightest flare spectrum in the 2014-Mar-29 flare. The purple and solid red spectra are taken from the decaying phase of the kernel in the 2024-Oct-03 flare. \label{fig:Mar29}}
\end{figure}

The nearby Fe I $\lambda 2814.115$ line follows a similar evolution to Fe II.  In particular, a redshifted component is clearly  in the observations (and the models below), and the red wing asymmetry attains a peak intensity that exceeds the intensity at the rest wavelength in the orange spectrum.  In Figure \ref{fig:FeIISpec}(c), I plot the LTE emissivity for the Fe II and Fe I lines as a function of temperature. Vertical lines indicate the temperatures at representative heights in the \texttt{m20s-5F11-25-4} model at $t=14$~s (which is $\Delta t = 5$~s after the maximum beam flux injection):  a location within the stationary flare layers at $z=600$ km ($T = 7000$ K, $n_e = 2 \times 10^{13}$ cm$^{-3}$), a location within the chromospheric condensation at $T = 8000$ K and $n_e = 7 \times 10^{13}$ cm$^{-3}$, and a location at the top of the chromospheric condensation at $T = 10,800$ K and $n_e = 5.4 \times 10^{14}$ cm$^{-3}$.   The emissivities of these transitions are comparable except at the higher temperatures of $T\ge 10^4$ K at the top of the chromospheric condensation where Fe II outshines Fe I by a factor of $\approx 15$.  
The contribution function to the emergent intensity is shown at $t=14$~s in Figure \ref{fig:FeIISpec}(d).  Both lines have contributions from the stationary flare layers and the chromospheric condensation, but neither line has any contribution from the photosphere or upper photosphere at this time in the model.   However, there are also some subtle differences between the two lines.  The Fe I contribution function extends slightly deeper into the cooler, less ionized stationary flare layers, as expected.  Furthermore, the Fe II line starts to become optically thick in the condensation, whereas the Fe I line is just building up opacity at this stage of the condensation evolution.  Fe I becomes optically thick only several seconds later: The rapid increase in optical depth at redshifted wavelengths by $t=15.0$~s is shown in Figure \ref{fig:FeIISpec}(e).  This difference results in an interesting effect in the model Fe I line:  the continuum-subtracted red-wing asymmetry peak continues to increase in brightness from the green to the orange spectra in panel (b)  while the Fe II starts to decrease.  As discussed for the H$\gamma$ and H$\alpha$ lines in Paper II, redshifted line spectra become fainter due to the decrease in temperature of the condensation at the $\tau(\lambda) = 1$ level \footnote{The decreasing injected beam flux density at these times also contributes to the decrease in line intensity;  Paper II shows that a constant injection can also lead to a decrease in line intensity as the condensation density increases and rapidly cools below $T \sim 10^4$ K.}.  The comparison of the solid red and purple spectra in Figure \ref{fig:Mar29}, which shows an expanded wavelength range, demonstrates that this difference between Fe I and Fe II is not as strikingly evident in the data evolution over $\Delta t \approx 7$~s.

An additional constraint is possible from the high-cadence of the spectral observations.  
In Figure \ref{fig:Mar29}, the dark blue, light blue, and light green spectra show three nearly consecutive spectra during the brightest C2815 values, which are indicated by horizontal dashed lines.  The progressive evolution of the red-wing asymmetry before it becomes very bright  (Fig. \ref{fig:FeIISpec}) convincingly shows that the most extremely redshifted intensities are due to the early stages of the chromospheric condensation evolution rather than a confluence of unidentified faint emission lines.  These spectra also inform the interpretation of  low-cadence ($\Delta t = 75$~s) spectra from the brightest footpoint (BFP1) in the 2014-Mar-29 solar flare, shown as the dashed red spectrum:  both BFP1 and BFP2 in the 2014-Mar-29 could have been significantly brighter between exposures.  The small blip at $\approx 25-50$ km s$^{-1}$ to the red of Fe II $\lambda 2814.445$ is now robustly identified as a low-level redshifted component.   The high-velocity faint tails of redshifted intensities extend out to at least $\approx 80-100$ km s$^{-1}$ in the observations of both flares\footnote{The effect is prominent in the Fe II 2832.39\AA\ line in the 2014-Mar-29 flare data.}.  The extreme red-wing asymmetries of Fe II are subtle but important shortcomings of the model chromospheric condensations.

\subsection{H$\alpha$ Line Broadening Analysis} \label{sec:haanalysis}

The high-flux densities of the \texttt{m20s-5F11-25-4} produce Fe II LTE profiles that are in better agreement with the 2024-Oct-03 flare than previous comparisons of high flux models to other flares \citep[e.g., Paper I,][]{Graham2020}.  A glaring inconsistency is that the red wing asymmetry is too bright in all comparisons at high-time resolution, which possibly suggests that the chromospheric condensations in RADYN are too dense (or that the heating in deeper layers is too weak). Unlike the Fe II lines, the hydrogen Balmer wing broadening is very sensitive to electron density and optical depth changes in the flare condensation (e.g., Paper II).   In this section, I compare the H$\alpha$ broadening predictions from Paper II to CHASE spectra of the 2024-Oct-03 flare.

\begin{figure}
	\gridline{\fig{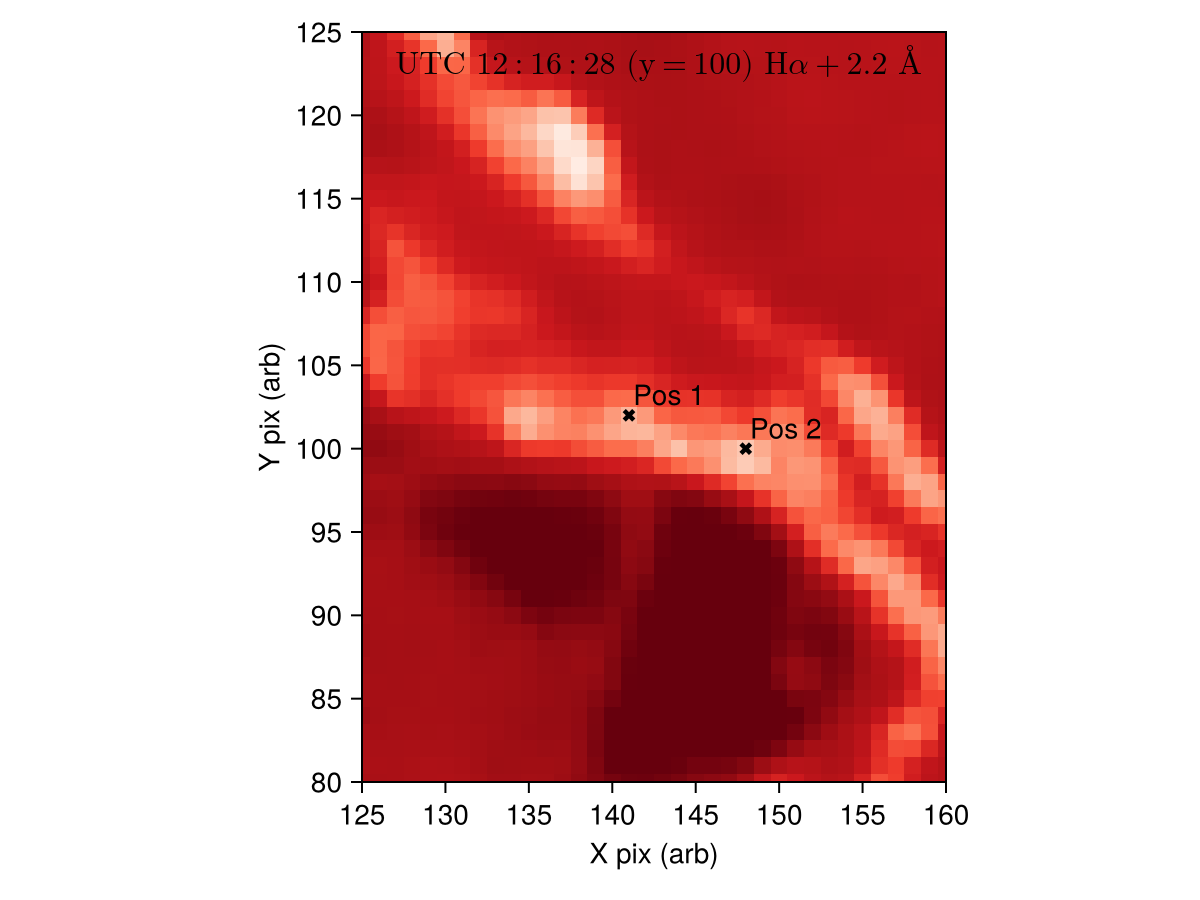}{0.5\textwidth}{(a)} \fig{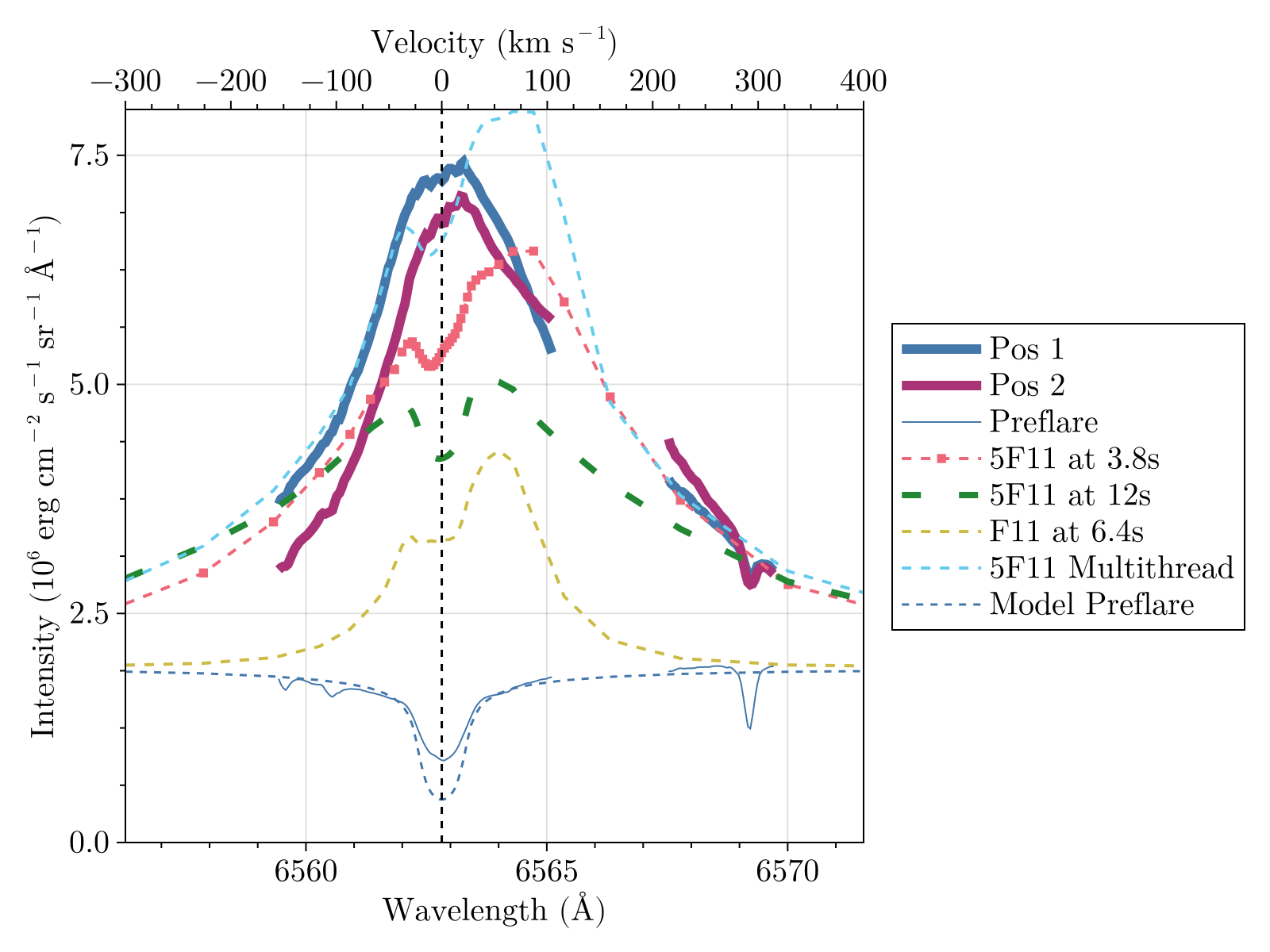}{0.5\textwidth}{(b)}}
	\caption{(a) CHASE context image formed from the spectra of H$\alpha$ at $\lambda - \lambda_0 = 2.2$ \AA.  The image shows several very bright kernels close to the locations of the bright NUV continuum source in IRIS (Figure \ref{fig:context}). The pixel scale is 1.04\arcsec\ per pixel, time spans $\sim 0.9$~s from top to bottom of this image, and the intensity scaling ranges linearly from 600 to 3500 DN. (b) CHASE spectra at ``Pos 1'' and ``Pos 2'' in (a) compared to four radiative-hydrodynamic model predictions.   Two snapshots are shown from the \texttt{c20s-5F11-25-4.2} model for a range of extreme broadening predictions.  The ``5F11 Multithread model'' is a superposition of several snapshots in the \texttt{c20s-5F11-25-4.2} model (see text).  The F11 model is presented in \citet{Graham2020}, and the line broadening is discussed in Paper II. All model intensities are convolved by the spatial resolution of CHASE. \label{fig:chasecontext}}
\end{figure}

 A CHASE context image from the red wing of H$\alpha$ is shown in Figure \ref{fig:chasecontext}(a).  The ribbons are evident with several kernels embedded. I choose two locations (Pos 1 and Pos 2) of bright kernels in the Southern ribbon around the times of prominent NUV continuum intensities and Fe line red wing asymmetries (Figure \ref{fig:FeIISpec}(a)). For reference, the location of the IRIS slit corresponds approximately to the solar $x$ location of Pos 2 in Figure \ref{fig:chasecontext}.  This is determined through comparison of the sunspot locations with known coordinates in SDO/HMI images \citep{Scherrer2012} to the same spot features in the H$\alpha$ red wing images (the spot features in the SJI 1330 context images are too faint to reliably align the IRIS data without the aid of SDO/HMI).   The spectra from Pos 1 and Pos 2 are shown in Figure \ref{fig:chasecontext}(b).
The spectra are very broad with wings extending beyond $-3.4$ \AA\ and $+6.9$ \AA\ from the rest wavelength.  Using the data in both spectral windows, I calculate the full width at 30\% maximum to be $\approx 8$ \AA\ for the spectrum at Pos 1.  The intensities are also very bright, and the broadening is asymmetric to the red.  Remarkably, the H$\alpha$ line wing intensity extends out to and beyond the Fe I CHASE window.  Without the H$\alpha$ line center spectra, it would be difficult to interpret the intensity slope within the Fe I window.  I note that the Fe I line is partially filled in during the flare, and it appears in emission in an excess spectrum.  Further analysis and modeling is outside the scope of this paper but will be included in future work that compares the Fe I lines in the NUV and in the optical.

To accurately test the model spectra, I convolve them to the spatial resolution of the CHASE spectra.   A synthetic spatial map is generated from the pre-flare spectrum, and the flare source is assumed to extend over three unbinned IRIS pixels.  Here, I prefer to present analysis of the \texttt{c15s-5F11-25-4.2} model rather than the \texttt{m20s-5F11-25-4} model because the predictions from the former are thoroughly discussed in Paper II.  After about $\approx 8$~s, the \texttt{m20s-5F11-25-4} model evolution is very similar to the $t=0-10$~s evolution of the \texttt{c15s-5F11-25-4.2}, as described in Paper II.  For reference, the $t=3.8$~s H$\alpha$ profile is very similar to the $t=8.8$~s profile in the \texttt{m20s-5F11-25-4} model, while the $t=10$~s H$\alpha$ spectrum in the \texttt{c15s-5F11-25-4.2} is very similar to the $t=14$~s spectrum in the \texttt{m20s-5F11-25-4} model.  The \texttt{c15s-5F11-25-4.2} model intensity at $t=12$~s is assigned to the flare source intensities in the spatial map. This snapshot is approximately  7~s after the maximum continuum intensity in the \texttt{c15s-5F11-25-4.2} model, thus matching the relative timing between the H$\alpha$ observation and the maximum of the IRIS NUV continuum intensity.  The synthetic spatial intensity map is convolved by a Gaussian\footnote{To my knowledge, the deconvolution of CHASE data is not yet provided by the mission.} with a FWHM of $1.\arcsec04$ and binned to $1.\arcsec04$ pix$^{-1}$ to arrive at the model intensities at the spectra in Figure \ref{fig:chasecontext}(b).  The same procedure is repeated for the spectrum at $t=3.8$~s to show a range of predictions during the evolution of this model.  No additional scale factors are applied.

The convolved intensities of the model H$\alpha$ wings agree with the data rather well in several ways.  The $t=3.8$~s model properly accounts for the near-wing shapes in the Fe I CHASE spectral window, and it predicts significant far-wing broadening at $\lambda > 6570$ \AA\  and at $\lambda < 6560$ \AA. The $t=12$~s model has much more pronounced wings due to the optical depth broadening factors increasing during the evolution of the chromospheric condensation (see Paper II).  Both models are broader than the spectrum at Pos 2, but the match to the wings at Pos 1 is more encouraging.   The wing shape in the $t=12$~s model is too flat compared to the observations in the Fe I window, which is clearly a powerful constraint on these model predictions.

 At the most optically thick wavelengths around the line center, the models are much fainter than the data.  The asymmetric dip in the model spectrum around the rest wavelength is due to mixing of the neighboring non-flaring model intensity and the flare model during the spatial convolution.  A perfect match with forward models should not be expected, but this is indeed a glaring discrepancy.  A more realistic comparison assigns one unbinned IRIS pixel size to each of the 5 snapshots ($t=1, 2, 3.8, 6.4, 10$~s) analyzed in this model in Paper II. I refer to this as a multithread model \citep[e.g.,][]{Warren2006, Litwicka}, which represents several unresolved flare locations that contribute to the intensity of a CHASE pixel. To be consistent with the direction of the apparent spreading motion of the ribbons, the bottom pixel is the earliest time while the top pixel corresponds to the latest time.
  The convolution of this model kernel (Figure \ref{fig:chasecontext}(b)) better explains the blue side of the spectral observation at Pos 1.  The agreement with the red wing in the Fe I window of Pos 1 remains, but the redshifted intensity around $50-100$ km s$^{-1}$ is too prominent in the model compared to the observations at Pos 1 and Pos 2.

The general consistencies in the near-wing broadening cannot be attained with lower beam flux models with less dense chromospheric condensations.  To confirm this, I repeat the same convolution procedures with the H$\alpha$ spectra (Paper II) from the F11 electron beam heating model that is described in \citet{Graham2020}.  The H$\alpha$ spectrum (shown in yellow) is too narrow and too faint.  This is not surprising because the excess NUV continuum intensities predicted by this model are about a factor of ten fainter than the 2024-Oct-03 observations.

\begin{figure}
 	\begin{center}
\gridline{
	\fig{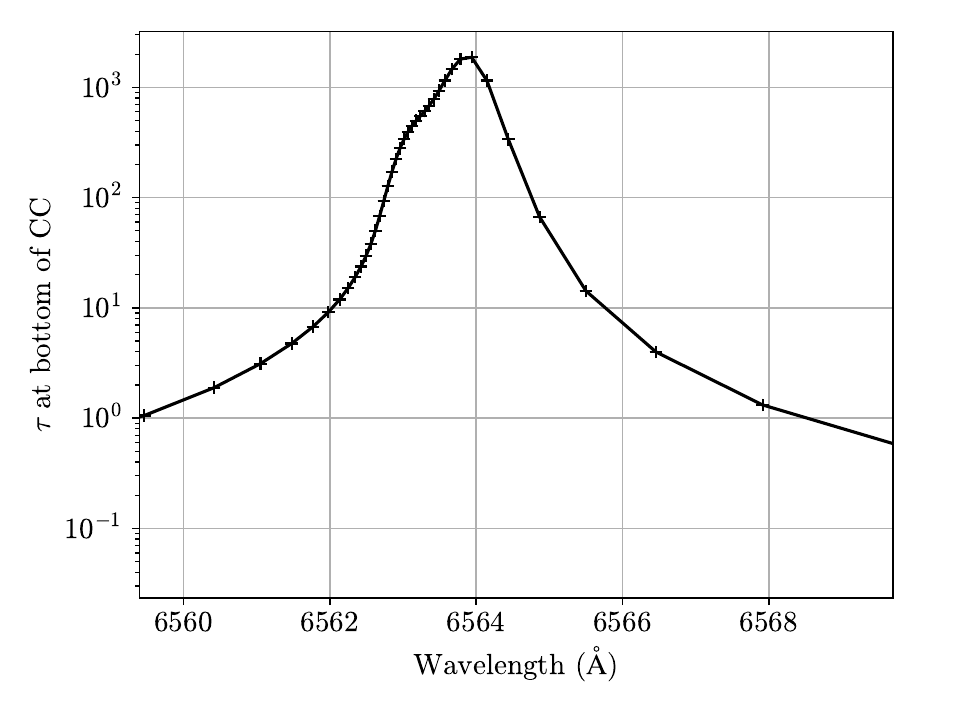}{0.5\textwidth}{(a)} 
	\fig{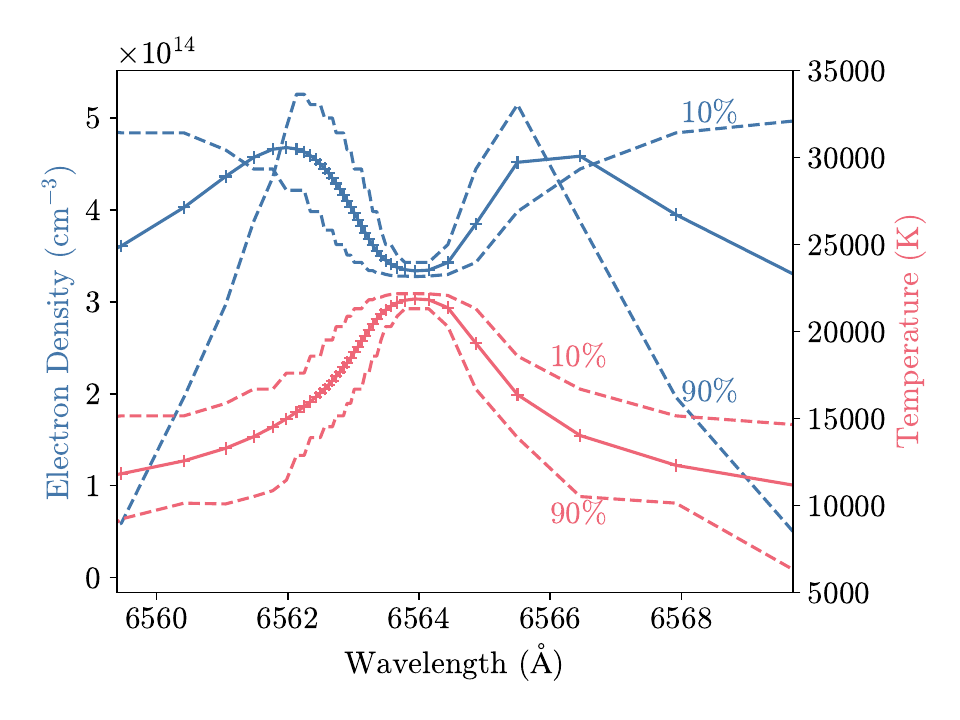}{0.5\textwidth}{(b)} }
 	 		\caption{ (a) The optical depth at the bottom of the chromospheric condensation (CC) varies over many orders of magnitude across the H$\alpha$ line spectrum.  Beyond the edges of the CHASE wavelength window, the line becomes optically thin within the chromospheric condensation. (b)  Electron densities and temperatures (solid lines) weighted by the contribution function to the emergent intensity at $t=3.8$~s in the \texttt{c20s-5F11-25-4.2} model.  The dashed lines correspond to the range of temperatures and densities at 10\% and 90\% of the cumulative contribution function; the 10\% line indicates the parameters that are higher up in the atmosphere.    \label{fig:haeldens}}
 	\end{center}
 \end{figure}

Paper II introduced an optical depth broadening factor that relates the spectral broadening to the electron densities over which a Balmer line forms.   Figure \ref{fig:haeldens}(a) shows the optical depth within the chromospheric condensation.   The line center wavelengths around $\lambda \approx 6564 $ \AA\ accrue a very large optical depth within the condensation. The optical depth broadening factors are very large for the H$\alpha$ profile in the 5F11 models (cf. Fig. 8(c) in Paper II) so it is necessary to investigate the range of atmospheric parameters over which the line forms in the atmosphere.  Figure \ref{fig:haeldens}(b) plots the contribution-function weighted electron densities and temperatures at $t=3.8$~s in the \texttt{c20s-5F11-25-4.2} model.  The wavelength range corresponds to the full wavelength extent of the CHASE data.  The dashed lines indicate the respective atmospheric variables at 10\% and 90\% of the cumulative contribution function (not shown).   Around the line center ($\lambda \approx 6564$ \AA), the line forms within a narrow range of electron densities, $n_e \approx 3.5 \times 10^{14}$ cm$^{-3}$, and temperatures, $T\approx 22,000$ K.  At wavelengths where the optical depth decreases to $\approx 10$ in the condensation, the line is formed over  a larger range of electron densities extending to as large as $\approx 5\times10^{14}$ cm$^{-3}$.  The edges of the CHASE wavelength range correspond to near-wing wavelengths  at which the H$\alpha$ line transitions from being optically thick within the chromospheric condensation to optically thin. Thus,  an even larger range of electrons densities ($n_e = 0.5 - 5 \times 10^{14}$ cm$^{-3}$) and temperatures (around $T\approx 10,000-15,000$ K) contribute to the intensity as the optical depth in the narrow chromospheric condensation drops.   At later times in the models, the electron densities are larger and the temperatures are cooler, but  otherwise the trends are similar.  For example, the line center is formed around temperatures of $T\approx 16,000$ K and electron densities of $n_e = 6-7 \times 10^{14}$ cm$^{-3}$ at $t=12$~s.
  
\subsection{NUV/FUV continuum ratio analysis}  \label{sec:ratioanalysis}
\begin{figure}
	\gridline{
		\fig{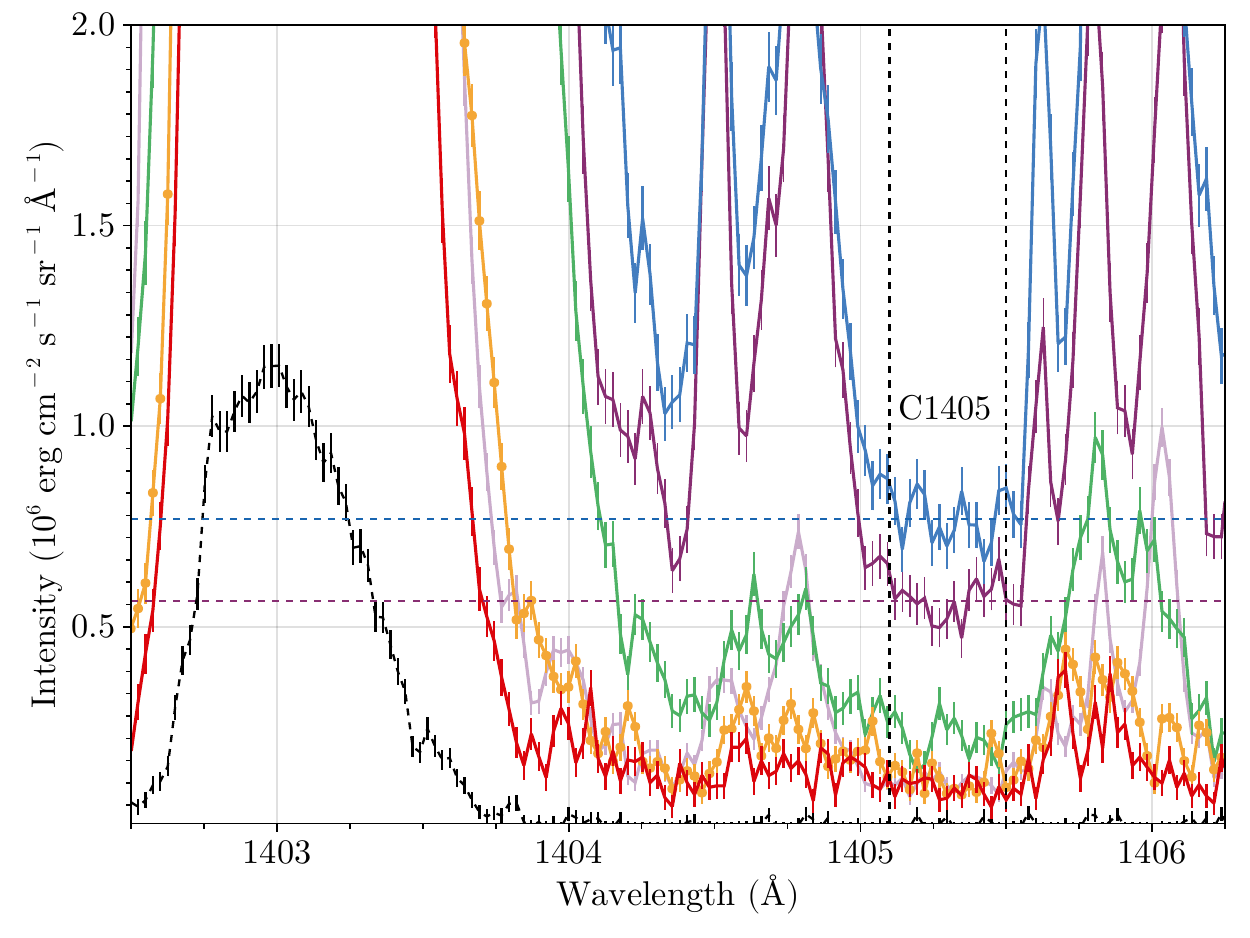}{0.5\textwidth}{(a)} }
\gridline{
	\fig{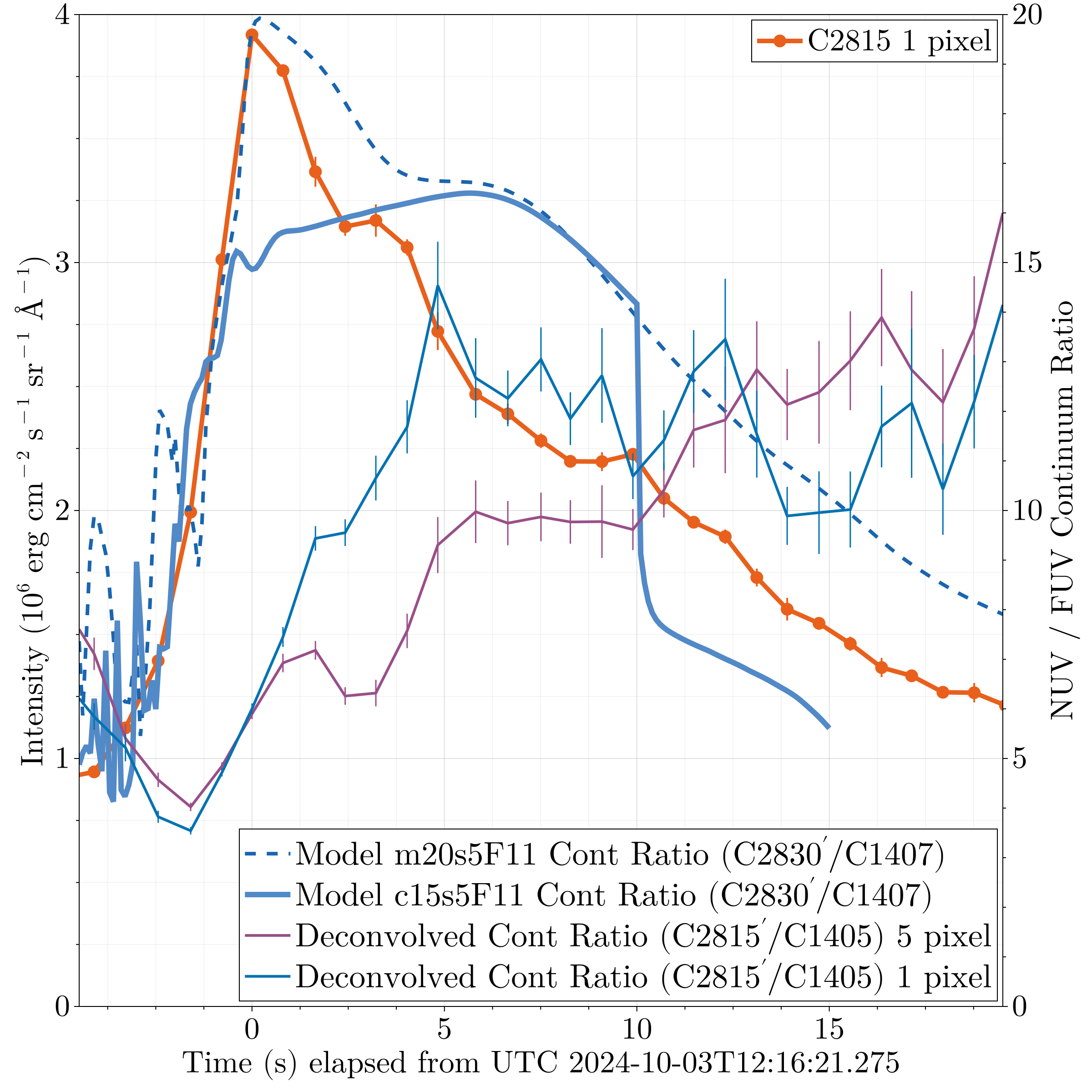}{0.5\textwidth}{(b)}}  
\caption{ (a) The IRIS/FUV flare spectra around Si IV at the same locations and times as in Figure \ref{fig:FeIISpec}(a).  The vertical dotted lines indicate the wavelength range over which the continuum intensity C1405 is calculated.  The features within C1405 window are consistent with random-noise variations.  The values of C1405 for the blue and purple spectra are indicated by horizontal dashed lines. (b) The ratio C2830$^{\prime}$/C1405 from the two 5F11 models compared to the NUV/FUV continuum ratios from BFP1.  At the peak of the C2815 light curve, the model ratios approach 20, while the data are significantly below 10.  The ratios are calculated from deconvolved IRIS intensities at 1 pixel and 5 pixel averages centered on BFP1.  The \texttt{m20s-5F11-25-4} model ratios are offset by -10~s as in Figure \ref{fig:lcs}, and the \texttt{c15s-5F11-25-4.2} model ratios are offset by -5~s.  These offsets  align the model NUV continuum intensity light curve evolution with the data. \label{fig:c1405}  }
\end{figure}

The peak NUV continuum intensity is well reproduced in the 5F11 electron beam heating models (Section \ref{sec:lcanalysis}). However, there is a discrepancy in the relative FUV continuum intensity.   The FUV continuum intensity is calculated over  the wavelength window $\lambda = 1405.1 - 1405.5$ \AA\ (hereafter, C1405) which is denoted in Figure \ref{fig:c1405}(a) for the same times as the NUV spectra in Figure \ref{fig:FeIISpec}(a).   In the IRIS FUV flare atlas from \citet{Jaeggli2024}, this region is free of major and minor emission lines.  The region is also clean in spectra of the 2024-Oct-03 flare (Figure \ref{fig:c1405}) except at some other locations within the ribbons where the Si IV line exhibits unusually large redshifts.  I test the null hypothesis that a constant intensity value explains the variations within the C1405 wavelength range.  The reduced $\chi^2$ values fall in the range of $0.9 - 1.2$.  The spectral variations in Figure \ref{fig:c1405}(a) also change among spectra, further suggesting that there are no significant emission lines systematically skewing the inferred continuum intensity in this window.

\begin{figure}
\gridline{
	\fig{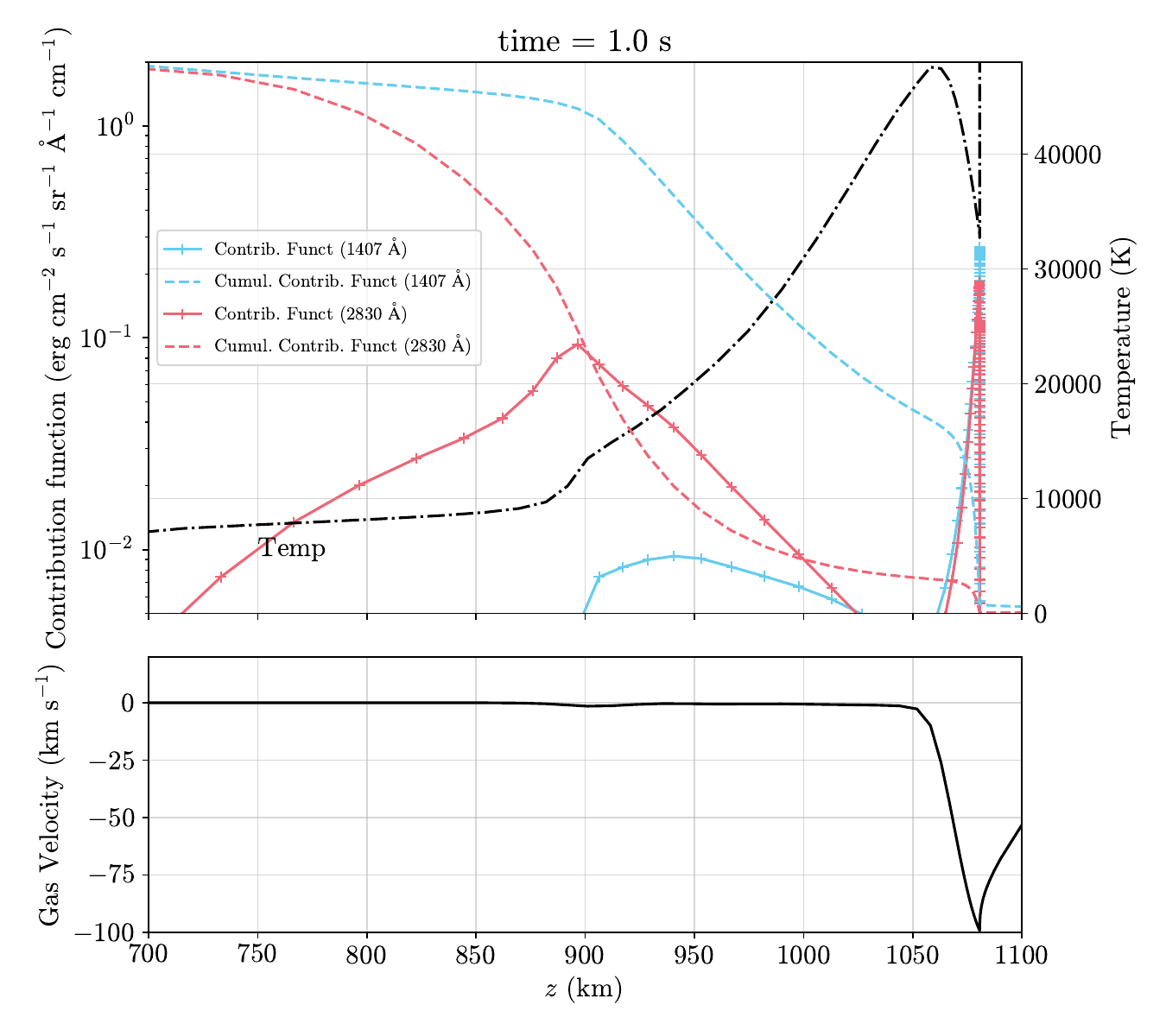}{0.5\textwidth}{(a)} }
\gridline{
	\fig{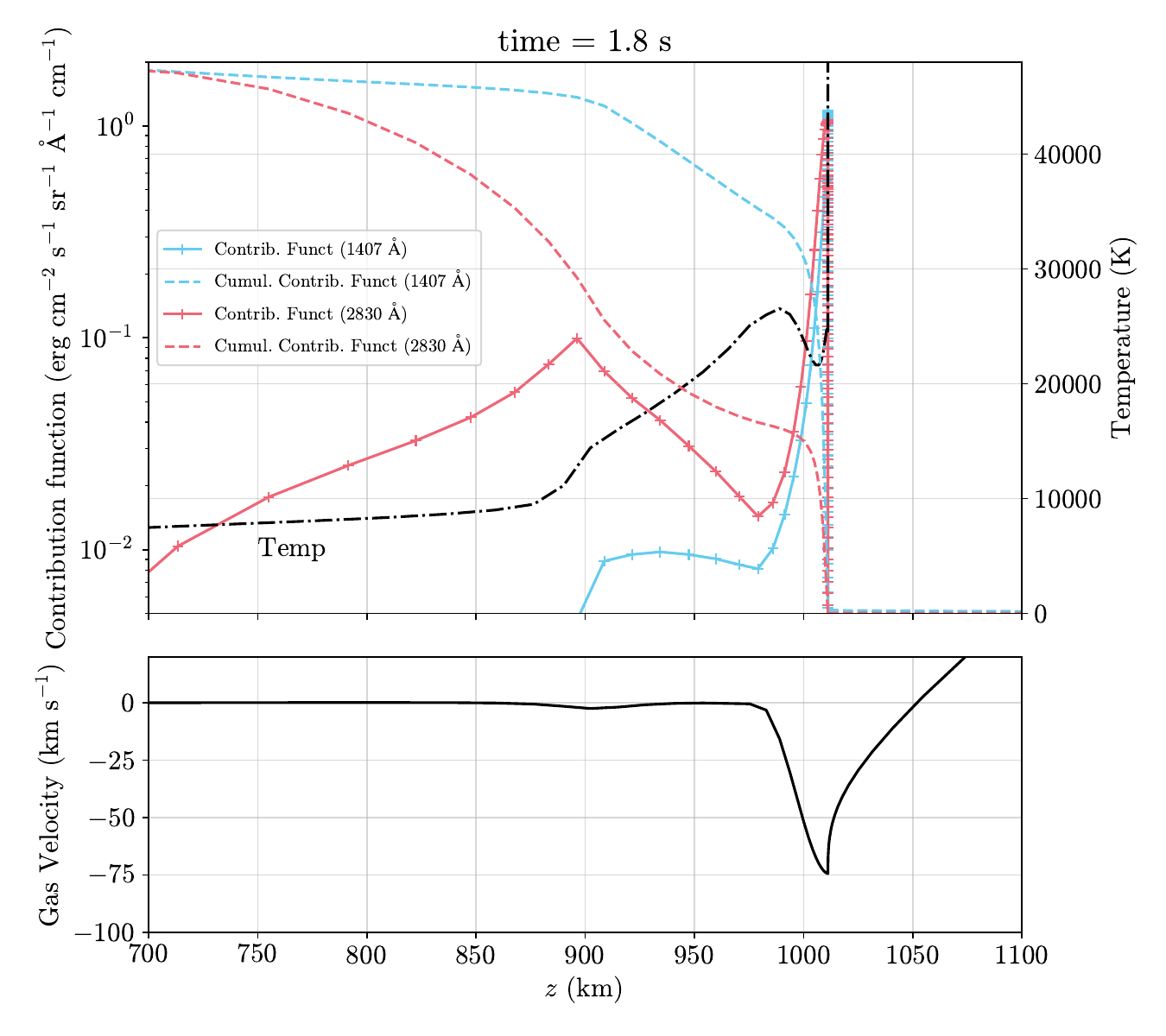}{0.5\textwidth}{(b)}}  
\caption{ Contribution functions (crosses) and cumulative contribution functions (dashed lines) for an FUV continuum wavelength (1407 $\rm{\AA}$) and an NUV continuum wavelength (2803 $\rm{\AA}$).  The snapshots are taken  from the early phases of the chromospheric condensation evolution in the \texttt{c15s-5F11-25-4.2} model.   The cumulative contribution functions are normalized to 1.0 at $z=500$ km and are plotted on a linear $y$-scaling from 0 to 1; the contribution function in the FUV is negligible below these heights. The downflowing velocities indicate the locations of the chromospheric condensation in the bottom panel of each figure; these locations  correspond to the dip in temperatures in the two upper panels. The FUV continuum has a much larger cumulative contribution at heights $z > 950 $ km  with $T\gtrsim 20,000$ K. Note, the $z$  coordinate corresponds to the  arc distance from the photosphere along the semi-circular model loop. \label{fig:contribci}  }
\end{figure}

The ratios of the flare excess NUV (C2815$^{\prime}$) to FUV (C1405) continuum intensities are shown in Figure \ref{fig:c1405}(b). Here, I ratio the deconvolved spectral intensities in the FUV and the NUV to account for the wavelength-dependent spatial resolution of IRIS \citep{DePontieu2014}.  Five-pixel averages and single-pixel sampling show similar behaviors throughout the C2815$^{\prime}$ evolution.  The ratio first attains significant values around $\sim 5$, which then decrease to a minimum of nearly $3.5-4.0$ in the early rise phase just before the main peak of the C2815$^{\prime}$ light curve.  At the peak of the light curve, the ratio has increased to $\sim 6-7$ and it attains a value of $\approx 10$ by the end of its impulsive phase. Thereafter, the ratios remain roughly constant at values of $\approx 10-15$. The \texttt{m20s-5F11-25-4.2} and \texttt{c15s-5F11-25-4.2} models attain NUV/FUV ratios as large as $15-20$ at peak  C2830 brightness, and they evolve in the opposite sense to the observations.  Thus, the 5F11 electron beam heating models of the NUV continuum intensity do not satisfactorily explain all of the observed FUV continuum intensity during the BFP1 evolution.  A similar discrepancy is found  comparing to the brightest ribbon sources in the 2014-Mar-29 flare, which exhibits a C2826$^{\prime}$/C1349$^{\prime}$ ratio of $\approx 5$ \citep[Paper 1; see also][]{Kleint2016}.

To determine the origin of the discrepancy, I analyze the contribution function to the emergent continuum intensities at $\lambda = 1407$ \AA\ in the FUV and $\lambda = 2830$ \AA\ in the NUV from the \texttt{c15s-5F11-25-4.2} model. The contribution functions and the cumulative contribution functions (normalized to 1 at $z=500$ km) at $t=1.0$ and $t=1.8$~s are shown in  Figure \ref{fig:contribci}(a) and Figure \ref{fig:contribci}(b), respectively.   The $\lambda = 2830$ \AA\ continuum intensity has a large contribution from heights with temperatures around $T=10^4$ K (see also Paper I). About 70\% and 50\% of the emergent NUV continuum intensity originates from these relatively cooler temperatures between $T=8200-20,000$ K at $t=1.0$ and $t=1.8$~s, respectively. The fraction of the emergent FUV continuum intensity that originates from $T=8200-20,000$ K is only 25\% and 13\% at $t=1.0$ and $t=1.8$~s, respectively.  At hotter temperatures of $T=20,000 -40,000$ K, the FUV contribution function is comparable to the NUV contribution function;  at even hotter temperatures around $T = 160,000$ K in the flare transition region, the FUV continuum contribution function exceeds the NUV. 
The locations of the chromospheric condensation (bottom panels) correspond to these hotter temperatures at these times.  In the condensation, there is a rise in the cumulative contribution function in the FUV to about 0.7 in these hot layers at $t=1.8$~s.  At the earlier time ($t=1$~s), most of the FUV continuum contribution function originates from the hotter layers just below the condensation with about 30\% formed within the condensation.  These snapshots are representative of early times during the chromospheric condensation evolution (cf. Fig 2 of Paper II) when it rapidly radiatively cools through temperatures of $T \approx 3 \times 10^4$ K as it is accrues mass.  
  The calculated ratios ($\sim 4.5-5$) of the emergent intensities at $t=1-1.8$~s are close to the minimum observed ratios ($\sim 3.5$) within the early fast rise phase of the light curve.   However, the condensation cools  to $T\approx 10^4$ K, and the calculated ratios of the emergent intensities become as large as $10-20$.   Thus, smaller NUV to FUV ratios are produced when the  condensation is hotter than $T\approx 10^4$ K.

A longer radiative cooling timescale through these higher temperatures present in the condensation at $t=1-1.8$~s may conceivably lead to smaller continuum ratios that extend another $\approx 5$~s further into the fast rise phase.   In RADYN, the optically thin radiative cooling is tabulated from the CHIANTI database for all elements except for those treated in detail (hydrogen, helium, and calcium II).  For the elements treated in detail, the radiative cooling is calculated from the transfer equation \citep[see, e.g.,][]{Allred2015}.   I analyze the terms in the conservation equation for specific internal energy, following previous RADYN papers.   At $t=1$~s, the optically thin radiative losses completely dominate over the detailed radiative losses at $z > 1000$ km and within the cooling condensation.  By $t=3.8$~s, the detailed radiative losses dominate over optically thin radiative losses as the condensation cools further.  \citet{KAC24} discuss that the amount of radiative losses at $T = 10^4$ K assumed to be optically thin (and thus included in the thin loss table) are very important in establishing the temperature in the deep chromosphere in stellar flare simulations.  Here, I speculate that the radiative losses assumed to be optically thin at $T = 20,000 - 80,000$ K could also be over-estimated in flare simulations, thus cooling the condensation too strongly.  The models of \citet{Kerr2019Si} show that opacity affects the formation of the Si IV resonance lines in a flare atmosphere with lower densities than the 5F11 models. Further study is needed to determine if a better treatment of the bound-bound transitions throughout the FUV can affect the evolution of chromospheric condensations appreciably.

\section{Discussion} \label{sec:discussion}

Only a few recent radiative-hydrodynamic models have considered large electron beam heating flux densities, $\gtrsim 5\times10^{11}$ erg cm$^{-2}$ s$^{-1}$ \citep[Paper I,][]{Kennedy2015}.  Most solar flare models use significantly smaller heating fluxes \citep[e.g.,][]{Simoes2024B, Lorincik2025} of  $\approx 10^{10} - 10^{11}$ erg cm$^{-2}$ s$^{-1}$, which is commonly reported as a fiducial range for solar flares \citep[e.g.,][]{Fletcher2008}.  Other static atmospheric models \citep{RC83, Heinzel2014, Kleint2016} can account for NUV continuum intensities in IRIS that are at most $1.74 \times 10^6$ erg cm$^{-2}$ s$^{-1}$ sr$^{-1}$ \AA$^{-1}$.  The 2024-Oct-03 flare produces a ``record'' NUV continuum intensity of at least $5-6\times 10^6$ \ilam.  Relatively low heating rates are thus not appropriate for the brightest kernels in solar flares, which severely challenge theories of nonthermal energy  transport to the chromosphere \citep{Lee2008,LeeBuchner2011,Li2014}. 

 In this paper, I have compared two fiducial simulations with large nonthermal electron energy flux densities to a comprehensive optical and ultraviolet dataset of an X9 solar flare.  Several predictions of high flux beam models are observed for the first time:  extremely bright NUV continuum intensities, red wing Fe I and Fe II line asymmetries that become brighter than the emission line components around the rest wavelength, and extreme near-wing broadening of H$\alpha$, which is roughly comparable to the broadening in the H$\alpha$ spectra of \citet{Ichimoto1984}.  To my knowledge,  all other solar flare models \citep[e.g.,][]{Abbett1998, Allred2005, Kasparova2009, Kuridze2015, Rubio2016, Graham2020, Simoes2024B, Yu2025, Litwicka} do not predict such line broadening and bright NUV continuum intensities. It was expected that the 5F11 model widths would far exceed observational constraints, since the  condensation densities (e.g., Figure \ref{fig:haeldens}) are an order of magnitude larger than electron densities that have been previously inferred from the Balmer lines in solar flares \citep[e.g.,][]{Svestka1976, Neidig1983, Donati1985}. They are nearly two orders of magnitude greater than chromospheric flare densities in many modern-day models \citep[e.g.,][]{Druett2018}.  My analysis leverages new deconvolution techniques when comparing model specific intensities to the observed intensities in IRIS and CHASE data, which do not spatially resolve the bright kernels in the 2024-Oct-03 flare.  
 
  There are several limitations to the modeling that warrant further discussion (see also Paper I).  An improved beam transport model that accounts for return current electric field Ohmic heating \citep{Allred2020} and possibly runaways \citep{Alaoui2021} is needed for spectral comparisons.  In a recent paper, \citet{KAC24} discuss a possible improvement to the pressure broadening of H$\alpha$.  Preliminary work suggests that neither improvements to the H$\alpha$ broadening nor energy loss from steady state return current electric fields largely affect the results herein (Kowalski \& Gomez in prep, Kowalski et al. 2025, in prep).  The models in Papers I and II, which include only Coulomb collisions in the energy loss during the beam transport, are useful benchmarks that relate to the assumptions in the widely used standard collisional thick target model of solar flare hard X-rays.  Discriminating among more sophisticated models would benefit from broader and more complete wavelength coverage in the wings of the Balmer lines.

  \begin{figure}
  	\begin{center}
  		\includegraphics[width=1.0\textwidth]{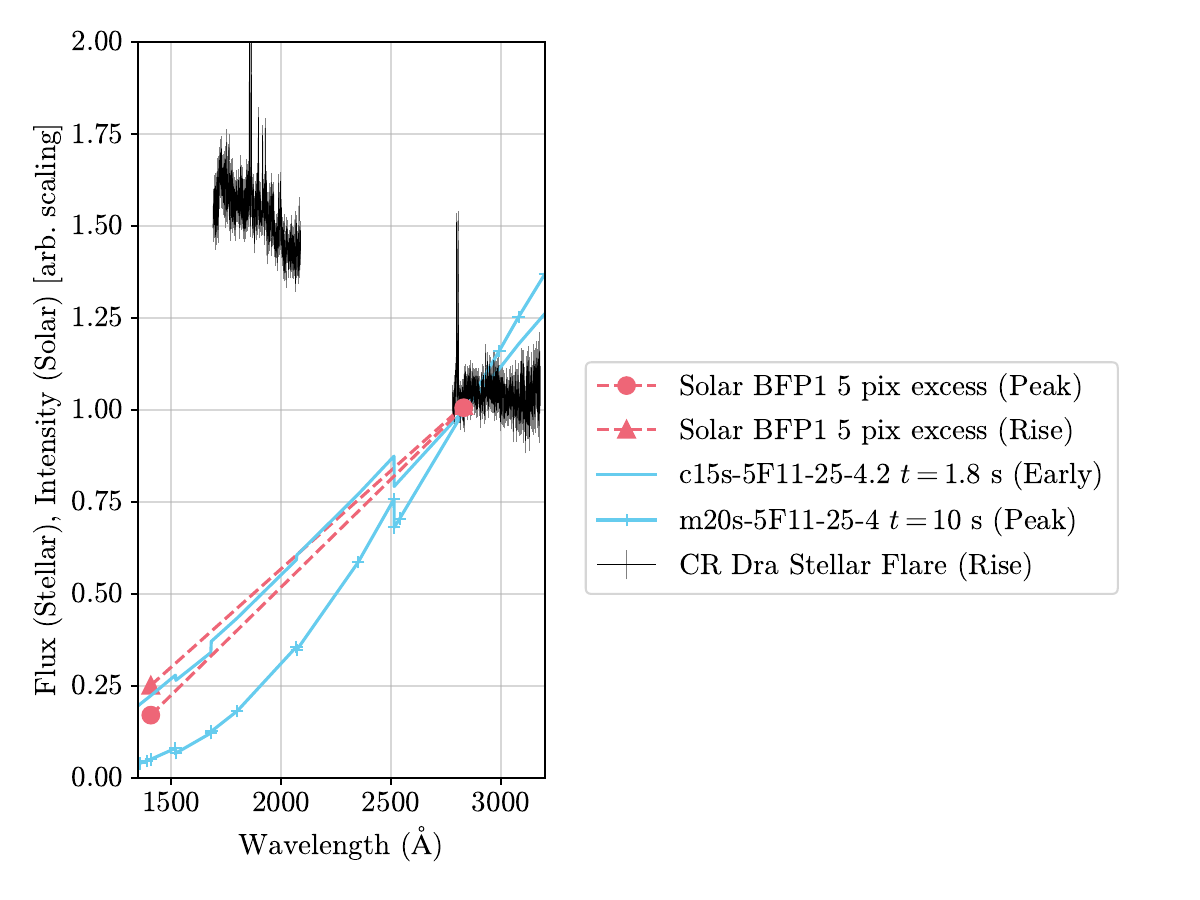}
  		\caption{ Comparison of a stellar megaflare spectrum \citep{Kowalski2025} from the Hubble Space Telescope to the constraints from BFP1 in the X9 solar flare.  Deconvolved intensities summed over 5 pixels are used in this comparison.  Several RADYN model continuum spectra generally reproduce the shapes implied by the IRIS solar flare constraints.  All spectra are scaled to a common wavelength around 2815 \AA.  
  			\label{fig:solarstellar}}
  	\end{center}
  \end{figure}
  
   Although the model beam-generated chromospheric condensation densities and optical depths give reasonable NUV continuum intensities and H$\alpha$ wing broadening extents, the redshifted component in the Fe II line is still too bright.  Also, there is a high velocity tail of redshifted Fe II intensity out to 80--100 km s$^{-1}$ that is not predicted by models.  The Mg II lines profiles (not shown) in this flare are too narrow, which is expected because no additional damping physics \citep{Zhu2019} has been added to the models in recent years.  The comparisons to Mg II will be presented in a future work (Kowalski et al 2026) with an update to the broadening theory (Gomez et al. 2025, in prep).  Further understanding of these  properties may ultimately require a multi-dimensional treatment of the chromospheric condensation evolution in the framework of radiative magnetohydrodynamics;  work along this line is in progress.
 
 Sustaining chromospheric condensations at higher temperatures (around $T \approx 20,000-30,000$ K) for longer durations may produce relatively fainter red-wing asymmetries.  The Fe II emissivity is an order of magnitude smaller at higher temperatures (Figure \ref{fig:FeIISpec}(c)), while the population densities of hydrogen in the $n=2$ stage are factors of $\approx 20$ smaller.
 Sustained temperatures may also facilitate longer durations of smaller NUV/FUV continuum ratios.  Previous modeling of the FUV continuum intensity in solar flares attributes FUV continuum enhancements to the increase of the photoionization of minor elements (namely Si I) by transition region lines \citep{Phillips1992,Doyle1992B}.  In contrast, the RADYN models suggest a similar origin as for the NUV continuum intensity: chromospheric condensations higher in the atmosphere.  Unlike  NUV continuum emissivity, however,  FUV continuum emission is larger at $T \approx 2-4\times 10^4$ K (and yet even stronger at $T \approx 10^5$ K) than at $T = 12,000-14,000$ K.  The small NUV/FUV continuum ratios thus probe the early phases of the  temperature evolution of chromospheric condensations.  The assumption that flare continuum radiation in the SDO/AIA 1600\AA\ and 1700\AA\ bands is due to a type of temperature minimum region heating, as in the quiet Sun \citep{Vernazza1981}, requires reconsideration. Isothermal slab approximations \citep[e.g.,][]{Donati1985,Potts2010,Dominique2018,Simoes2024,Yang2025}  are further limited in their predictions for the FUV continuum intensity, which shows a relatively strong response in the early evolution of a hot ($T=20,000 - 40,000$ K) chromospheric condensation in the RADYN models.  %in the brightest (and thus most strongly heated) footpoint sources.

 In recent stellar megaflare observations from the Hubble Space Telescope, the short-wavelength NUV continuum fluxes at $\lambda \approx 1680-2100$ \AA\ surprisingly rise above continuum fluxes spanning $\lambda \approx 2750-3150$ \AA\ \citep{Kowalski2025}.  In the solar flare data, however, the NUV continuum intensity is brighter than the FUV continuum intensity by factors of at least $\approx 3-10$ at all times.   This constrains the peak of the brightest sources in solar flares to longer wavelengths than the FUV and has important implications for the broadband energetics of solar flares \citep[e.g.,][]{Woods2004,Fletcher2007,WarmuthMann2016}.  Figure \ref{fig:solarstellar} compares the brightest NUV kernel (BFP1) in the 2024-Oct-03 solar flare to the rise phase spectrum of the stellar ``Flare Event 2'' from \citet{Kowalski2025}.
  The radically different spectra suggest different strengths and depths of atmospheric heating in powerful solar flares Sun and stellar megaflares.   
  
  Models of the stellar flare UV radiation place its origin deeper in the atmosphere, which is heated directly by extremely high energy electron or proton beams \citep{Kowalski2025}. In Figure \ref{fig:solarstellar}, several snapshots from the two solar 5F11 detailed continua are scaled to the observed C2815$^{\prime}$ intensity at the times of the early fast rise phase and the maximum continuum brightness of BFP1 (e.g., Figure \ref{fig:lcs}). The model spectra include continua from the three elements (H, He, Ca II) calculated in detail and background bound-free opacities in LTE from several minor elements \citep[see][]{KAC24}.  The \texttt{m20s-5F11} model at the time of its maximum continuum brightness under-predicts the FUV continuum intensity C1405 (Section \ref{sec:ratioanalysis}), whereas the spectrum from an early phase of the condensation in the \texttt{c20s-5F11} model decreases from the NUV to the FUV in the same way as the data (as described above).  Model chromospheric condensations of stellar flares \citep{Kowalski2015} generally are so dense that the hydrogen wings become broader than available observational constraints \citep{HP91, Kowalski2022Frontiers}. I speculate that reconciling the evolution of chromospheric  condensations with the observed NUV/FUV continuum ratios in solar flares may help align denser stellar flare chromospheric condensation models with Hubble Space Telescope observations.

\section{Summary \& Conclusions} \label{sec:conclusions}
The IRIS and CHASE satellites observed several bright kernels in one of the largest X-class solar flares in decades.  In this unique dataset, the brightest observed NUV continuum source is much brighter than the 2014-Sept-10 X1 flare \citep{Graham2020}, and  the IRIS cadence is much faster than the well-studied 2014-Mar-29 X1 flare (Paper I).  The new data provide comprehensive tests to the current state of 1D radiative-hydrodynamic flare models, including two fiducial models with large electron beam heating rates from the literature that are analyzed here.  A summary of the tests is outlined in Table \ref{table:summary}.  Many salient properties of bright solar flare kernels can be explained by high-flux electron beam heating.  However, there are several constraints that motivate model improvements.

\begin{deluxetable}{lll}[ht!]
	\tablecaption{Summary of Model Comparisons} \label{table:summary}
	\tablewidth{700pt}
	\tabletypesize{\scriptsize}
	\tablehead{
		\colhead{Constraint} & \colhead{Grade} & \colhead{Reference Figure}} 
	\startdata
	Deconvolved NUV continuum peak intensity & $++$ & \ref{fig:lcs} \\
	Deconvolved NUV continuum intensity evolution & $+$ & \ref{fig:lcs} \\
	Ratio of NUV to FUV continuum intensity & $+$ & \ref{fig:c1405} \\
	Temporal evolution of ratio of NUV to FUV continuum intensity & $-$ & \ref{fig:c1405} \\
	Brighter Fe II red-wing asymmetry than rest-wavelength & $+$ & \ref{fig:FeIISpec} \\
	Fe II red-wing and rest-wavelength components equal after peak NUV continuum intensity & $-$ & \ref{fig:FeIISpec} \\
	High velocity tail $>50-100$ km s$^{-1}$ to Fe II red-wing asymmetry & $-$ & \ref{fig:FeIISpec}, \ref{fig:Mar29} \\
	$\lesssim 7$ km s$^{-1}$ blueshift of Fe II rest wavelength component & $-$ & \ref{fig:FeIISpec} \\
	Two components in the Fe I $\lambda2814.11$ line & $+$ & \ref{fig:FeIISpec} \\
	Nearly identical red-wing asymmetry evolution in Fe I and Fe II & $+$ & \ref{fig:FeIISpec} \\
	H$\alpha$ line maximum intensity value & $+$ & \ref{fig:chasecontext} \\
	H$\alpha$ near-wing broadening & $++$& \ref{fig:chasecontext} \\
	H$\alpha$ red-wing asymmetry spectral peak & $-$ & \ref{fig:chasecontext}\\
	\enddata
	\tablecomments{Comparisons between the spectra of BFP1 and 5F11 model predictions. $++$ indicates that the quantity is well-reproduced quantitatively, $+$ indicates that the property is well reproduced qualitatively, and $-$ indicates that the property is not well reproduced.  The assessments for the H$\alpha$ near-wing broadening and line maximum intensity are with respect to Pos 1 in the CHASE data.}
\end{deluxetable}

 The chromospheric condensation attains maximum electron densities of $n_e \approx 5 \times 10^{14}$ cm$^{-3}$ and optical depths $\tau(\lambda) \approx 1$ in the near wings of H$\alpha$.  These conditions generate large amounts of pressure and optical depth broadening.  The observed H$\alpha$ spectra show remarkable wing broadening to at least $\lambda - \lambda_0 = _{-3.4}^{+6.9}$ \AA, which is satisfactorily reproduced by a dense chromospheric condensation model. The spectral shape of the blue side  as well as the spectral shape and intensity at $\lambda - \lambda_0 = +4.7$ to $+6.9$ \AA\ in the CHASE Fe I window are especially well matched with the observations of Pos 1 (Figure \ref{fig:chasecontext}).  However, detailed temporal comparisons of the $t=12$~s model to the observations of Pos 2 reveal that the H$\alpha$ line is too broad on the blue side and too flat on the red side in the Fe I window. The large electron densities in the chromospheric condensations produce very bright NUV continuum intensities, which agree with the observations to within $\approx$ 20\% at their respective maxima but show larger discrepancies ($\approx 60-80$\%) in the decay.

At the time of the peak NUV continuum light curve, the FUV continuum intensities in the high beam flux models are too faint by factors of $\approx 2-3$.   The disagreement with the RADYN models (e.g., Figure \ref{fig:solarstellar}) implies that hydrogen recombination continuum spectra formed over low optical depth at $T=10^4$ K from the upper chromosphere and recombination continua of minor elements from the temperature minimum do not solely explain the broadband UV energetics of solar flares. A hotter ($T=20,000 -40,000$ K) contribution from a chromospheric condensation produces additional FUV continuum intensity that better agrees with the data. Nonetheless, the solar flare kernel FUV continuum intensities are not nearly bright enough to explain the short-wavelength NUV rising continua in stellar flares. This reveals a major solar-stellar disconnection between flare heating mechanisms in the Sun and in very active M-dwarf flare stars.

The observed Fe I and Fe II spectra show prominent redshifted emission line components that brighten over the intensity around the respective rest wavelengths.  This is a notable property that has not been seen before.  These spectra are interpreted with chromospheric condensation models from Papers I and II, but the Fe lines in the IRIS/NUV are also sensitive to heating in the stationary layers at $T \approx 7000-9000$ K that are just below the condensation (see also Paper I).  
 The model chromospheric condensations produce optically thick Fe II and Fe I emitting layers; although in qualitative  agreement with the observations, the optical depths generate redshifted components that are too bright too early in the evolution of the NUV continuum intensity.  This discrepancy could in part be due to the model condensations cooling too quickly because large electron densities are important for explaining some properties of the H$\alpha$ wing broadening and the bright peak of the NUV continuum intensity.  Detailed calculations are needed to verify this hypothesis.  Spectral coverage spanning the FUV and NUV \citep[e.g.,][]{Cook1979,Simoes2019} and further into the optically thin wings of the Balmer lines \citep[e.g.,][]{Namekata2022} at high spatial resolution would further guide improvements to models of chromospheric condensations in bright solar flare kernels.

\begin{acknowledgements}
AFK gratefully acknowledges an anonymous referee for their comments and critiques, which greatly improved the clarity of the paper and presentation of the results.
IRIS is a NASA small explorer mission developed and operated by LMSAL with mission operations executed at NASA Ames Research Center and major contributions to downlink communications funded by ESA and the Norwegian Space Centre.  This work uses the data from CHASE mission supported by China National Space Administration.  Most figures were made in the Julia programming language with the Makie plotting software \citep{Makie}.  I thank Dr. Kevin Reardon for the Hamburg-Neckel solar intensity spectrum. I thank Dr. Gianna Cauzzi for referring me to CHASE H$\alpha$ data of solar flares.

\end{acknowledgements}

\bibliography{main.bib}{}

\begin{thebibliography}{}
\expandafter\ifx\csname natexlab\endcsname\relax\def\natexlab#1{#1}\fi
\providecommand{\url}[1]{\href{#1}{#1}}
\providecommand{\dodoi}[1]{doi:~\href{http://doi.org/#1}{\nolinkurl{#1}}}
\providecommand{\doeprint}[1]{\href{http://ascl.net/#1}{\nolinkurl{http://ascl.net/#1}}}
\providecommand{\doarXiv}[1]{\href{https://arxiv.org/abs/#1}{\nolinkurl{https://arxiv.org/abs/#1}}}

% type= phdthesis
\bibitem[{W.~P. {Abbett}(1998){Abbett}}]{Abbett1998}
{Abbett}, W.~P. 1998, \bibinfo{title}{{A Theoretical Investigation of Optical
  Emission in Solar Flares},} PhD thesis, MICHIGAN STATE UNIVERSITY

% type= article
\bibitem[{E. {Acampa} {et~al.}(1982){Acampa}, {Smaldone}, {Sambuco}, \&
  {Falciani}}]{Acampa1982}
{Acampa}, E., {Smaldone}, L.~A., {Sambuco}, A.~M., \& {Falciani}, R. 1982,
  \bibinfo{title}{{Analysis of the optical spectra of solar flares. I - The
  flare of April 30, 1976},} \aaps, 47, 485

% type= article
\bibitem[{M. {Alaoui} {et~al.}(2021){Alaoui}, {Holman}, {Allred}, \&
  {Eufrasio}}]{Alaoui2021}
{Alaoui}, M., {Holman}, G.~D., {Allred}, J.~C., \& {Eufrasio}, R.~T. 2021,
  \bibinfo{title}{{Role of Suprathermal Runaway Electrons Returning to the
  Acceleration Region in Solar Flares},} \apj, 917, 74,
  \dodoi{10.3847/1538-4357/ac0820}

% type= article
\bibitem[{J.~C. {Allred} {et~al.}(2020){Allred}, {Alaoui}, {Kowalski}, \&
  {Kerr}}]{Allred2020}
{Allred}, J.~C., {Alaoui}, M., {Kowalski}, A.~F., \& {Kerr}, G.~S. 2020,
  \bibinfo{title}{{Modeling the Transport of Nonthermal Particles in Flares
  Using Fokker-Planck Kinetic Theory},} \apj, 902, 16,
  \dodoi{10.3847/1538-4357/abb239}

% type= article
\bibitem[{J.~C. {Allred} {et~al.}(2005){Allred}, {Hawley}, {Abbett}, \&
  {Carlsson}}]{Allred2005}
{Allred}, J.~C., {Hawley}, S.~L., {Abbett}, W.~P., \& {Carlsson}, M. 2005,
  \bibinfo{title}{{Radiative Hydrodynamic Models of the Optical and Ultraviolet
  Emission from Solar Flares},} \apj, 630, 573, \dodoi{10.1086/431751}

% type= article
\bibitem[{J.~C. {Allred} {et~al.}(2006){Allred}, {Hawley}, {Abbett}, \&
  {Carlsson}}]{Allred2006}
{Allred}, J.~C., {Hawley}, S.~L., {Abbett}, W.~P., \& {Carlsson}, M. 2006,
  \bibinfo{title}{{Radiative Hydrodynamic Models of Optical and Ultraviolet
  Emission from M Dwarf Flares},} \apj, 644, 484, \dodoi{10.1086/503314}

% type= article
\bibitem[{J.~C. {Allred} {et~al.}(2015){Allred}, {Kowalski}, \&
  {Carlsson}}]{Allred2015}
{Allred}, J.~C., {Kowalski}, A.~F., \& {Carlsson}, M. 2015, \bibinfo{title}{{A
  Unified Computational Model for Solar and Stellar Flares},} \apj, 809, 104,
  \dodoi{10.1088/0004-637X/809/1/104}

% type= article
\bibitem[{A. {Asai} {et~al.}(2002){Asai}, {Masuda}, {Yokoyama}, {Shimojo},
  {Isobe}, {Kurokawa}, \& {Shibata}}]{Asai2002}
{Asai}, A., {Masuda}, S., {Yokoyama}, T., {et~al.} 2002,
  \bibinfo{title}{{Difference between Spatial Distributions of the
  H{\ensuremath{\alpha}} Kernels and Hard X-Ray Sources in a Solar Flare},}
  \apjl, 578, L91, \dodoi{10.1086/344566}

% type= article
\bibitem[{M.~J. {Aschwanden}(2004){Aschwanden}}]{Aschwanden2004}
{Aschwanden}, M.~J. 2004, \bibinfo{title}{{Pulsed Particle Injection in a
  Reconnection-Driven Dynamic Trap Model in Solar Flares},} \apj, 608, 554,
  \dodoi{10.1086/392494}

% type= article
\bibitem[{W.~H. {Ashfield} \& D.~W. {Longcope}(2021){Ashfield} \&
  {Longcope}}]{Ashfield2021}
{Ashfield}, W.~H., \& {Longcope}, D.~W. 2021, \bibinfo{title}{{Relating the
  Properties of Chromospheric Condensation to Flare Energy Transported by
  Thermal Conduction},} \apj, 912, 25, \dodoi{10.3847/1538-4357/abedb4}

% type= article
\bibitem[{A.~F. {Battaglia} \& S. {Krucker}(2025){Battaglia} \&
  {Krucker}}]{Battaglia2025}
{Battaglia}, A.~F., \& {Krucker}, S. 2025, \bibinfo{title}{{New insights into
  the proton precipitation sites in solar flares},} \aap, 694, A58,
  \dodoi{10.1051/0004-6361/202453144}

% type= phdthesis
\bibitem[{E. {Butler}(2022){Butler}}]{Butler2022PhD}
{Butler}, E. 2022, \bibinfo{title}{{On Timely Behavior: Improving the Space
  Weather Research/Operations Pipeline and Investigating Solar Flare
  Ultraviolet Time Series},} PhD thesis, University of Colorado, Boulder

% type= article
\bibitem[{E.~C. {Butler} \& A.~F. {Kowalski}(2024){Butler} \&
  {Kowalski}}]{Butler2024}
{Butler}, E.~C., \& {Kowalski}, A.~F. 2024, \bibinfo{title}{{Decay Timescales
  of Chromospheric Condensations in Solar Flare Footpoints},} \apj, 970, 33,
  \dodoi{10.3847/1538-4357/ad3dfb}

% type= article
\bibitem[{R.~C. {Canfield} {et~al.}(1984){Canfield}, {Gunkler}, \&
  {Ricchiazzi}}]{Canfield1984}
{Canfield}, R.~C., {Gunkler}, T.~A., \& {Ricchiazzi}, P.~J. 1984,
  \bibinfo{title}{{The H-alpha spectral signatures of solar flare nonthermal
  electrons, conductive flux, and coronal pressure},} \apj, 282, 296,
  \dodoi{10.1086/162203}

% type= article
\bibitem[{R.~C. {Canfield} {et~al.}(1990){Canfield}, {Penn}, {Wulser}, \&
  {Kiplinger}}]{Canfield1990}
{Canfield}, R.~C., {Penn}, M.~J., {Wulser}, J.-P., \& {Kiplinger}, A.~L. 1990,
  \bibinfo{title}{{H alpha Spectra of Dynamic Chromospheric Processes in Five
  Well-observed X-Ray Flares},} \apj, 363, 318, \dodoi{10.1086/169345}

% type= article
\bibitem[{M. {Carlsson}(1986){Carlsson}}]{MULTI1}
{Carlsson}, M. 1986, \bibinfo{title}{{A computer program for solving
  multi-level non-LTE radiative transferproblems in moving or static
  atmospheres.},} Uppsala Astronomical Observatory Reports, 33

% type= article
\bibitem[{M. {Carlsson} \& R.~F. {Stein}(1992){Carlsson} \&
  {Stein}}]{Carlsson1992B}
{Carlsson}, M., \& {Stein}, R.~F. 1992, \bibinfo{title}{{Non-LTE Radiating
  Acoustic Shocks and CA II K2V Bright Points},} \apjl, 397, L59,
  \dodoi{10.1086/186544}

% type= article
\bibitem[{M. {Carlsson} \& R.~F. {Stein}(1995){Carlsson} \&
  {Stein}}]{Carlsson1995}
{Carlsson}, M., \& {Stein}, R.~F. 1995, \bibinfo{title}{{Does a Nonmagnetic
  Solar Chromosphere Exist?},} \apjl, 440, L29, \dodoi{10.1086/187753}

% type= article
\bibitem[{M. {Carlsson} \& R.~F. {Stein}(1997){Carlsson} \&
  {Stein}}]{Carlsson1997}
{Carlsson}, M., \& {Stein}, R.~F. 1997, \bibinfo{title}{{Formation of Solar
  Calcium H and K Bright Grains},} \apj, 481, 500

% type= article
\bibitem[{M. {Carlsson} {et~al.}(2023){Carlsson}, {Fletcher}, {Allred},
  {Heinzel}, {Ka{\v{s}}parov{\'a}}, {Kowalski}, {Mathioudakis}, {Reid}, \&
  {Sim{\~o}es}}]{Carlsson2023}
{Carlsson}, M., {Fletcher}, L., {Allred}, J., {et~al.} 2023,
  \bibinfo{title}{{The F-CHROMA grid of 1D RADYN flare models},} \aap, 673,
  A150, \dodoi{10.1051/0004-6361/202346087}

% type= article
\bibitem[{J.~W. {Cook} \& G.~E. {Brueckner}(1979){Cook} \&
  {Brueckner}}]{Cook1979}
{Cook}, J.~W., \& {Brueckner}, G.~E. 1979, \bibinfo{title}{{EUV continua of
  solar flares 1420 - 1960 {\r{A}}.},} \apj, 227, 645, \dodoi{10.1086/156775}

% type= article
\bibitem[{H. {Courrier} {et~al.}(2018){Courrier}, {Kankelborg}, {De Pontieu},
  \& {W{\"u}lser}}]{Courrier2018}
{Courrier}, H., {Kankelborg}, C., {De Pontieu}, B., \& {W{\"u}lser}, J.-P.
  2018, \bibinfo{title}{{An on Orbit Determination of Point Spread Functions
  for the Interface Region Imaging Spectrograph},} \solphys, 293, 125,
  \dodoi{10.1007/s11207-018-1347-9}

% type= article
\bibitem[{A.~J. {Coyner} \& D. {Alexander}(2009){Coyner} \&
  {Alexander}}]{Coyner2009}
{Coyner}, A.~J., \& {Alexander}, D. 2009, \bibinfo{title}{{Implications of
  Temporal Development of Localized Ultraviolet and Hard X-ray Emission for
  Large Solar Flares},} \apj, 705, 554, \dodoi{10.1088/0004-637X/705/1/554}

% type= article
\bibitem[{S. Danisch \& J. Krumbiegel(2021)Danisch \& Krumbiegel}]{Makie}
Danisch, S., \& Krumbiegel, J. 2021, \bibinfo{title}{{Makie.jl}: Flexible
  high-performance data visualization for {Julia},} Journal of Open Source
  Software, 6, 3349, \dodoi{10.21105/joss.03349}

% type= article
\bibitem[{B. {De Pontieu} {et~al.}(2014){De Pontieu}, {Title}, {Lemen},
  {Kushner}, {Akin}, {Allard}, {Berger}, {Boerner}, {Cheung}, {Chou}, {Drake},
  {Duncan}, {Freeland}, {Heyman}, {Hoffman}, {Hurlburt}, {Lindgren}, {Mathur},
  {Rehse}, {Sabolish}, {Seguin}, {Schrijver}, {Tarbell}, {W{\"u}lser},
  {Wolfson}, {Yanari}, {Mudge}, {Nguyen-Phuc}, {Timmons}, {van Bezooijen},
  {Weingrod}, {Brookner}, {Butcher}, {Dougherty}, {Eder}, {Knagenhjelm},
  {Larsen}, {Mansir}, {Phan}, {Boyle}, {Cheimets}, {DeLuca}, {Golub}, {Gates},
  {Hertz}, {McKillop}, {Park}, {Perry}, {Podgorski}, {Reeves}, {Saar}, {Testa},
  {Tian}, {Weber}, {Dunn}, {Eccles}, {Jaeggli}, {Kankelborg}, {Mashburn},
  {Pust}, {Springer}, {Carvalho}, {Kleint}, {Marmie}, {Mazmanian}, {Pereira},
  {Sawyer}, {Strong}, {Worden}, {Carlsson}, {Hansteen}, {Leenaarts},
  {Wiesmann}, {Aloise}, {Chu}, {Bush}, {Scherrer}, {Brekke}, {Martinez-Sykora},
  {Lites}, {McIntosh}, {Uitenbroek}, {Okamoto}, {Gummin}, {Auker}, {Jerram},
  {Pool}, \& {Waltham}}]{DePontieu2014}
{De Pontieu}, B., {Title}, A.~M., {Lemen}, J.~R., {et~al.} 2014,
  \bibinfo{title}{{The Interface Region Imaging Spectrograph (IRIS)},}
  \solphys, 289, 2733, \dodoi{10.1007/s11207-014-0485-y}

% type= article
\bibitem[{B.~R. {Dennis} \& R.~L. {Pernak}(2009){Dennis} \&
  {Pernak}}]{Dennis2009}
{Dennis}, B.~R., \& {Pernak}, R.~L. 2009, \bibinfo{title}{{Hard X-Ray Flare
  Source Sizes Measured with the Ramaty High Energy Solar Spectroscopic
  Imager},} \apj, 698, 2131, \dodoi{10.1088/0004-637X/698/2/2131}

% type= article
\bibitem[{T. {Ding} {et~al.}(2025){Ding}, {Zhang}, \& {Hou}}]{Ding25}
{Ding}, T., {Zhang}, J., \& {Hou}, Y. 2025, \bibinfo{title}{{The Giant Eruption
  in Solar Cycle 25 Caused by Collisional Shearing},} \apjl, 985, L16,
  \dodoi{10.3847/2041-8213/add32c}

% type= article
\bibitem[{M. {Dominique} {et~al.}(2018){Dominique}, {Zhukov}, {Heinzel},
  {Dammasch}, {Wauters}, {Dolla}, {Shestov}, {Kretzschmar}, {Machol},
  {Lapenta}, \& {Schmutz}}]{Dominique2018}
{Dominique}, M., {Zhukov}, A.~N., {Heinzel}, P., {et~al.} 2018,
  \bibinfo{title}{{First Detection of Solar Flare Emission in Mid-ultraviolet
  Balmer Continuum},} \apjl, 867, L24, \dodoi{10.3847/2041-8213/aaeace}

% type= article
\bibitem[{A. {Donati-Falchi} {et~al.}(1985){Donati-Falchi}, {Falciani}, \&
  {Smaldone}}]{Donati1985}
{Donati-Falchi}, A., {Falciani}, R., \& {Smaldone}, L.~A. 1985,
  \bibinfo{title}{{Analysis of the optical spectra of solar flares. IV - The
  'blue' continuum of white light flares},} \aap, 152, 165

% type= article
\bibitem[{J.~G. {Doyle} \& K.~J.~H. {Phillips}(1992){Doyle} \&
  {Phillips}}]{Doyle1992B}
{Doyle}, J.~G., \& {Phillips}, K.~J.~H. 1992, \bibinfo{title}{{Excitation of
  the solar flare far-ultraviolet continuum by line irradiation},} \aap, 257,
  773

% type= article
\bibitem[{M.~K. {Druett} \& V.~V. {Zharkova}(2018){Druett} \&
  {Zharkova}}]{Druett2018}
{Druett}, M.~K., \& {Zharkova}, V.~V. 2018, \bibinfo{title}{{HYDRO2GEN:
  Non-thermal hydrogen Balmer and Paschen emission in solar flares generated by
  electron beams},} \aap, 610, A68, \dodoi{10.1051/0004-6361/201731053}

% type= article
\bibitem[{P. {Feautrier}(1964){Feautrier}}]{Feautrier1964}
{Feautrier}, P. 1964, \bibinfo{title}{{Sur la resolution numerique de
  l'equation de transfert.},} Comptes Rendus Academie des Sciences (serie non
  specifiee), 258, 3189

% type= article
\bibitem[{G.~H. {Fisher}(1989){Fisher}}]{Fisher1989}
{Fisher}, G.~H. 1989, \bibinfo{title}{{Dynamics of flare-driven chromospheric
  condensations},} \apj, 346, 1019, \dodoi{10.1086/168084}

% type= article
\bibitem[{G.~H. {Fisher} {et~al.}(1985{\natexlab{a}}){Fisher}, {Canfield}, \&
  {McClymont}}]{Fisher1985V}
{Fisher}, G.~H., {Canfield}, R.~C., \& {McClymont}, A.~N. 1985{\natexlab{a}},
  \bibinfo{title}{{Flare loop radiative hydrodynamics. V - Response to
  thick-target heating. VI - Chromospheric evaporation due to heating by
  nonthermal electrons. VII - Dynamics of the thick-target heated
  chromosphere},} \apj, 289, 414, \dodoi{10.1086/162901}

% type= article
\bibitem[{G.~H. {Fisher} {et~al.}(1985{\natexlab{b}}){Fisher}, {Canfield}, \&
  {McClymont}}]{Fisher1985VI}
{Fisher}, G.~H., {Canfield}, R.~C., \& {McClymont}, A.~N. 1985{\natexlab{b}},
  \bibinfo{title}{{Flare Loop Radiative Hydrodynamics - Part Six -
  Chromospheric Evaporation due to Heating by Nonthermal Electrons},} \apj,
  289, 425, \dodoi{10.1086/162902}

% type= article
\bibitem[{G.~H. {Fisher} {et~al.}(1985{\natexlab{c}}){Fisher}, {Canfield}, \&
  {McClymont}}]{Fisher1985VII}
{Fisher}, G.~H., {Canfield}, R.~C., \& {McClymont}, A.~N. 1985{\natexlab{c}},
  \bibinfo{title}{{Flare Loop Radiative Hydrodynamics - Part Seven - Dynamics
  of the Thick Target Heated Chromosphere},} \apj, 289, 434,
  \dodoi{10.1086/162903}

% type= article
\bibitem[{L. {Fletcher} {et~al.}(2007){Fletcher}, {Hannah}, {Hudson}, \&
  {Metcalf}}]{Fletcher2007}
{Fletcher}, L., {Hannah}, I.~G., {Hudson}, H.~S., \& {Metcalf}, T.~R. 2007,
  \bibinfo{title}{{A TRACE White Light and RHESSI Hard X-Ray Study of Flare
  Energetics},} \apj, 656, 1187, \dodoi{10.1086/510446}

% type= article
\bibitem[{L. {Fletcher} \& H.~S. {Hudson}(2008){Fletcher} \&
  {Hudson}}]{Fletcher2008}
{Fletcher}, L., \& {Hudson}, H.~S. 2008, \bibinfo{title}{{Impulsive Phase Flare
  Energy Transport by Large-Scale Alfv{\'e}n Waves and the Electron
  Acceleration Problem},} \apj, 675, 1645, \dodoi{10.1086/527044}

% type= article
\bibitem[{L. {Fletcher} {et~al.}(2011){Fletcher}, {Dennis}, {Hudson},
  {Krucker}, {Phillips}, {Veronig}, {Battaglia}, {Bone}, {Caspi}, {Chen},
  {Gallagher}, {Grigis}, {Ji}, {Liu}, {Milligan}, \& {Temmer}}]{Fletcher2011}
{Fletcher}, L., {Dennis}, B.~R., {Hudson}, H.~S., {et~al.} 2011,
  \bibinfo{title}{{An Observational Overview of Solar Flares},} \ssr, 159, 19,
  \dodoi{10.1007/s11214-010-9701-8}

% type= article
\bibitem[{W.-Q. {Gan} {et~al.}(2019){Gan}, {Zhu}, {Deng}, {Li}, {Su}, {Zhang},
  {Chen}, {Zhang}, {Wu}, {Deng}, {Huang}, {Yang}, {Cui}, {Chang}, {Wang}, {Wu},
  {Yin}, {Chen}, {Fang}, {Yan}, {Lin}, {Xiong}, {Chen}, {Bao}, {Cao}, {Bai},
  {Wang}, {Chen}, {Li}, {Zhang}, {Feng}, {Su}, {Li}, {Chen}, {Li}, {Su}, {Wu},
  {Gu}, {Huang}, \& {Tang}}]{Gan2019}
{Gan}, W.-Q., {Zhu}, C., {Deng}, Y.-Y., {et~al.} 2019,
  \bibinfo{title}{{Advanced Space-based Solar Observatory (ASO-S): an
  overview},} Research in Astronomy and Astrophysics, 19, 156,
  \dodoi{10.1088/1674-4527/19/11/156}

% type= article
\bibitem[{M. {Garc{\'\i}a-Rivas} {et~al.}(2024){Garc{\'\i}a-Rivas},
  {Ka{\v{s}}parov{\'a}}, {Berlicki}, {{\v{S}}vanda}, {Dud{\'\i}k},
  {{\v{C}}tvrte{\v{c}}ka}, {Zapi{\'o}r}, {Liu}, {Sobotka}, {Pavelkov{\'a}}, \&
  {Motorina}}]{GarciaRivas2024}
{Garc{\'\i}a-Rivas}, M., {Ka{\v{s}}parov{\'a}}, J., {Berlicki}, A., {et~al.}
  2024, \bibinfo{title}{{Flare heating of the chromosphere: Observations of
  flare continuum from GREGOR and IRIS},} \aap, 690, A254,
  \dodoi{10.1051/0004-6361/202451219}

% type= article
\bibitem[{D.~R. {Graham} \& G. {Cauzzi}(2015){Graham} \& {Cauzzi}}]{Graham2015}
{Graham}, D.~R., \& {Cauzzi}, G. 2015, \bibinfo{title}{{Temporal Evolution of
  Multiple Evaporating Ribbon Sources in a Solar Flare},} \apjl, 807, L22,
  \dodoi{10.1088/2041-8205/807/2/L22}

% type= article
\bibitem[{D.~R. {Graham} {et~al.}(2020){Graham}, {Cauzzi}, {Zangrilli},
  {Kowalski}, {Sim{\~o}es}, \& {Allred}}]{Graham2020}
{Graham}, D.~R., {Cauzzi}, G., {Zangrilli}, L., {et~al.} 2020,
  \bibinfo{title}{{Spectral Signatures of Chromospheric Condensation in a Major
  Solar Flare},} \apj, 895, 6, \dodoi{10.3847/1538-4357/ab88ad}

% type= article
\bibitem[{C.~C. {Haggerty} {et~al.}(2015){Haggerty}, {Shay}, {Drake}, {Phan},
  \& {McHugh}}]{Haggerty2015}
{Haggerty}, C.~C., {Shay}, M.~A., {Drake}, J.~F., {Phan}, T.~D., \& {McHugh},
  C.~T. 2015, \bibinfo{title}{{The competition of electron and ion heating
  during magnetic reconnection},} \grl, 42, 9657, \dodoi{10.1002/2015GL065961}

% type= article
\bibitem[{J. {Halenka} \& B. {Grabowski}(1984){Halenka} \&
  {Grabowski}}]{Halenka1984}
{Halenka}, J., \& {Grabowski}, B. 1984, \bibinfo{title}{{Atomic partition
  functions for iron},} \aaps, 57, 43

% type= article
\bibitem[{S.~L. {Hawley} \& B.~R. {Pettersen}(1991){Hawley} \&
  {Pettersen}}]{HP91}
{Hawley}, S.~L., \& {Pettersen}, B.~R. 1991, \bibinfo{title}{{The great flare
  of 1985 April 12 on AD Leonis},} \apj, 378, 725, \dodoi{10.1086/170474}

% type= article
\bibitem[{P. {Heinzel} \& L. {Kleint}(2014){Heinzel} \& {Kleint}}]{Heinzel2014}
{Heinzel}, P., \& {Kleint}, L. 2014, \bibinfo{title}{{Hydrogen Balmer Continuum
  in Solar Flares Detected by the Interface Region Imaging Spectrograph
  (IRIS)},} \apjl, 794, L23, \dodoi{10.1088/2041-8205/794/2/L23}

% type= article
\bibitem[{G.~D. {Holman} {et~al.}(2011){Holman}, {Aschwanden}, {Aurass},
  {Battaglia}, {Grigis}, {Kontar}, {Liu}, {Saint-Hilaire}, \&
  {Zharkova}}]{Holman2011}
{Holman}, G.~D., {Aschwanden}, M.~J., {Aurass}, H., {et~al.} 2011,
  \bibinfo{title}{{Implications of X-ray Observations for Electron Acceleration
  and Propagation in Solar Flares},} \ssr, 159, 107,
  \dodoi{10.1007/s11214-010-9680-9}

% type= article
\bibitem[{H.~S. {Hudson}(2016){Hudson}}]{Hudson2016}
{Hudson}, H.~S. 2016, \bibinfo{title}{{Chasing White-Light Flares},} \solphys,
  291, 1273, \dodoi{10.1007/s11207-016-0904-3}

% type= article
\bibitem[{K. {Ichimoto} \& H. {Kurokawa}(1984){Ichimoto} \&
  {Kurokawa}}]{Ichimoto1984}
{Ichimoto}, K., \& {Kurokawa}, H. 1984, \bibinfo{title}{{H-alpha red asymmetry
  of solar flares},} \solphys, 93, 105, \dodoi{10.1007/BF00156656}

% type= article
\bibitem[{S.~A. {Jaeggli} \& A.~N. {Daw}(2024){Jaeggli} \& {Daw}}]{Jaeggli2024}
{Jaeggli}, S.~A., \& {Daw}, A.~N. 2024, \bibinfo{title}{{Molecular Hydrogen
  Line Identifications in Solar Flares Observed by IRIS: Lower Atmospheric
  Structure from Radiometric Analysis},} \apj, 976, 18,
  \dodoi{10.3847/1538-4357/ad8445}

% type= article
\bibitem[{D.~B. {Jess} {et~al.}(2008){Jess}, {Mathioudakis}, {Crockett}, \&
  {Keenan}}]{Jess2008}
{Jess}, D.~B., {Mathioudakis}, M., {Crockett}, P.~J., \& {Keenan}, F.~P. 2008,
  \bibinfo{title}{{Do All Flares Have White-Light Emission?},} \apjl, 688,
  L119, \dodoi{10.1086/595588}

% type= article
\bibitem[{J. {Ka{\v{s}}parov{\'a}} {et~al.}(2009){Ka{\v{s}}parov{\'a}},
  {Varady}, {Heinzel}, {Karlick{\'y}}, \& {Moravec}}]{Kasparova2009}
{Ka{\v{s}}parov{\'a}}, J., {Varady}, M., {Heinzel}, P., {Karlick{\'y}}, M., \&
  {Moravec}, Z. 2009, \bibinfo{title}{{Response of optical hydrogen lines to
  beam heating. I. Electron beams},} \aap, 499, 923,
  \dodoi{10.1051/0004-6361/200811559}

% type= article
\bibitem[{M.~D. {Kazachenko} {et~al.}(2022){Kazachenko}, {Albelo-Corchado},
  {Tamburri}, \& {Welsch}}]{Kazachenko2022}
{Kazachenko}, M.~D., {Albelo-Corchado}, M.~F., {Tamburri}, C.~A., \& {Welsch},
  B.~T. 2022, \bibinfo{title}{{Invited Review: Short-term Variability with the
  Observations from the Helioseismic and Magnetic Imager (HMI) Onboard the
  Solar Dynamics Observatory (SDO): Insights into Flare Magnetism},} \solphys,
  297, 59, \dodoi{10.1007/s11207-022-01987-6}

% type= article
\bibitem[{M.~B. {Kennedy} {et~al.}(2015){Kennedy}, {Milligan}, {Allred},
  {Mathioudakis}, \& {Keenan}}]{Kennedy2015}
{Kennedy}, M.~B., {Milligan}, R.~O., {Allred}, J.~C., {Mathioudakis}, M., \&
  {Keenan}, F.~P. 2015, \bibinfo{title}{{Radiative hydrodynamic modelling and
  observations of the X-class solar flare on 2011 March 9},} \aap, 578, A72,
  \dodoi{10.1051/0004-6361/201425144}

% type= article
\bibitem[{G.~S. {Kerr} {et~al.}(2019){Kerr}, {Carlsson}, {Allred}, {Young}, \&
  {Daw}}]{Kerr2019Si}
{Kerr}, G.~S., {Carlsson}, M., {Allred}, J.~C., {Young}, P.~R., \& {Daw}, A.~N.
  2019, \bibinfo{title}{{SI IV Resonance Line Emission during Solar Flares:
  Non-LTE, Nonequilibrium, Radiation Transfer Simulations},} \apj, 871, 23,
  \dodoi{10.3847/1538-4357/aaf46e}

% type= article
\bibitem[{G.~S. {Kerr} {et~al.}(2024{\natexlab{a}}){Kerr}, {Kowalski},
  {Allred}, {Daw}, \& {Kane}}]{Kerr2024A}
{Kerr}, G.~S., {Kowalski}, A.~F., {Allred}, J.~C., {Daw}, A.~N., \& {Kane},
  M.~R. 2024{\natexlab{a}}, \bibinfo{title}{{An optically thin view of the
  flaring chromosphere: non-thermal widths in a chromospheric condensation
  during an X-class solar flare},} \mnras, 527, 2523,
  \dodoi{10.1093/mnras/stad3135}

% type= article
\bibitem[{G.~S. {Kerr} {et~al.}(2024{\natexlab{b}}){Kerr}, {Polito}, {Xu}, \&
  {Allred}}]{Kerr2024RibbonsII}
{Kerr}, G.~S., {Polito}, V., {Xu}, Y., \& {Allred}, J.~C. 2024{\natexlab{b}},
  \bibinfo{title}{{Solar Flare Ribbon Fronts. II. Evolution of Heating Rates in
  Individual Flare Footpoints},} \apj, 970, 21,
  \dodoi{10.3847/1538-4357/ad47e1}

% type= article
\bibitem[{L. {Kleint} {et~al.}(2016){Kleint}, {Heinzel}, {Judge}, \&
  {Krucker}}]{Kleint2016}
{Kleint}, L., {Heinzel}, P., {Judge}, P., \& {Krucker}, S. 2016,
  \bibinfo{title}{{Continuum Enhancements in the Ultraviolet, the Visible and
  the Infrared during the X1 Flare on 2014 March 29},} \apj, 816, 88,
  \dodoi{10.3847/0004-637X/816/2/88}

% type= article
\bibitem[{L. {Kleint} {et~al.}(2017){Kleint}, {Heinzel}, \&
  {Krucker}}]{Kleint2017}
{Kleint}, L., {Heinzel}, P., \& {Krucker}, S. 2017, \bibinfo{title}{{On the
  Origin of the Flare Emission in IRIS{\textquoteright} SJI 2832 Filter:Balmer
  Continuum or Spectral Lines?},} \apj, 837, 160,
  \dodoi{10.3847/1538-4357/aa62fe}

% type= article
\bibitem[{A.~G. {Kosovichev} \& V.~V. {Zharkova}(2001){Kosovichev} \&
  {Zharkova}}]{Kosovichev2001}
{Kosovichev}, A.~G., \& {Zharkova}, V.~V. 2001, \bibinfo{title}{{Magnetic
  Energy Release and Transients in the Solar Flare of 2000 July 14},} \apjl,
  550, L105, \dodoi{10.1086/319484}

% type= article
\bibitem[{A.~F. {Kowalski}(2022){Kowalski}}]{Kowalski2022Frontiers}
{Kowalski}, A.~F. 2022, \bibinfo{title}{{Near-ultraviolet continuum modeling of
  the 1985 April 12 great flare of AD Leo},} Frontiers in Astronomy and Space
  Sciences, 9, 1034458, \dodoi{10.3389/fspas.2022.1034458}

% type= article
\bibitem[{A.~F. {Kowalski} {et~al.}(2024){Kowalski}, {Allred}, \&
  {Carlsson}}]{KAC24}
{Kowalski}, A.~F., {Allred}, J.~C., \& {Carlsson}, M. 2024,
  \bibinfo{title}{{Time-dependent Stellar Flare Models of Deep Atmospheric
  Heating},} \apj, 969, 121, \dodoi{10.3847/1538-4357/ad4148}

% type= article
\bibitem[{A.~F. {Kowalski} {et~al.}(2022){Kowalski}, {Allred}, {Carlsson},
  {Kerr}, {Tremblay}, {Namekata}, {Kuridze}, \& {Uitenbroek}}]{Kowalski2022}
{Kowalski}, A.~F., {Allred}, J.~C., {Carlsson}, M., {et~al.} 2022,
  \bibinfo{title}{{The Atmospheric Response to High Nonthermal Electron-beam
  Fluxes in Solar Flares. II. Hydrogen-broadening Predictions for Solar Flare
  Observations with the Daniel K. Inouye Solar Telescope},} \apj, 928, 190,
  \dodoi{10.3847/1538-4357/ac5174}

% type= article
\bibitem[{A.~F. {Kowalski} {et~al.}(2017{\natexlab{a}}){Kowalski}, {Allred},
  {Daw}, {Cauzzi}, \& {Carlsson}}]{Kowalski2017Mar29}
{Kowalski}, A.~F., {Allred}, J.~C., {Daw}, A., {Cauzzi}, G., \& {Carlsson}, M.
  2017{\natexlab{a}}, \bibinfo{title}{{The Atmospheric Response to High
  Nonthermal Electron Beam Fluxes in Solar Flares. I. Modeling the Brightest
  NUV Footpoints in the X1 Solar Flare of 2014 March 29},} \apj, 836, 12,
  \dodoi{10.3847/1538-4357/836/1/12}

% type= article
\bibitem[{A.~F. {Kowalski} {et~al.}(2019){Kowalski}, {Butler}, {Daw},
  {Fletcher}, {Allred}, {De Pontieu}, {Kerr}, \& {Cauzzi}}]{Kowalski2019IRIS}
{Kowalski}, A.~F., {Butler}, E., {Daw}, A.~N., {et~al.} 2019,
  \bibinfo{title}{{Spectral Evidence for Heating at Large Column Mass in Umbral
  Solar Flare Kernels. I. IRIS Near-UV Spectra of the X1 Solar Flare of 2014
  October 25},} \apj, 878, 135, \dodoi{10.3847/1538-4357/ab1f8b}

% type= article
\bibitem[{A.~F. {Kowalski} {et~al.}(2015{\natexlab{a}}){Kowalski}, {Cauzzi}, \&
  {Fletcher}}]{KCF15}
{Kowalski}, A.~F., {Cauzzi}, G., \& {Fletcher}, L. 2015{\natexlab{a}},
  \bibinfo{title}{{Optical Spectral Observations of a Flickering White-light
  Kernel in a C1 Solar Flare},} \apj, 798, 107,
  \dodoi{10.1088/0004-637X/798/2/107}

% type= article
\bibitem[{A.~F. {Kowalski} {et~al.}(2015{\natexlab{b}}){Kowalski}, {Hawley},
  {Carlsson}, {Allred}, {Uitenbroek}, {Osten}, \& {Holman}}]{Kowalski2015}
{Kowalski}, A.~F., {Hawley}, S.~L., {Carlsson}, M., {et~al.}
  2015{\natexlab{b}}, \bibinfo{title}{{New Insights into White-Light Flare
  Emission from Radiative-Hydrodynamic Modeling of a Chromospheric
  Condensation},} \solphys, 290, 3487, \dodoi{10.1007/s11207-015-0708-x}

% type= article
\bibitem[{A.~F. {Kowalski} {et~al.}(2025){Kowalski}, {Osten}, {Notsu},
  {Tristan}, {Segura}, {Maehara}, {Namekata}, \& {Inoue}}]{Kowalski2025}
{Kowalski}, A.~F., {Osten}, R.~A., {Notsu}, Y., {et~al.} 2025,
  \bibinfo{title}{{Rising Near-ultraviolet Spectra in Stellar Megaflares},}
  \apj, 978, 81, \dodoi{10.3847/1538-4357/ad9395}

% type= article
\bibitem[{A.~F. {Kowalski} {et~al.}(2017{\natexlab{b}}){Kowalski}, {Allred},
  {Uitenbroek}, {Tremblay}, {Brown}, {Carlsson}, {Osten}, {Wisniewski}, \&
  {Hawley}}]{Kowalski2017Broadening}
{Kowalski}, A.~F., {Allred}, J.~C., {Uitenbroek}, H., {et~al.}
  2017{\natexlab{b}}, \bibinfo{title}{{Hydrogen Balmer Line Broadening in Solar
  and Stellar Flares},} \apj, 837, 125, \dodoi{10.3847/1538-4357/aa603e}

% type= article
\bibitem[{M. {Kretzschmar}(2011){Kretzschmar}}]{Kretzschmar2011}
{Kretzschmar}, M. 2011, \bibinfo{title}{{The Sun as a star: observations of
  white-light flares},} \aap, 530, A84, \dodoi{10.1051/0004-6361/201015930}

% type= article
\bibitem[{S. {Krucker} {et~al.}(2011){Krucker}, {Hudson}, {Jeffrey},
  {Battaglia}, {Kontar}, {Benz}, {Csillaghy}, \& {Lin}}]{Krucker2011}
{Krucker}, S., {Hudson}, H.~S., {Jeffrey}, N.~L.~S., {et~al.} 2011,
  \bibinfo{title}{{High-resolution Imaging of Solar Flare Ribbons and Its
  Implication on the Thick-target Beam Model},} \apj, 739, 96,
  \dodoi{10.1088/0004-637X/739/2/96}

% type= article
\bibitem[{S. {Krucker} {et~al.}(2020){Krucker}, {Hurford}, {Grimm}, {K{\"o}gl},
  {Gr{\"o}belbauer}, {Etesi}, {Casadei}, {Csillaghy}, {Benz}, {Arnold},
  {Molendini}, {Orleanski}, {Schori}, {Xiao}, {Kuhar}, {Hochmuth}, {Felix},
  {Schramka}, {Marcin}, {Kobler}, {Iseli}, {Dreier}, {Wiehl}, {Kleint},
  {Battaglia}, {Lastufka}, {Sathiapal}, {Lapadula}, {Bednarzik}, {Birrer},
  {Stutz}, {Wild}, {Marone}, {Skup}, {Cichocki}, {Ber}, {Rutkowski}, {Bujwan},
  {Juchnikowski}, {Winkler}, {Darmetko}, {Michalska}, {Seweryn}, {Bia{\l}ek},
  {Osica}, {Sylwester}, {Kowalinski}, {{\'S}cis{\l}owski}, {Siarkowski},
  {St{\k{e}}{\'s}licki}, {Mrozek}, {Podg{\'o}rski}, {Meuris}, {Limousin},
  {Gevin}, {Le Mer}, {Brun}, {Strugarek}, {Vilmer}, {Musset}, {Maksimovi{\'c}},
  {F{\'a}rn{\'\i}k}, {Koz{\'a}{\v{c}}ek}, {Ka{\v{s}}parov{\'a}}, {Mann},
  {{\"O}nel}, {Warmuth}, {Rendtel}, {Anderson}, {Bauer}, {Dionies}, {Paschke},
  {Pl{\"u}schke}, {Woche}, {Schuller}, {Veronig}, {Dickson}, {Gallagher},
  {Maloney}, {Bloomfield}, {Piana}, {Massone}, {Benvenuto}, {Massa},
  {Schwartz}, {Dennis}, {van Beek}, {Rodr{\'\i}guez-Pacheco}, \& {Lin}}]{STIX1}
{Krucker}, S., {Hurford}, G.~J., {Grimm}, O., {et~al.} 2020,
  \bibinfo{title}{{The Spectrometer/Telescope for Imaging X-rays (STIX)},}
  \aap, 642, A15, \dodoi{10.1051/0004-6361/201937362}

% type= article
\bibitem[{D. {Kuridze} {et~al.}(2015){Kuridze}, {Mathioudakis}, {Sim{\~o}es},
  {Rouppe van der Voort}, {Carlsson}, {Jafarzadeh}, {Allred}, {Kowalski},
  {Kennedy}, {Fletcher}, {Graham}, \& {Keenan}}]{Kuridze2015}
{Kuridze}, D., {Mathioudakis}, M., {Sim{\~o}es}, P.~J.~A., {et~al.} 2015,
  \bibinfo{title}{{H{$\alpha$} Line Profile Asymmetries and the Chromospheric
  Flare Velocity Field},} \apj, 813, 125, \dodoi{10.1088/0004-637X/813/2/125}

% type= article
\bibitem[{K.~W. {Lee} \& J. {B{\"u}chner}(2011){Lee} \&
  {B{\"u}chner}}]{LeeBuchner2011}
{Lee}, K.~W., \& {B{\"u}chner}, J. 2011, \bibinfo{title}{{Collisionless
  turbulent transport and anisotropic electron heating in coronal flare
  loops},} \aap, 535, A61, \dodoi{10.1051/0004-6361/201117186}

% type= article
\bibitem[{K.~W. {Lee} {et~al.}(2008){Lee}, {B{\"u}chner}, \&
  {Elkina}}]{Lee2008}
{Lee}, K.~W., {B{\"u}chner}, J., \& {Elkina}, N. 2008,
  \bibinfo{title}{{Collisionless transport of energetic electrons in the solar
  corona at current-free double layers},} \aap, 478, 889,
  \dodoi{10.1051/0004-6361:20078419}

% type= article
\bibitem[{J.~R. {Lemen} {et~al.}(2012){Lemen}, {Title}, {Akin}, {Boerner},
  {Chou}, {Drake}, {Duncan}, {Edwards}, {Friedlaender}, {Heyman}, {Hurlburt},
  {Katz}, {Kushner}, {Levay}, {Lindgren}, {Mathur}, {McFeaters}, {Mitchell},
  {Rehse}, {Schrijver}, {Springer}, {Stern}, {Tarbell}, {Wuelser}, {Wolfson},
  {Yanari}, {Bookbinder}, {Cheimets}, {Caldwell}, {Deluca}, {Gates}, {Golub},
  {Park}, {Podgorski}, {Bush}, {Scherrer}, {Gummin}, {Smith}, {Auker},
  {Jerram}, {Pool}, {Soufli}, {Windt}, {Beardsley}, {Clapp}, {Lang}, \&
  {Waltham}}]{Lemen2012}
{Lemen}, J.~R., {Title}, A.~M., {Akin}, D.~J., {et~al.} 2012,
  \bibinfo{title}{{The Atmospheric Imaging Assembly (AIA) on the Solar Dynamics
  Observatory (SDO)},} \solphys, 275, 17, \dodoi{10.1007/s11207-011-9776-8}

% type= article
\bibitem[{C. {Li} {et~al.}(2019){Li}, {Fang}, {Li}, {Ding}, {Chen}, {Chen},
  {Lin}, {Chen}, {Chen}, {Tao}, {You}, {Hao}, {Dai}, {Cheng}, {Guo}, {Hong},
  {An}, {Cheng}, {Chen}, {Wang}, \& {Zhang}}]{CHASEOverview1}
{Li}, C., {Fang}, C., {Li}, Z., {et~al.} 2019, \bibinfo{title}{{Chinese
  H{\ensuremath{\alpha}} Solar Explorer (CHASE) - a complementary space mission
  to the ASO-S},} Research in Astronomy and Astrophysics, 19, 165,
  \dodoi{10.1088/1674-4527/19/11/165}

% type= article
\bibitem[{C. {Li} {et~al.}(2022){Li}, {Fang}, {Li}, {Ding}, {Chen}, {Qiu},
  {You}, {Yuan}, {An}, {Tao}, {Li}, {Chen}, {Liu}, {Mei}, {Yang}, {Zhang},
  {Cheng}, {Chen}, {Chen}, {Gu}, {Huang}, {Liu}, {Han}, {Xin}, {Chen}, {Ni},
  {Wang}, {Rao}, {Li}, {Lu}, {Wang}, {Lin}, {Jiang}, {Meng}, \&
  {Zhao}}]{CHASEOverview2}
{Li}, C., {Fang}, C., {Li}, Z., {et~al.} 2022, \bibinfo{title}{{The Chinese
  H{\ensuremath{\alpha}} Solar Explorer (CHASE) mission: An overview},} Science
  China Physics, Mechanics, and Astronomy, 65, 289602,
  \dodoi{10.1007/s11433-022-1893-3}

% type= article
\bibitem[{D. {Li}(2025){Li}}]{Li2025}
{Li}, D. 2025, \bibinfo{title}{{Localizing short-period pulsations in hard
  X-rays and {\ensuremath{\gamma}}-rays during an X9.0 flare},} \aap, 695, L4,
  \dodoi{10.1051/0004-6361/202453613}

% type= article
\bibitem[{T.~C. {Li} {et~al.}(2014){Li}, {Drake}, \& {Swisdak}}]{Li2014}
{Li}, T.~C., {Drake}, J.~F., \& {Swisdak}, M. 2014, \bibinfo{title}{{Dynamics
  of Double Layers, Ion Acceleration, and Heat Flux Suppression during Solar
  Flares},} \apj, 793, 7, \dodoi{10.1088/0004-637X/793/1/7}

% type= article
\bibitem[{T. {Libbrecht} {et~al.}(2019){Libbrecht}, {de la Cruz
  Rodr{\'\i}guez}, {Danilovic}, {Leenaarts}, \& {Pazira}}]{Libbrecht}
{Libbrecht}, T., {de la Cruz Rodr{\'\i}guez}, J., {Danilovic}, S., {Leenaarts},
  J., \& {Pazira}, H. 2019, \bibinfo{title}{{Chromospheric condensations and
  magnetic field in a C3.6-class flare studied via He I D$_{3}$
  spectro-polarimetry},} \aap, 621, A35, \dodoi{10.1051/0004-6361/201833610}

% type= article
\bibitem[{M. {Litwicka} {et~al.}(2025){Litwicka}, {Heinzel}, \&
  {Ka{\v{s}}parov{\'a}}}]{Litwicka}
{Litwicka}, M., {Heinzel}, P., \& {Ka{\v{s}}parov{\'a}}, J. 2025,
  \bibinfo{title}{{Filamentary Electron-beam Heating in Solar Flares},} \apj,
  983, 155, \dodoi{10.3847/1538-4357/adc393}

% type= article
\bibitem[{C. {Liu} {et~al.}(2007){Liu}, {Lee}, {Gary}, \& {Wang}}]{Liu2007}
{Liu}, C., {Lee}, J., {Gary}, D.~E., \& {Wang}, H. 2007, \bibinfo{title}{{The
  Ribbon-like Hard X-Ray Emission in a Sigmoidal Solar Active Region},} \apjl,
  658, L127, \dodoi{10.1086/513739}

% type= article
\bibitem[{Q. {Liu} {et~al.}(2022){Liu}, {Tao}, {Chen}, {Han}, {Chen}, {Mei},
  {Yang}, {Hu}, {Xin}, {Li}, {Guan}, {Xue}, {Zhu}, {Hu}, {Ha}, {He}, {Fang},
  {Li}, \& {Li}}]{CHASEOverview3}
{Liu}, Q., {Tao}, H., {Chen}, C., {et~al.} 2022, \bibinfo{title}{{On the
  technologies of H{\ensuremath{\alpha}} imaging spectrograph for the CHASE
  mission},} Science China Physics, Mechanics, and Astronomy, 65, 289605,
  \dodoi{10.1007/s11433-022-1917-1}

% type= article
\bibitem[{M.~A. {Livshits} {et~al.}(1981){Livshits}, {Badalian}, {Kosovichev},
  \& {Katsova}}]{Livshits1981}
{Livshits}, M.~A., {Badalian}, O.~G., {Kosovichev}, A.~G., \& {Katsova}, M.~M.
  1981, \bibinfo{title}{{The Optical Continuum of Solar and Stellar Flares},}
  \solphys, 73, 269, \dodoi{10.1007/BF00151682}

% type= article
\bibitem[{D.~W. {Longcope} \& S.~E. {Guidoni}(2011){Longcope} \&
  {Guidoni}}]{Longcope2011}
{Longcope}, D.~W., \& {Guidoni}, S.~E. 2011, \bibinfo{title}{{A Model for the
  Origin of High Density in Looptop X-Ray Sources},} \apj, 740, 73,
  \dodoi{10.1088/0004-637X/740/2/73}

% type= article
\bibitem[{J. {L{\"o}rin{\v{c}}{\'\i}k} {et~al.}(2025){L{\"o}rin{\v{c}}{\'\i}k},
  {Polito}, {Kerr}, {Hayes}, \& {Russell}}]{Lorincik2025}
{L{\"o}rin{\v{c}}{\'\i}k}, J., {Polito}, V., {Kerr}, G.~S., {Hayes}, L.~A., \&
  {Russell}, A. J.~B. 2025, \bibinfo{title}{{Probing Progression of Heating
  Through the Lower Flare Atmosphere via High-cadence IRIS Spectroscopy},}
  \apj, 986, 73, \dodoi{10.3847/1538-4357/adccc8}

% type= article
\bibitem[{R.~A. {Maurya} \& A. {Ambastha}(2009){Maurya} \&
  {Ambastha}}]{Maurya2009}
{Maurya}, R.~A., \& {Ambastha}, A. 2009, \bibinfo{title}{{Transient Magnetic
  and Doppler Features Related to the White-Light Flares in NOAA 10486},}
  \solphys, 258, 31, \dodoi{10.1007/s11207-009-9397-7}

% type= book
\bibitem[{D. {Mihalas}(1978){Mihalas}}]{Mihalas1978}
{Mihalas}, D. 1978, {Stellar atmospheres /2nd edition/}

% type= article
\bibitem[{D. {Mihalas}(1985){Mihalas}}]{Mihalas1985}
{Mihalas}, D. 1985, \bibinfo{title}{{The computation of radiation transport
  using Feautrier variables. I - Static media},} Journal of Computational
  Physics, 57, 1, \dodoi{10.1016/0021-9991(85)90050-6}

% type= article
\bibitem[{D. {M{\"u}ller} {et~al.}(2020){M{\"u}ller}, {St. Cyr}, {Zouganelis},
  {Gilbert}, {Marsden}, {Nieves-Chinchilla}, {Antonucci}, {Auch{\`e}re},
  {Berghmans}, {Horbury}, {Howard}, {Krucker}, {Maksimovic}, {Owen}, {Rochus},
  {Rodriguez-Pacheco}, {Romoli}, {Solanki}, {Bruno}, {Carlsson}, {Fludra},
  {Harra}, {Hassler}, {Livi}, {Louarn}, {Peter}, {Sch{\"u}hle}, {Teriaca}, {del
  Toro Iniesta}, {Wimmer-Schweingruber}, {Marsch}, {Velli}, {De Groof},
  {Walsh}, \& {Williams}}]{Solo}
{M{\"u}ller}, D., {St. Cyr}, O.~C., {Zouganelis}, I., {et~al.} 2020,
  \bibinfo{title}{{The Solar Orbiter mission. Science overview},} \aap, 642,
  A1, \dodoi{10.1051/0004-6361/202038467}

% type= article
\bibitem[{K. {Namekata} {et~al.}(2022){Namekata}, {Ichimoto}, {Ishii}, \&
  {Shibata}}]{Namekata2022}
{Namekata}, K., {Ichimoto}, K., {Ishii}, T.~T., \& {Shibata}, K. 2022,
  \bibinfo{title}{{Sun-as-a-star Analysis of H{\ensuremath{\alpha}} Spectra of
  a Solar Flare Observed by SMART/SDDI: Time Evolution of Red Asymmetry and
  Line Broadening},} \apj, 933, 209, \dodoi{10.3847/1538-4357/ac75cd}

% type= article
\bibitem[{G. Nave \& S. Johansson(2013)Nave \& Johansson}]{Nave}
Nave, G., \& Johansson, S. 2013, \bibinfo{title}{The Spectrum of Fe II,} The
  Astrophysical Journal Supplement Series, 204, 1.
\newblock \url{http://stacks.iop.org/0067-0049/204/i=1/a=1}

% type= article
\bibitem[{G. {Nave} {et~al.}(1994){Nave}, {Johansson}, {Learner}, {Thorne}, \&
  {Brault}}]{Nave1994}
{Nave}, G., {Johansson}, S., {Learner}, R.~C.~M., {Thorne}, A.~P., \& {Brault},
  J.~W. 1994, \bibinfo{title}{{A New Multiplet Table for Fe i},} \apjs, 94,
  221, \dodoi{10.1086/192079}

% type= article
\bibitem[{H. {Neckel}(1999){Neckel}}]{Neckel1999}
{Neckel}, H. 1999, \bibinfo{title}{{Announcement},} \solphys, 184, 421,
  \dodoi{10.1023/A:1017165208013}

% type= article
\bibitem[{D.~F. {Neidig}(1983){Neidig}}]{Neidig1983}
{Neidig}, D.~F. 1983, \bibinfo{title}{{Spectral analysis of the optical
  continuum in the 24 April 1981 flare},} \solphys, 85, 285,
  \dodoi{10.1007/BF00148655}

% type= article
\bibitem[{D.~F. {Neidig} {et~al.}(1994){Neidig}, {Grosser}, \&
  {Hrovat}}]{Neidig1994}
{Neidig}, D.~F., {Grosser}, H., \& {Hrovat}, M. 1994, \bibinfo{title}{{Optical
  output of the 24 April 1984 white-light flare},} \solphys, 155, 199,
  \dodoi{10.1007/BF00670740}

% type= article
\bibitem[{D.~F. {Neidig} {et~al.}(1993){Neidig}, {Kiplinger}, {Cohl}, \&
  {Wiborg}}]{Neidig1993}
{Neidig}, D.~F., {Kiplinger}, A.~L., {Cohl}, H.~S., \& {Wiborg}, P.~H. 1993,
  \bibinfo{title}{{The Solar White-Light Flare of 1989 March 7: Simultaneous
  Multiwavelength Observations at High Time Resolution},} \apj, 406, 306,
  \dodoi{10.1086/172442}

% type= article
\bibitem[{W.~D. {Pesnell} {et~al.}(2012){Pesnell}, {Thompson}, \&
  {Chamberlin}}]{Pesnell2012}
{Pesnell}, W.~D., {Thompson}, B.~J., \& {Chamberlin}, P.~C. 2012,
  \bibinfo{title}{{The Solar Dynamics Observatory (SDO)},} \solphys, 275, 3,
  \dodoi{10.1007/s11207-011-9841-3}

% type= article
\bibitem[{K.~J.~H. {Phillips} {et~al.}(1992){Phillips}, {Bromage}, \&
  {Doyle}}]{Phillips1992}
{Phillips}, K.~J.~H., {Bromage}, G.~E., \& {Doyle}, J.~G. 1992,
  \bibinfo{title}{{The Origin of the Far-Ultraviolet Continuum in Solar and
  Stellar Flares},} \apj, 385, 731, \dodoi{10.1086/170979}

% type= article
\bibitem[{A.~G.~M. {Pietrow} {et~al.}(2024){Pietrow}, {Druett}, \&
  {Singh}}]{Pietrow2024}
{Pietrow}, A.~G.~M., {Druett}, M.~K., \& {Singh}, V. 2024,
  \bibinfo{title}{{Spectral variations within solar flare ribbons},} \aap, 685,
  A137, \dodoi{10.1051/0004-6361/202348839}

% type= article
\bibitem[{H. {Potts} {et~al.}(2010){Potts}, {Hudson}, {Fletcher}, \&
  {Diver}}]{Potts2010}
{Potts}, H., {Hudson}, H., {Fletcher}, L., \& {Diver}, D. 2010,
  \bibinfo{title}{{The Optical Depth of White-light Flare Continuum},} \apj,
  722, 1514, \dodoi{10.1088/0004-637X/722/2/1514}

% type= article
\bibitem[{O. {Proch{\'a}zka} {et~al.}(2017){Proch{\'a}zka}, {Milligan},
  {Allred}, {Kowalski}, {Kotr{\v c}}, \& {Mathioudakis}}]{Ondrej1}
{Proch{\'a}zka}, O., {Milligan}, R.~O., {Allred}, J.~C., {et~al.} 2017,
  \bibinfo{title}{{Suppression of Hydrogen Emission in an X-class White-light
  Solar Flare},} \apj, 837, 46, \dodoi{10.3847/1538-4357/aa5da8}

% type= article
\bibitem[{J. {Qiu} {et~al.}(2010){Qiu}, {Liu}, {Hill}, \&
  {Kazachenko}}]{Qiu2010}
{Qiu}, J., {Liu}, W., {Hill}, N., \& {Kazachenko}, M. 2010,
  \bibinfo{title}{{Reconnection and Energetics in Two-ribbon Flares: A Revisit
  of the Bastille-day Flare},} \apj, 725, 319,
  \dodoi{10.1088/0004-637X/725/1/319}

% type= article
\bibitem[{J. {Qiu} {et~al.}(2017){Qiu}, {Longcope}, {Cassak}, \&
  {Priest}}]{Qiu2017}
{Qiu}, J., {Longcope}, D.~W., {Cassak}, P.~A., \& {Priest}, E.~R. 2017,
  \bibinfo{title}{{Elongation of Flare Ribbons},} \apj, 838, 17,
  \dodoi{10.3847/1538-4357/aa6341}

% type= article
\bibitem[{Y. {Qiu} {et~al.}(2022){Qiu}, {Rao}, {Li}, {Fang}, {Ding}, {Li},
  {Ni}, {Wang}, {Hong}, {Hao}, {Dai}, {Chen}, {Wan}, {Xu}, {You}, {Yuan},
  {Tao}, {Li}, {He}, \& {Liu}}]{CHASECal}
{Qiu}, Y., {Rao}, S., {Li}, C., {et~al.} 2022, \bibinfo{title}{{Calibration
  procedures for the CHASE/HIS science data},} Science China Physics,
  Mechanics, and Astronomy, 65, 289603, \dodoi{10.1007/s11433-022-1900-5}

% type= article
\bibitem[{K. {Radziszewski} {et~al.}(2007){Radziszewski}, {Rudawy}, \&
  {Phillips}}]{Radz2007}
{Radziszewski}, K., {Rudawy}, P., \& {Phillips}, K.~J.~H. 2007,
  \bibinfo{title}{{High time resolution observations of solar
  H{\ensuremath{\alpha}} flares. I.},} \aap, 461, 303,
  \dodoi{10.1051/0004-6361:20053460}

% type= article
\bibitem[{P.~J. {Ricchiazzi} \& R.~C. {Canfield}(1983){Ricchiazzi} \&
  {Canfield}}]{RC83}
{Ricchiazzi}, P.~J., \& {Canfield}, R.~C. 1983, \bibinfo{title}{{A static model
  of chromospheric heating in solar flares},} \apj, 272, 739,
  \dodoi{10.1086/161336}

% type= article
\bibitem[{F. {Rubio da Costa} {et~al.}(2016){Rubio da Costa}, {Kleint},
  {Petrosian}, {Liu}, \& {Allred}}]{Rubio2016}
{Rubio da Costa}, F., {Kleint}, L., {Petrosian}, V., {Liu}, W., \& {Allred},
  J.~C. 2016, \bibinfo{title}{{Data-driven Radiative Hydrodynamic Modeling of
  the 2014 March 29 X1.0 Solar Flare},} \apj, 827, 38,
  \dodoi{10.3847/0004-637X/827/1/38}

% type= inproceedings
\bibitem[{A.~J.~B. {Russell}(2024){Russell}}]{Russell2024}
{Russell}, A. J.~B. 2024, \bibinfo{title}{{Alfv{\'e}n Waves in Solar Flares},}
  in Alfv{\'e}n Waves Across Heliophysics: Progress, Challenges, and
  Opportunities, ed. A.~{Keiling}, Vol. 285, 39--73,
  \dodoi{10.1002/9781394195985.ch3}

% type= article
\bibitem[{A. {Sainz Dalda} \& B. {De Pontieu}(2023){Sainz Dalda} \& {De
  Pontieu}}]{SainzDalda23}
{Sainz Dalda}, A., \& {De Pontieu}, B. 2023, \bibinfo{title}{{Chromospheric
  thermodynamic conditions from inversions of complex Mg II h \& k profiles
  observed in flares},} Frontiers in Astronomy and Space Sciences, 10, 1133429,
  \dodoi{10.3389/fspas.2023.1133429}

% type= article
\bibitem[{P.~H. {Scherrer} {et~al.}(2012){Scherrer}, {Schou}, {Bush},
  {Kosovichev}, {Bogart}, {Hoeksema}, {Liu}, {Duvall}, {Zhao}, {Title},
  {Schrijver}, {Tarbell}, \& {Tomczyk}}]{Scherrer2012}
{Scherrer}, P.~H., {Schou}, J., {Bush}, R.~I., {et~al.} 2012,
  \bibinfo{title}{{The Helioseismic and Magnetic Imager (HMI) Investigation for
  the Solar Dynamics Observatory (SDO)},} \solphys, 275, 207,
  \dodoi{10.1007/s11207-011-9834-2}

% type= article
\bibitem[{P.~J.~A. {Sim{\~o}es} {et~al.}(2024{\natexlab{a}}){Sim{\~o}es},
  {Ara{\'u}jo}, {V{\'a}lio}, \& {Fletcher}}]{Simoes2024}
{Sim{\~o}es}, P. J.~A., {Ara{\'u}jo}, A., {V{\'a}lio}, A., \& {Fletcher}, L.
  2024{\natexlab{a}}, \bibinfo{title}{{Hydrogen recombination continuum as the
  radiative model for stellar optical flares},} \mnras, 528, 2562,
  \dodoi{10.1093/mnras/stae186}

% type= article
\bibitem[{P.~J.~A. {Sim{\~o}es} {et~al.}(2024{\natexlab{b}}){Sim{\~o}es},
  {Fletcher}, {Hudson}, {Kerr}, {Penn}, \& {Lopez}}]{Simoes2024B}
{Sim{\~o}es}, P. J.~A., {Fletcher}, L., {Hudson}, H.~S., {et~al.}
  2024{\natexlab{b}}, \bibinfo{title}{{Precise timing of solar flare footpoint
  sources from mid-infrared observations},} \mnras, 532, 705,
  \dodoi{10.1093/mnras/stae1511}

% type= article
\bibitem[{P.~J.~A. {Sim{\~o}es} {et~al.}(2019){Sim{\~o}es}, {Reid}, {Milligan},
  \& {Fletcher}}]{Simoes2019}
{Sim{\~o}es}, P. J.~A., {Reid}, H. A.~S., {Milligan}, R.~O., \& {Fletcher}, L.
  2019, \bibinfo{title}{{The Spectral Content of SDO/AIA 1600 and 1700 {\r{A}}
  Filters from Flare and Plage Observations},} \apj, 870, 114,
  \dodoi{10.3847/1538-4357/aaf28d}

% type= article
\bibitem[{Y. {Su} {et~al.}(2019){Su}, {Liu}, {Li}, {Zhang}, {Hurford}, {Chen},
  {Huang}, {Li}, {Jiang}, {Wang}, {Xia}, {Chen}, {Yu}, {Yu}, {Wu}, \&
  {Gan}}]{Su2019}
{Su}, Y., {Liu}, W., {Li}, Y.-P., {et~al.} 2019, \bibinfo{title}{{Simulations
  and software development for the Hard X-ray Imager onboard ASO-S},} Research
  in Astronomy and Astrophysics, 19, 163, \dodoi{10.1088/1674-4527/19/11/163}

% type= book
\bibitem[{Z. {Svestka}(1976){Svestka}}]{Svestka1976}
{Svestka}, Z. 1976, {Solar Flares}

% type= article
\bibitem[{J. {Thoen Faber} {et~al.}(2025){Thoen Faber}, {Joshi}, {Rouppe van
  der Voort}, {Wedemeyer}, {Fletcher}, {Aulanier}, \&
  {N{\'o}brega-Siverio}}]{Faber2025}
{Thoen Faber}, J., {Joshi}, R., {Rouppe van der Voort}, L., {et~al.} 2025,
  \bibinfo{title}{{High-resolution observational analysis of flare ribbon fine
  structures},} \aap, 693, A8, \dodoi{10.1051/0004-6361/202452370}

% type= article
\bibitem[{S. {Toriumi} \& H. {Wang}(2019){Toriumi} \& {Wang}}]{Toriumi2019}
{Toriumi}, S., \& {Wang}, H. 2019, \bibinfo{title}{{Flare-productive active
  regions},} Living Reviews in Solar Physics, 16, 3,
  \dodoi{10.1007/s41116-019-0019-7}

% type= article
\bibitem[{P.-E. {Tremblay} \& P. {Bergeron}(2009){Tremblay} \&
  {Bergeron}}]{Tremblay2009}
{Tremblay}, P.-E., \& {Bergeron}, P. 2009, \bibinfo{title}{{Spectroscopic
  Analysis of DA White Dwarfs: Stark Broadening of Hydrogen Lines Including
  Nonideal Effects},} \apj, 696, 1755, \dodoi{10.1088/0004-637X/696/2/1755}

% type= article
\bibitem[{J.~E. {Vernazza} {et~al.}(1981){Vernazza}, {Avrett}, \&
  {Loeser}}]{Vernazza1981}
{Vernazza}, J.~E., {Avrett}, E.~H., \& {Loeser}, R. 1981,
  \bibinfo{title}{{Structure of the solar chromosphere. III. Models of the EUV
  brightness components of the quiet sun.},} \apjs, 45, 635,
  \dodoi{10.1086/190731}

% type= article
\bibitem[{A. {Warmuth} \& G. {Mann}(2016){Warmuth} \& {Mann}}]{WarmuthMann2016}
{Warmuth}, A., \& {Mann}, G. 2016, \bibinfo{title}{{Constraints on energy
  release in solar flares from RHESSI and GOES X-ray observations. II.
  Energetics and energy partition},} \aap, 588, A116,
  \dodoi{10.1051/0004-6361/201527475}

% type= article
\bibitem[{H.~P. {Warren}(2006){Warren}}]{Warren2006}
{Warren}, H.~P. 2006, \bibinfo{title}{{Multithread Hydrodynamic Modeling of a
  Solar Flare},} \apj, 637, 522, \dodoi{10.1086/497904}

% type= article
\bibitem[{T.~N. {Woods} {et~al.}(2004){Woods}, {Eparvier}, {Fontenla},
  {Harder}, {Kopp}, {McClintock}, {Rottman}, {Smiley}, \& {Snow}}]{Woods2004}
{Woods}, T.~N., {Eparvier}, F.~G., {Fontenla}, J., {et~al.} 2004,
  \bibinfo{title}{{Solar irradiance variability during the October 2003 solar
  storm period},} \grl, 31, L10802, \dodoi{10.1029/2004GL019571}

% type= article
\bibitem[{J.-P. {Wuelser} \& H. {Marti}(1989){Wuelser} \&
  {Marti}}]{Wuelser1989}
{Wuelser}, J.-P., \& {Marti}, H. 1989, \bibinfo{title}{{High Time Resolution
  Observations of H alpha Line Profiles during the Impulsive Phase of a Solar
  Flare},} \apj, 341, 1088, \dodoi{10.1086/167567}

% type= article
\bibitem[{J.-P. {Wuelser} {et~al.}(1994){Wuelser}, {Canfield}, {Acton},
  {Culhane}, {Phillips}, {Fludra}, {Sakao}, {Masuda}, {Kosugi}, \&
  {Tsuneta}}]{Wuelser1994}
{Wuelser}, J.-P., {Canfield}, R.~C., {Acton}, L.~W., {et~al.} 1994,
  \bibinfo{title}{{Multispectral Observations of Chromospheric Evaporation in
  the 1991 November 15 X-Class Solar Flare},} \apj, 424, 459,
  \dodoi{10.1086/173903}

% type= article
\bibitem[{J.~P. {W{\"u}lser} {et~al.}(2018){W{\"u}lser}, {Jaeggli}, {De
  Pontieu}, {Tarbell}, {Boerner}, {Freeland}, {Liu}, {Timmons}, {Brannon},
  {Kankelborg}, {Madsen}, {McKillop}, {Prchlik}, {Saar}, {Schanche}, {Testa},
  {Bryans}, \& {Wiesmann}}]{Wulser2018}
{W{\"u}lser}, J.~P., {Jaeggli}, S., {De Pontieu}, B., {et~al.} 2018,
  \bibinfo{title}{{Instrument Calibration of the Interface Region Imaging
  Spectrograph (IRIS) Mission},} \solphys, 293, 149,
  \dodoi{10.1007/s11207-018-1364-8}

% type= article
\bibitem[{H. {Xiao} {et~al.}(2023){Xiao}, {Maloney}, {Krucker}, {Dickson},
  {Massa}, {Lastufka}, {Francesco Battaglia}, {Etesi}, {Hochmuth}, {Schuller},
  {Ryan}, {Limousin}, {Collier}, {Warmuth}, \& {Piana}}]{STIX2}
{Xiao}, H., {Maloney}, S., {Krucker}, S., {et~al.} 2023, \bibinfo{title}{{The
  data center for the Spectrometer and Telescope for Imaging X-rays (STIX) on
  board Solar Orbiter},} \aap, 673, A142, \dodoi{10.1051/0004-6361/202346031}

% type= article
\bibitem[{X. {Yang} {et~al.}(2025){Yang}, {Cao}, {Wang}, {Jennings}, {Qiu},
  {He}, {Perriyil}, {Yurchyshyn}, {Fletcher}, {Sim{\~o}es}, {Jhabvala},
  {Lunsford}, {Chen}, \& {Hudson}}]{Yang2025}
{Yang}, X., {Cao}, W., {Wang}, M., {et~al.} 2025,
  \bibinfo{title}{{High-resolution Observations of an X6.4 Solar Flare in the
  Mid-infrared},} \apjl, 988, L56, \dodoi{10.3847/2041-8213/adee95}

% type= article
\bibitem[{H.~C. {Yu} {et~al.}(2025){Yu}, {Hong}, \& {Ding}}]{Yu2025}
{Yu}, H.~C., {Hong}, J., \& {Ding}, M.~D. 2025, \bibinfo{title}{{Sun-as-a-star
  analysis of simulated solar flares},} \aap, 694, A315,
  \dodoi{10.1051/0004-6361/202451706}

% type= article
\bibitem[{Y. {Zhu} {et~al.}(2019){Zhu}, {Kowalski}, {Tian}, {Uitenbroek},
  {Carlsson}, \& {Allred}}]{Zhu2019}
{Zhu}, Y., {Kowalski}, A.~F., {Tian}, H., {et~al.} 2019,
  \bibinfo{title}{{Modeling Mg II h, k and Triplet Lines at Solar Flare
  Ribbons},} \apj, 879, 19, \dodoi{10.3847/1538-4357/ab2238}

\end{thebibliography}
\bibliographystyle{aasjournalv7}

%% This command is needed to show the entire author+affiliation list when
%% the collaboration and author truncation commands are used.  It has to
%% go at the end of the manuscript.
%\allauthors

%% Include this line if you are using the \added, \replaced, \deleted
%% commands to see a summary list of all changes at the end of the article.
%\listofchanges

\end{document}